\documentclass[revtex4]{emulateapj}
\usepackage{color}
\usepackage{courier}
\usepackage{natbib} 
\usepackage{amsmath}
\usepackage[caption=false]{subfig}

\def\simlt{\lower.5ex\hbox{$\; \buildrel < \over \sim \;$}}
\def\simgt{\lower.5ex\hbox{$\; \buildrel > \over \sim \;$}}

\def\gsim{\lower 2pt \hbox{$\, \buildrel {\scriptstyle >}\over
{\scriptstyle \sim}\,$}}
\def\lsim{\lower 2pt \hbox{$\, \buildrel {\scriptstyle <}\over
{\scriptstyle \sim}\,$}}

\def\deg{\ifmmode ^{\circ}
         \else $^{\circ}$\fi}
\def\pdeg{\ifmmode
           $\setbox0=\hbox{$^{\circ}$}\rlap{\hskip.11\wd0 .}$^{\circ}
     \else \setbox0=\hbox{$^{\circ}$}\rlap{\hskip.11\wd0 .}$^{\circ}$\fi}
     
\def\pc{\ifmmode \mathrm{pc} \else $\mathrm{pc}$ \fi}
\def\mpc{\ifmmode \mathrm{Mpc} \else $\mathrm{Mpc}$\fi}
\def\mpcthree{\ifmmode \mathrm{Mpc}^{-3} \else $\mathrm{Mpc}^{-3}$\fi}
\def\gpcthree{\ifmmode \mathrm{Gpc}^{-3} \else $\mathrm{Gpc}^{-3}$\fi}

\def\kelvin{\ifmmode \mathrm{K} \else {$\mathrm{K}$}\fi}
\def\kev{\ifmmode \mathrm{keV} \else $\mathrm{keV}$ \fi}

\def\lsun{\ifmmode {L_\odot} \else $L_\odot$\fi}
\def\msun{\ifmmode M_\odot \else $M_\odot$\fi}
\def\msunyr{\ifmmode M_\odot~\mathrm{yr}^{-1} \else $M_\odot~\mathrm{yr}^{-1}$\fi}

\def\cosi{\ifmmode {\cos\,i} \else $\cos\,i$\fi}

\def\heii{\ifmmode {\rm He{\sc ii}} \else He~{\sc ii}\fi}
\def\mgii{\ifmmode {\rm Mg{\sc ii}} \else Mg~{\sc ii}\fi}
\def\caii{\ifmmode {\rm Ca{\sc ii}} \else Ca~{\sc ii}\fi}
\def\ciii{\ifmmode {\rm C{\sc iii}]} \else C~{\sc iii}]\fi}
\def\civ{\ifmmode {\rm C{\sc iv}} \else C~{\sc iv}\fi}
\def\mgii{\ifmmode {\rm Mg{\sc ii}} \else Mg~{\sc ii}\fi}

\def\teff{\ifmmode {T_{\rm eff}} \else $T_{\rm eff}$\fi}
\def\tmax{\ifmmode {T_{\rm max}} \else $T_{\rm max}$\fi}

\def\mbh{\ifmmode {M_{\rm BH}} \else $M_{\rm BH}$\fi}
\def\led{\ifmmode L_{\mathrm{Ed}} \else $L_{\mathrm{Ed}}$\fi}
\def\lbolflare{\ifmmode L_{\mathrm{bol,flare}} \else $L_{\mathrm{bol,flare}}$\fi}
\def\lagn{\ifmmode L_{\mathrm{agn}} \else $L_{\mathrm{agn}}$\fi}
\def\lbolagn{\ifmmode L_{\mathrm{bol,agn}} \else $L_{\mathrm{bol,agn}}$\fi}
\def\lbol{\ifmmode L_{\mathrm{bol}} \else $L_{\mathrm{bol}}$\fi}
\def\mdot{\ifmmode {\dot M} \else $\dot M$\fi}
\def\mdoto{\ifmmode {\dot{M}_0} \else  $\dot{M}_0$\fi}
\def\mdotf{\ifmmode {\dot{M}_\mathrm{flare}} \else  $\dot{M}_\mathrm{flare}$\fi}

\def\hnot{\ifmmode H_0 \else H$_0$ \fi}

\def\vkep{\ifmmode v_\mathrm{Kep} \else $v_\mathrm{Kep}$ \fi}
\def\vc{\ifmmode v_\mathrm{c} \else $v_\mathrm{c}$ \fi}

\def\vthree{\ifmmode v_{1000} \else $v_{1000}$ \fi}
\def\vrel{\ifmmode v_\mathrm{rel} \else $v_\mathrm{rel}$ \fi}
\def\vkick{\ifmmode v_\mathrm{kick} \else $v_\mathrm{kick}$ \fi}
\def\vkickz{\ifmmode v_{\mathrm{kick},z} \else $v_{\mathrm{kick},z} $ \fi}
\def\vkicky{\ifmmode v_{\mathrm{kick},y} \else $v_{\mathrm{kick},y} $ \fi}
\def\vchar{\ifmmode v_\mathrm{char} \else $v_\mathrm{char}$ \fi}
\def\eflare{\ifmmode E_\mathrm{flare} \else $E_\mathrm{flare}$ \fi}
\def\ekick{\ifmmode E_\mathrm{kick} \else $E_\mathrm{kick}$ \fi}
\def\ecoll{\ifmmode E_\mathrm{coll} \else $E_\mathrm{coll}$ \fi}
\def\ezero{\ifmmode E_\mathrm{0} \else $E_\mathrm{0}$ \fi}
\def\efac{\ifmmode \xi_\mathrm{E} \else $\xi_\mathrm{E}$ \fi}
\def\tqso{\ifmmode t_\mathrm{QSO} \else $t_\mathrm{QSO}$ \fi}
\def\tflare{\ifmmode t_\mathrm{flare} \else $t_\mathrm{flare}$ \fi}
\def\tzero{\ifmmode t_\mathrm{0} \else $t_\mathrm{0}$ \fi}
\def\tfac{\ifmmode \xi_\mathrm{t} \else $\xi_\mathrm{t}$ \fi}
\def\gfac{\ifmmode f_\mathrm{g} \else $f_\mathrm{g}$ \fi}
\def\lflare{\ifmmode L_\mathrm{flare} \else $L_\mathrm{flare}$ \fi}
\def\fflare{\ifmmode F_\mathrm{flare} \else $F_\mathrm{flare}$ \fi}
\def\nflare{\ifmmode N_\mathrm{flare} \else $N_\mathrm{flare}$ \fi}
\def\tshock{\ifmmode T_\mathrm{shock} \else $T_\mathrm{shock}$ \fi}
\def\rmin{\ifmmode R_\mathrm{1} \else $R_\mathrm{1}$ \fi}
\def\rmax{\ifmmode R_\mathrm{2} \else $R_\mathrm{2}$ \fi}
\def\rbound{\ifmmode R_\mathrm{b} \else $R_\mathrm{b}$ \fi}
\def\pbound{\ifmmode P_\mathrm{b} \else $P_\mathrm{b}$ \fi}
\def\mbound{\ifmmode M_\mathrm{b} \else $M_\mathrm{b}$ \fi}
\def\mbo{\ifmmode M_{\mathrm{b}0} \else $M_{\mathrm{b}0} $ \fi}
\def\ebo{\ifmmode E_{\mathrm{b}0} \else $E_{\mathrm{b}0} $ \fi}
\def\efinal{\ifmmode E_\mathrm{final} \else $E_\mathrm{final} $ \fi}
\def\tbound{\ifmmode t_\mathrm{b} \else $t_\mathrm{b}$ \fi}
\def\tagn{\ifmmode t_\mathrm{AGN} \else $t_\mathrm{AGN}$ \fi}
\def\torb{\ifmmode t_\mathrm{orb} \else $t_\mathrm{orb}$ \fi}
\def\tdf{\ifmmode t_\mathrm{df} \else $t_\mathrm{df}$ \fi}
\def\rlim{\ifmmode R_\mathrm{lim} \else $R_\mathrm{lim}$ \fi}
\def\vlim{\ifmmode v_\mathrm{lim} \else $v_\mathrm{lim}$ \fi}
\def\vphi{\ifmmode v_\phi \else $v_\phi$ \fi}
\def\mlim{\ifmmode M_\mathrm{lim} \else $M_\mathrm{lim}$ \fi}
\def\tlim{\ifmmode t_\mathrm{lim} \else $t_\mathrm{lim}$ \fi}
\def\llim{\ifmmode L_\mathrm{lim} \else $L_\mathrm{lim}$ \fi}
\def\fqso{\ifmmode f_\mathrm{QSO} \else $f_\mathrm{QSO}$ \fi}

\def\hbeta{\ifmmode \rm{H}\beta \else H$\beta$\fi}
\def\hbetan{\ifmmode \rm{H}\beta_{\rm n} \else H$\beta_{\rm n}$\fi}
\def\hgamma{\ifmmode \rm{H}\gamma \else H$\gamma$\fi}
\def\hdelta{\ifmmode \rm{H}\delta \else H$\delta$\fi}
\def\hepsilon{\ifmmode \rm{H}\epsilon \else H$\epsilon$\fi}
\def\hzeta{\ifmmode \rm{H}\zeta \else H$\zeta$\fi}
\def\halpha{\ifmmode \rm{H}\alpha \else H$\alpha$\fi}
\def\lalpha{\ifmmode \rm{Ly}\alpha \else Ly$\alpha$}

\def\dvhb{\ifmmode \Delta v_{\hbeta} \else $\Delta v_{\hbeta}$\fi}
\def\dvmg{\ifmmode \Delta v_{\rm{Mg}} \else $\Delta v_{\rm{Mg}}$\fi}

\def\muobs{\ifmmode {\mu_{o}} \else  $\mu_{o}$ \fi}
\def\cosi{\ifmmode {\mathrm{cos}\,i} \else $\mathrm{cos}\,i$\fi}

\def\teff{\ifmmode {T_{eff}} \else $T_{eff}$ \fi}
\def\tmax{\ifmmode {T_{max}} \else $T_{max}$ \fi}

\def\tauh{\ifmmode {\tau_{\rm H}} \else $\tau_{\rm H}$ \fi}

\def\yr{\ifmmode {\rm yr} \else  yr \fi}
\def\kms{\ifmmode \rm km~s^{-1}\else $\rm km~s^{-1}$\fi}
\def\cm{\ifmmode {\rm cm} \else  cm \fi}
\def\cmmitwo{\ifmmode \rm cm^{-2} \else $\rm cm^{-2}$\fi}
\def\cmmithree{\ifmmode \rm cm^{-3} \else $\rm cm^{-3}$\fi}
\def\cmps{\ifmmode \rm cm~s^{-1}\else $\rm cm~s^{-1}$\fi}
\def\cmpsps{\ifmmode \rm cm~s^{-2}\else $\rm cm~s^{-2}$\fi}
\def\kmps{\ifmmode \rm km~s^{-1}\else $\rm km~s^{-1}$\fi}
\def\kmpspmpc{\ifmmode \rm km~s^{-1}~Mpc^{-1} \else
    $\rm km~s^{-1}~Mpc^{-1}$\fi}
  
\def\gcmthree{\ifmmode \rm g~cm^{-3} \else $\rm g~cm^{-3}$\fi}
\def\gcmtwo{\ifmmode \rm g~cm^{-2} \else $\rm g~cm^{-2}$\fi}
   
\def\erg{\ifmmode {\rm erg} \else $\rm erg$ \fi}
\def\ergps{\ifmmode {\rm erg~s^{-1}} \else $\rm erg~s^{-1}$ \fi}
\def\ergcms{\ifmmode \rm erg~cm^{-2}~s^{-1} \else $\rm erg~cm^{-2}~s^{-1}$ \fi}
\def\ergcmshz{\ifmmode \rm erg~s^{-1}~cm^{-2}~Hz^{-1} \else $\rm
erg~cm^{-2}~s^{-1}~Hz^{-1}$ \fi}
\def\ergcmsa{\ifmmode \rm erg~cm^{-2}~s^{-1}~\AA^{-1} \else $\rm
erg~cm^{-2}~s^{-1}~\AA^{-1}$ \fi}
\def\ergshz{\ifmmode \rm erg s^{-1} Hz^{-1} \else
   $\rm erg s^{-1} Hz^{-1}$ \fi}

\def\lam{\ifmmode {\lambda} \else {$\lambda$} \fi}
\def\llam{\ifmmode {L_\lambda} \else  $L_\lambda$ \fi}
\def\lamLlam{\ifmmode \lambda L_{\lambda}(5100) \else {$\lambda L_{\lambda}(5100)$} \fi}
\def\nuLnu{\ifmmode \nu L_{\nu}(5100) \else {$\nu L_{\nu}(5100)$} \fi}
\def\ilam{\ifmmode {I_\lambda} \else  $I_\lambda$ \fi}
\def\flam{\ifmmode {F_\lambda} \else  $F_\lambda$ \fi}
\def\inu{\ifmmode {I_\nu} \else  $I_\nu$ \fi}
\def\fnu{\ifmmode {F_\nu} \else  $F_\nu$ \fi}
\def\bnu{\ifmmode {B_\nu} \else  $B_\nu$ \fi}

\def\msigma{\ifmmode M_{\sigma} \else $M_{\sigma}$\fi}
\def\mbulge{\ifmmode M_{\mathrm{bulge}} \else $M_{\mathrm{bulge}}$\fi}
\def\mgal{\ifmmode M_{\mathrm{gal}} \else $M_{\mathrm{gal}}$\fi}
\def\lgal{\ifmmode L_{\mathrm{gal}} \else $L_{\mathrm{gal}}$\fi}
\def\lbulge{\ifmmode L_{\mathrm{bulge}} \else $L_{\mathrm{bulge}}$\fi}
\def\mgalstar{\ifmmode M^*_{\mathrm{gal}} \else $M^*_{\mathrm{gal}}$\fi}

\def\mbhsigstar{\ifmmode M_{\mathrm{BH}} - \sigma_* \else $M_{\mathrm{BH}} - \sigma_*$ \fi}
\def\deltalogmbh{\ifmmode \Delta~{\mathrm{log}}~M_{\mathrm{BH}} \else $\Delta$~log~$M_{\mathrm{BH}}$\fi}

\def\sigstar{\ifmmode \sigma_* \else $\sigma_*$\fi}
\def\sigthree{\ifmmode \sigma_{\mathrm{[O~III]}} \else $\sigma_{\mathrm{[O~III]}}$\fi}
\def\sigtwo{\ifmmode \sigma_{\mathrm{[O~II]}} \else $\sigma_{\mathrm{[O~II]}}$\fi}
\def\signl{\ifmmode \sigma_{\mathrm{NL}} \else $\sigma_{\mathrm{NL}}$\fi}
\def\wthree{\ifmmode {\rm FWHM({[O~III]})} \else $FWHM({[O~III]})$ \fi}
\def\wtwo{\ifmmode {\rm FWHM({[O~II]})} \else $FWHM({[O~II]})$ \fi}
\def\mthree{\ifmmode M_{\mathrm [O~III]} \else $M_{\mathrm [O~III]}$ \fi}
\def\mtwo{\ifmmode M_{\mathrm [O II]} \else $M_{\mathrm [O II]}$ \fi}
\def\lbreak{\ifmmode L_{\mathrm{break}} \else $L_{\mathrm{break}}$\fi}
\def\lcut{\ifmmode L_{\mathrm{cut}} \else $L_{\mathrm{cut}}$\fi}

\slugcomment{Accepted to ApJ}

\shortauthors{Smith et al.}
\shorttitle{\emph{Kepler} Timing of AGN}

\begin{document}

\title{The \emph{Kepler} Light Curves of AGN: A Detailed Analysis}

\author{Krista Lynne Smith\altaffilmark{1,2}, Richard~F.~Mushotzky\altaffilmark{3}, Patricia~T.~Boyd\altaffilmark{4}, Matt Malkan\altaffilmark{5}, Steve B. Howell\altaffilmark{6} \& Dawn M. Gelino\altaffilmark{7}}

\altaffiltext{1}{Stanford University KIPAC, SLAC, Menlo Park, CA 94025}

\altaffiltext{2}{Einstein Fellow}

\altaffiltext{3}{University of Maryland, College Park, MD}

\altaffiltext{4}{NASA/GSFC, Greenbelt, MD 20771, USA}

\altaffiltext{5}{Department of Physics and Astronomy, University of California Los Angeles}

\altaffiltext{6}{NASA Ames Research Center, Moffett Field, CA}

\altaffiltext{7}{NASA Exoplanet Science Institute, Caltech, Pasadena, CA}

\begin{abstract}

We present a comprehensive analysis of 21 light curves of Type~1 AGN from the \emph{Kepler} spacecraft. First, we describe the necessity and development of a customized pipeline for treating \emph{Kepler} data of stochastically variable sources like AGN. We then present the light curves, power spectral density functions (PSDs), and flux histograms. The light curves display an astonishing variety of behaviors, many of which would not be detected in ground-based studies, including switching between distinct flux levels. Six objects exhibit PSD flattening at characteristic timescales which roughly correlate with black hole mass. These timescales are consistent with orbital timescales or freefall accretion timescales. We check for correlations of variability and high-frequency PSD slope with accretion rate, black hole mass, redshift and luminosity. We find that bolometric luminosity is anticorrelated with both variability and steepness of the PSD slope. We do not find evidence of the linear rms-flux relationships or lognormal flux distributions found in X-ray AGN light curves, indicating that reprocessing is not a significant contributor to optical variability at the 0.1$-$10\% level. 

\end{abstract}

\section{Introduction}
\label{sec:intro}

Active galactic nuclei (AGN) are the most luminous non-transient objects in the universe, powered by accretion onto a central supermassive black hole. The fueling required to ignite the AGN phase can be caused by gravitational tidal torques in major mergers or by secular processes. Simulations of this fueling are well understood on kiloparsec scales; however, the situation becomes obscure on the small scales where the accretion is actually taking place. The accretion disk of an AGN, limited to scales on the order of hundreds to thousands of AU, is too small to image directly at extragalactic distances. It can, however, be studied by taking advantage of the ubiquitous variability of optical AGN light.

The optical continuum light in AGN is primarily supplied by the accreting matter itself, frequently assumed to be thermal emission from the standard geometrically-thin \citet{Shakura1973} disk. Although the disk geometry may vary from object to object based on, for example, accretion rate, the optical variability must come from highly nuclear regions based on the relatively fast timescales on which it is observed, on order hours to days to months \citep[e.g.,][]{Pica1983}. 

Several theoretical models have been proposed to explain the observed optical variability. These include magnetohydrodynamic (MHD) turbulence driving the magneto-rotational instability \citep[MRI,][]{Balbus1991,Reynolds2009}, Poissonian flares \citep{Cid-Fernandes2000}, microlensing \citep{Hawkins1993}, starburst activity in the host \citep{Aretxaga1997}, and a damped random walk of thermal flux within the disk \citep{Kelly2013, Zu2013}. Observational studies of optical variability in AGN have been obtained on a large variety of baselines and with many different sampling patterns, photometric sensitivities and parent samples. There have also been reconstructions of quasar light curves from multi-epoch archival data, such as those obtained from the SDSS Stripe 82 survey by \citet{MacLeod2010} and from Pan-STARRS by \citet{Simm2015}, as well as studies of ensemble AGN variability \citep{Wold2007}. The conclusions from these many studies have often been contradictory regarding the correlation of AGN parameters with variability properties. With the exception of a small handful of agreed-upon relationships, the state of our current understanding of the optical flux variations is confused at best.

Very rich studies of AGN variability have been conducted in the X-ray band, with a number of important results. Characteristic timescales and candidate quasi-periodic oscillations have been detected in the power spectral density functions (PSDs) of X-ray AGN light curves \citep{Papadakis1993, Papadakis1995, Uttley2002, Markowitz2003, Vaughan2003, Uttley2005, Gonzalez-Martin2012}. These characteristic timescales, defined as the point at which the PSD ``breaks," or flattens, towards low frequencies, have been found to correlate with the black hole mass in AGN \citep{McHardy2004}. Recent work by \citet{Scaringi2015} has shown that across a wide range of accreting objects including AGN, the break frequency scales most closely with the radius of the accretor (in the case of black holes, the innermost stable circular orbit). 

Ground-based optical AGN timing has struggled to make the same progress as X-ray variability studies owing both to poorer photometric sensitivity from the ground and long, irregular gaps in sampling which hamper traditional PSD-analysis approaches. Unfortunately, the X-ray emission in AGN is still far more mysterious than the optical emission. The geometry and location of the X-ray emitter and whether it is a corona, the base of a jet, or some other source is still under contention. It would therefore be desirable to have optical light curves with many of the same properties (e.g., continuous sampling and high photometric precision) as X-ray light curves. The \emph{Kepler} space telescope has lately put this goal within reach.

\emph{Kepler} was launched to detect exoplanets in the habitable zone by searching for repeating transits in stellar light curves. In order to detect such transits for planets with orbital periods $\geq$1 year, \emph{Kepler} remained continuously pointed at a region of the sky in the constellations Cygnus and Lyra, chosen for its high density of observable dwarf stars. The \emph{Kepler} mission provided 30-minute sampling across a $\sim4$~year baseline for $\sim160,000$~exoplanet search target stars and across various partial baselines for Guest Observer proposed targets. Initially, only two AGN were known to exist in the \emph{Kepler} field of view (FOV). Using the infrared photometric selection technique of \citet{Edelson2012} and X-ray selection from the \emph{Kepler-Swift} Active Galaxies and Stars survey \citep{Smith2015}, we have discovered dozens of new AGN in this field with a wide range of accretion rates and black hole masses as measured from single-epoch optical spectra. 

Some work has been done on these AGN in recent years. \citet{Mushotzky2011} and \citet{Kasliwal2015} have found that \emph{Kepler} PSD slopes are too steep to be consistent with the predictions of the damped random walk model, and \citet{Carini2012} and \citet{Edelson2014} both reported optical characteristic timescales in the \emph{Kepler}-monitored AGN Zw~229-015. 

We present here a comprehensive analysis of this sample of \emph{Kepler}-monitored AGN, with light curves extracted from a custom AGN-optimized pipeline and Fourier-derived PSD results. We examine the data for correlations with various physical parameters, similarities with X-ray observations, and characteristic timescales. The paper is organized as follows. In Section~\ref{sec:selection}, we discuss the selection and optical properties of the sample. Section~\ref{sec:reduction} describes the development of a special pipeline for AGN science with \emph{Kepler} light curves. Sections~\ref{lightcurves} and \ref{powspec} discuss the light curves and power spectra. In Section~\ref{sec:results} we present our results, and in Section~\ref{sec:implications} we discuss their physical implications.

\section{The \emph{Kepler} AGN Sample}
\label{sec:selection}
 
 \subsection{Sample Selection}
The majority of our objects were selected using the infrared photometric algorithm developed by \citet{Edelson2012}. Their statistic, $S_I$, is based on photometric fluxes from the 2-Micron All-Sky Survey \citep[2MASS; ][]{Skrutskie2006} and the Wide-field Infrared Survey Explorer \citep[WISE; ][]{Wright2010}. The distribution of $S_I$ among $\sim5000$ sources with 2MASS/WISE photometry and SDSS spectra is bimodal, showing separation surrounding the value of $S_I$=0.888. Selection below this value indicates a high likelihood of the object being a Type 1 AGN, quasar, or blazar. There is still some small chance that the object is stellar, and so optical spectra are required for positive identification (see next section). In the end, 41 objects met the infrared and spectroscopic criteria for AGN classification. Three objects were hard X-ray selected in the survey by \citet{Smith2015}, which covered four modules of the \emph{Kepler} field. The survey detected approximately 30 new AGN confirmed with optical spectral follow-up, but only four were requested for monitoring before \emph{Kepler}'s second reaction wheel failure, which prevented the spacecraft from maintaining the necessary pointing precision to continue operating in the original field. Two of these, Zw~229-15 and KIC~7610713, overlap with the previous infrared-selected sample. The other two X-ray selected targets are the BL Lac object BZB~J1848+4245 (KIC~7175757) and the radio galaxy Cygnus~A, which is excluded from this analysis because it is a Seyfert~2. 

The original \emph{Kepler} mission spanned $\sim$4 years, broken up into 17 individual quarters, and ended when the spacecraft's second reaction wheel failed. Each quarter lasts approximately 90 days; exceptions are the initial Quarter~0 (ten days), Quarter~1 (one month) and Quarter~17 (32 days, due to the failure). Between quarters, the spacecraft rotated in order to preserve the sunward pointing of the solar panels. This resulted in a flux discontinuity across quarters due to a variety of factors resulting from a given source landing on a different part of the CCD, including quantum efficiency variations, different readout electronics, and possible variations in the point spread function. \emph{Kepler}'s single monitoring bandpass is broad white light (4200$-$9000\AA), preventing any comparison of variability across colors. The \emph{Kepler} detector is divided into 21 modules, each of which has four output channels. Module~3 failed during Quarter~4; thus, any source in that position in the FOV will have one quarter-long gap every four spacecraft rotations. 

In order to ensure a reasonably consistent set of light curves, we have imposed several criteria for rejection of data from our analysis: 1) light curves shorter than 3 quarters, 2) light curves with any quarter landing on Module~3, 3) an overly-crowded field near the target that would unavoidably include stars in the extraction aperture, and 4) unacceptable levels of rolling band noise in the extraction region. The latter two are further described in Section~\ref{sec:reduction}. 

The final sample consists of 21 spectroscopically-confirmed AGN, listed in Table~\ref{t:tab1}.


\begin{table*}
\caption{The \emph{Kepler} AGN}
\centering
\footnotesize
{\renewcommand{\arraystretch}{0.5}
\begin{tabular}{lcccccccccc}
\hline\hline
KIC \# & RA & DEC & $z$ & Kep. Mag. & log M$_\mathrm{BH}$ & log L$_\mathrm{Bol}$ & L / L$_\mathrm{Edd}$ & PSD Slope & $\tau_\mathrm{char}$ & Quarters \\
  &   &   &   &   &  M$_\odot$ & erg s$^{-1}$ &  & & days &  \\
\hline
10841941	&	18 45 59.578	&	+48 16 47.57	&	0.152	&	17.30	&		&	44.95	&		&	-2.1	& &	7	\\
10645722	&	18 47 22.340	&	+47 56 16.13	&	0.068	&	15.69	&	7.73	&	44.18	&	0.023	&	-2.4	& &	8	\\
7175757	&	18 48 47.117	&	+42 45 39.54	&		&	18.13	&		&		&		&	-2.5	&	& 4	\\
2694186	&	19 04 58.674	&	+37 55 41.09	&	0.089	&	13.46	&	7.66	&	44.71	&	0.089	&	-2.5	&	& 10	\\
6932990	&	19 05 25.969	&	+42 27 40.07	&	0.025	&	11.13	&	6.91	&	44.11	&	0.125	&	-3.4	& 16.0 &	14	\\
2837332	&	19 10 02.496	&	+38 00 09.47	&	0.130	&	17.62	&	7.52	&	44.23	&	0.040	&	-2.5	& &	6	\\
9145961	&	19 11 32.813	&	+45 34 51.35	&	0.546	&	17.11	&	8.59	&	45.78	&	0.124	&	-1.7	&	& 4	\\
12401487	&	19 11 43.365	&	+51 17 56.94	&	0.067	&	19.42	&	7.8	&	44.32	&	0.026	&	-2.8	& 35.7	& 4	\\
5781475	&	19 15 09.127	&	+41 02 39.08	&	0.222	&	17.62	&		&		&		&	-2.2	&	 & 	4\\
8946433	&	19 17 34.883	&	+45 13 37.57	&	0.078	&	14.29	&	7.58	&	44.77	&	0.124	&	-2.4	&	& 4	\\
11606854	&	19 18 45.617	&	+49 37 55.06	&	0.918	&	17.75	&		&	46.94	&		&	-2	&	& 12	\\
12010193	&	19 19 21.644	&	+50 26 46.25	&	0.067	&	16.82	&		&	44.80	&		&	-2.7	& 46.2 &	4	\\
9215110	&	19 22 11.234	&	+45 38 06.16	&	0.115	&	15.63	&	7.3	&	44.14	&	0.055	&	-3	& 9.6 &	8	\\
7523720	&	19 22 19.963	&	+43 11 29.76	&	0.132	&	17.63	&	7.37	&	44.40	&	0.085	&	-2.3	& &	4	\\
12158940	&	19 25 02.181	&	+50 43 13.95	&	0.067	&	14.85	&	8.04	&	44.25	&	0.013	&	-3.3	& 31.6	&12	\\
12208602	&	19 26 06.318	&	+50 52 57.14	&	1.090	&	18.45	&	8.94	&	46.13	&	0.123	&	-1.9	&	& 12	\\
9650712	&	19 29 50.490	&	+46 22 23.59	&	0.128	&	16.64	&	8.17	&	45.62	&	0.226	&	-2.9	&	53.0 & 12	\\
10798894	&	19 30 10.409	&	+48 08 25.69	&	0.091	&	18.23	&	7.38	&	44.36	&	0.076	&	-2.4	& & 	3	\\
7610713	&	19 31 12.566	&	+43 13 27.62	&	0.439	&	16.74	&	8.49	&	45.74	&	0.140	&	-2.5	&	& 8	\\
3347632	&	19 31 15.485	&	+38 28 17.29	&	0.158	&	17.65	&	7.43	&	44.66	&	0.135	&	-2.4	&	& 7	\\
11413175	&	19 46 05.549	&	+49 15 03.89	&	0.161	&	17.07	&	7.9	&	45.01	&	0.101	&	-2.8	&	& 3	\\[1ex]

\hline
\end{tabular}}
\label{t:tab1}
\\[10pt]
The physical properties of the \emph{Kepler} AGN, sorted by right ascension. The Kep. Mag. is the generic optical ``Kepler magnitude" used in the KIC and calculated in \citet{Brown2011}. The final column denotes the number of quarters for which long-cadence \emph{Kepler} monitoring is available. 
\end{table*}


\begin{figure*}
    \centering
    \includegraphics[width=\textwidth]{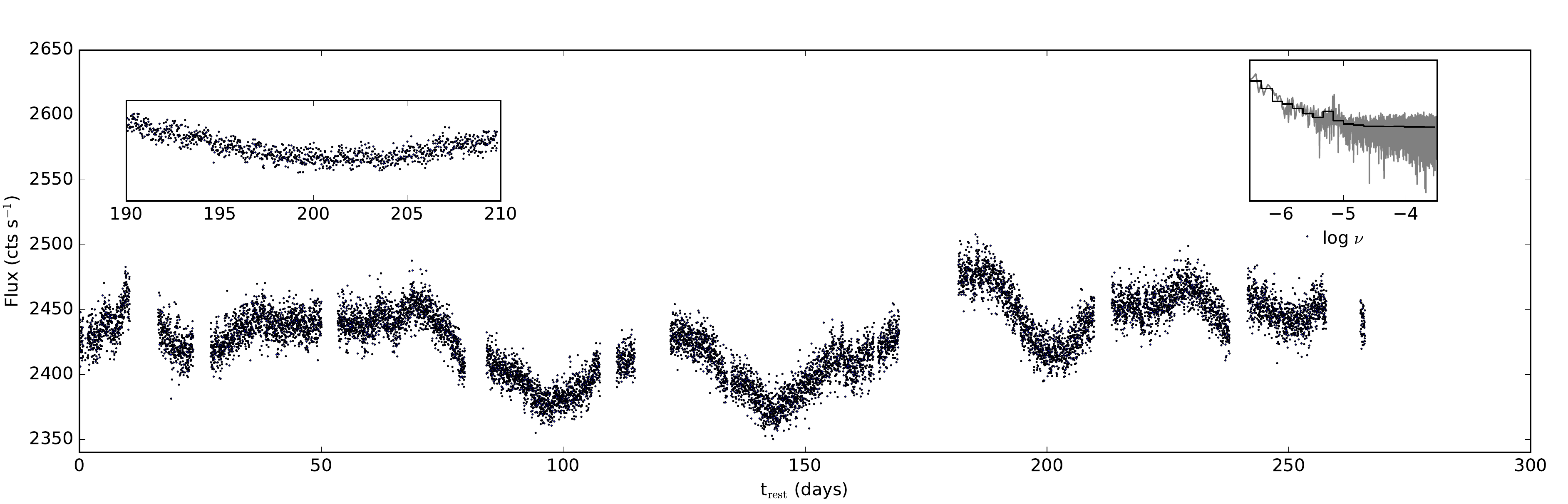}
    \caption{Light curve of spectroscopic AGN KIC~11614932, an object with stellar contamination within the extraction aperture. An excerpt of the light curve (left) and the power spectrum of the full light curve with the periodic signal visible (right) are shown as insets.}
    \label{fig:cautionlc}
\end{figure*}


\subsection{Optical Spectra and Physical Parameters}
\label{sec:spectra}

In order to positively identify IR or X-ray selected sources as AGN, optical spectroscopy is required. We obtained spectra for all targets across four observing runs: August 2011 and June 2012 using the KAST double spectrograph on the Shane 3-m telescope at Lick Observatory, August 2014 at Palomar Observatory using the double-beam spectrograph, and June 2015 using the DeVeny spectrograph on the Discovery Channel Telescope at Lowell Observatory. We only requested \emph{Kepler} monitoring for confirmed Type 1 AGN, as these are the most likely to exhibit optical variability. Type 2 AGN suffer from dust obscuration, and have an optical continuum contributed mostly by non-varying galaxy starlight.

Although the spectrographs used had a variety of resolving powers, all were sufficient to measure with confidence the FWHM of the H$\beta$~or \mgii~lines. These lines are frequently-used single-epoch tracers of the supermassive black hole mass, $M_{\mathrm{BH}}$, which we calculate using the calibrated formulae from \citet{Wang2009} (see their Equations~10 and~11). 

The calculation of Eddington ratio requires a proxy for the bolometric luminosity. Although X-ray luminosity is the most reliable, not all objects in our sample have archival X-ray fluxes. For consistency, we therefore use the luminosity at 5100\AA~and the updated bolometric luminosity corrections by \citet{Runnoe2012}. 

The parameters calculated from our optical spectra for the full sample are given in Table~\ref{t:tab1}. KIC~7175757 is a BL~Lac and has the characteristic featureless optical continuum spectrum, so no values are given for it. Four objects (KIC~10841941, KIC~5781475, KIC~12010193 and KIC~11606854) had H$\beta$~or \mgii~profiles too noisy or contaminated by the dichroic break to allow a confident estimation of the FWHM. 

We note that we are aware of an upcoming paper, Tsan et al. (in preparation), that treats the spectra of the \emph{Kepler} AGN in detail, including analysis of velocity line widths and line flux ratios.

\section{Data Reduction of \emph{Kepler} Light Curves}
\label{sec:reduction}

We recognized early in our analysis that the pipeline-processed archival \emph{Kepler} light curves were unsuitable for AGN science. The mission's original goal was to find periodic signals in point sources. This is fundamentally different from the signal of interest in AGN: the variability is stochastic, and the AGN resides in an extended host galaxy (although luminous quasars are typically point sources that outshine their host, most of our sample and most AGN in general are of the less luminous Sy1 type). Using \emph{Kepler} light curves for AGN analysis requires several steps, which we have cultivated after much trial and error. The general outline of the steps described in this section is as follows: 1) modifying the apertures for photometric light curve extraction, 2) assessing and removing objects badly affected by rolling band noise, 3) carefully removing long-term systematics due to spacecraft effects, 4) stitching across observing quarters and interpolation over gaps, and 5) removing spurious behavior during thermal recovery periods. 

\subsection{Customized Extraction Apertures}
\label{aperture}
\emph{Kepler} extracts its light curves using aperture photometry from a postage-stamp image of the sky surrounding the target, called a Target Pixel File (TPF). Originally designed for stellar extractions, the default apertures are nearly always too small for AGN science because the host galaxies extend beyond the mask. This causes artificial rising and falling of the light curve as the aperture encompasses more or less of the source due to spacecraft drift effects. Thus, the first step in adapting the light curves for AGN is to create larger custom extraction apertures. The software package PyKe \citep{Still2012} includes several useful tools which are utilized in this analysis. The source code for these tools is readily available, allowing us to include the relevant packages in our pipeline. Enlarging the extraction apertures can be achieved with \texttt{kepmask}, which allows the user to select pixels by hand to add to the extraction aperture, and \texttt{kepextract}, which compiles the light curve including the selected pixels. Some locations on the detector are more subject to drift effects than others, so this and host galaxy shape/size, as well as crowdedness of the field, requires an individual approach to creating the extraction mask. An ideal extraction mask is as small as possible to minimize background noise in the light curve, while large enough to fully encompass the galaxy drift throughout the entire duration of the quarter. In order to determine the optimal aperture, we made animations of every 10 frames in each quarter's TPF to assess the maximal extent of the drift. The aperture for each source is large enough for a 1-pixel buffer zone around this maximal drift extent. Several objects were necessarily excluded if this optimal aperture happened to include another object in the field. Some examples of these apertures for the previously-studied AGN Zw~229-015 can be seen in \citet{Edelson2014}. The shape of each object's extraction aperture (i.e., the dimensions and number of total pixels) remained the same for every quarter. 

\begin{figure}
    \centering
    \includegraphics[width=0.3\textwidth]{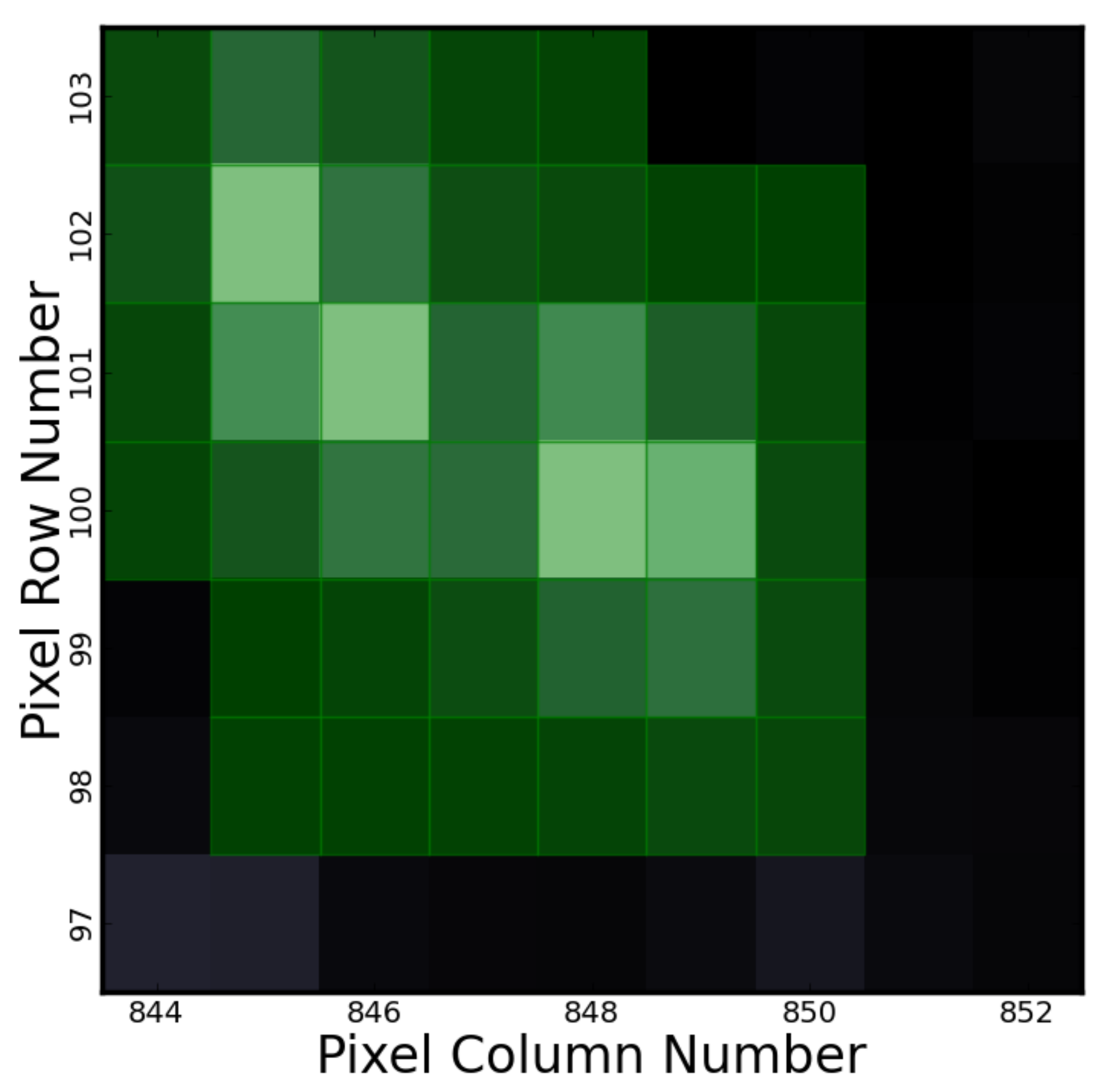}
    \caption{The \emph{Kepler} view of the region around KIC~11614932 (shown in the \texttt{kepmask} user interface), which clearly includes a nearby contaminating star, with the only aperture possible following our extraction requirements shown in green. This object was rejected from the sample.\\}
    \label{fig:starmask}
\end{figure}

As a cautionary tale, we show the case of our spectroscopically-confirmed AGN KIC~11614932. Figure~\ref{fig:cautionlc} shows the full light curve, which exhibits the stochastic variability expected for an AGN. However, a closer inspection reveals a periodic stellar variability signature (easiest to see by eye starting at approximately 160 days). The $\sim1.6$~day periodicity of this star can also be seen in the PSD (inset in the figure). The star in question was close enough to the extended AGN host galaxy that it was impossible to remove from the mask while still encompassing the drift effects discussed above. Additionally, only partially encompassing the star will also result in drift effects. The only possible aperture is shown in Figure~\ref{fig:starmask}, and so the object is excluded from our sample. Our recommendation is to always compare one's extraction aperture with DSS images of the sky in the vicinity of the object of interest to ensure that such contamination is unlikely. 
\\
\\

\subsection{Rolling Band Noise}
\label{sec:mpd}

\begin{figure*}
\caption{Rolling Band Severity Levels for Representative Cases}
\includegraphics[width=\textwidth]{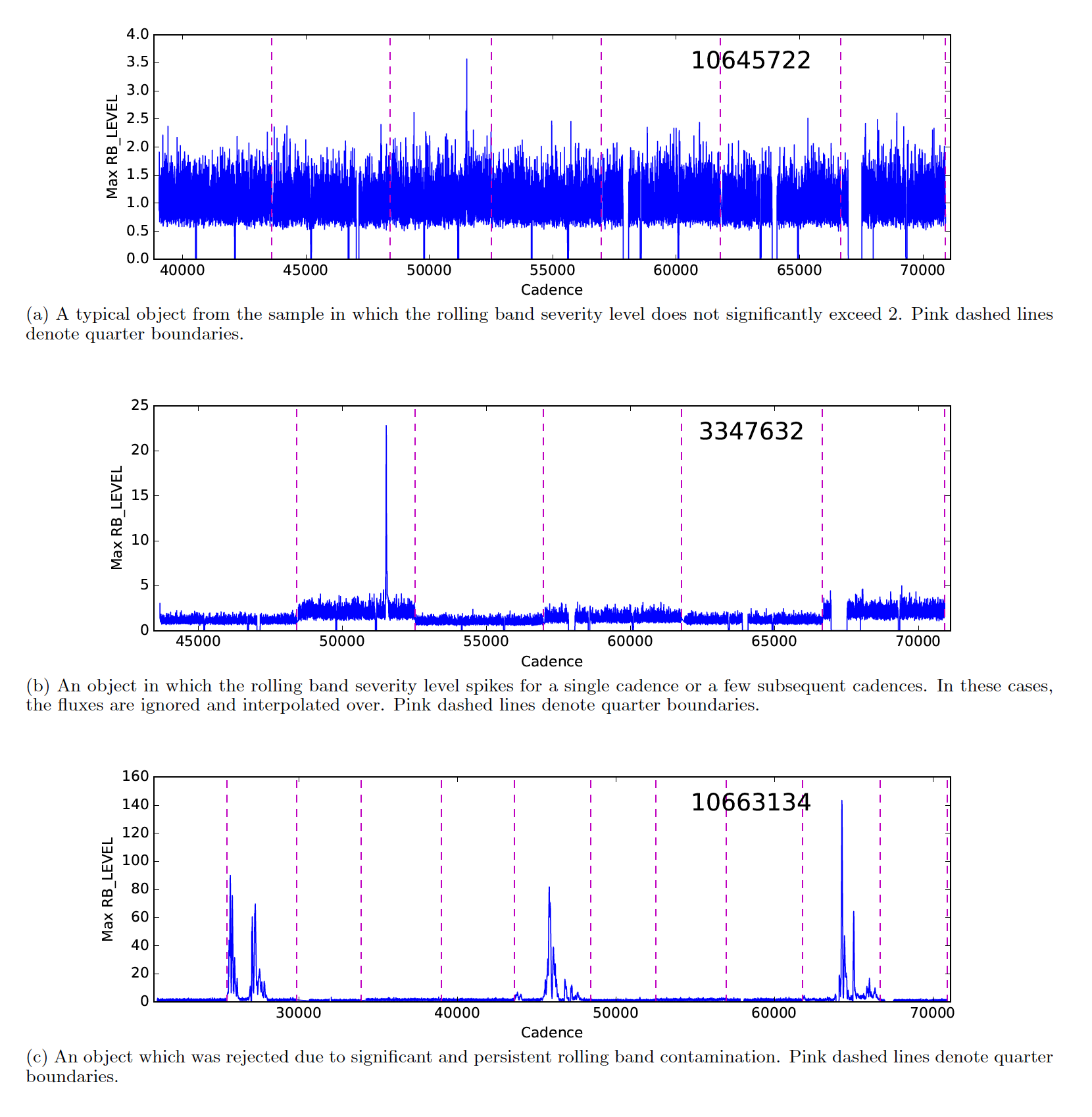}
\label{fig:mpd}
\end{figure*}

\begin{figure*}
    \centering
    \includegraphics[width=\textwidth]{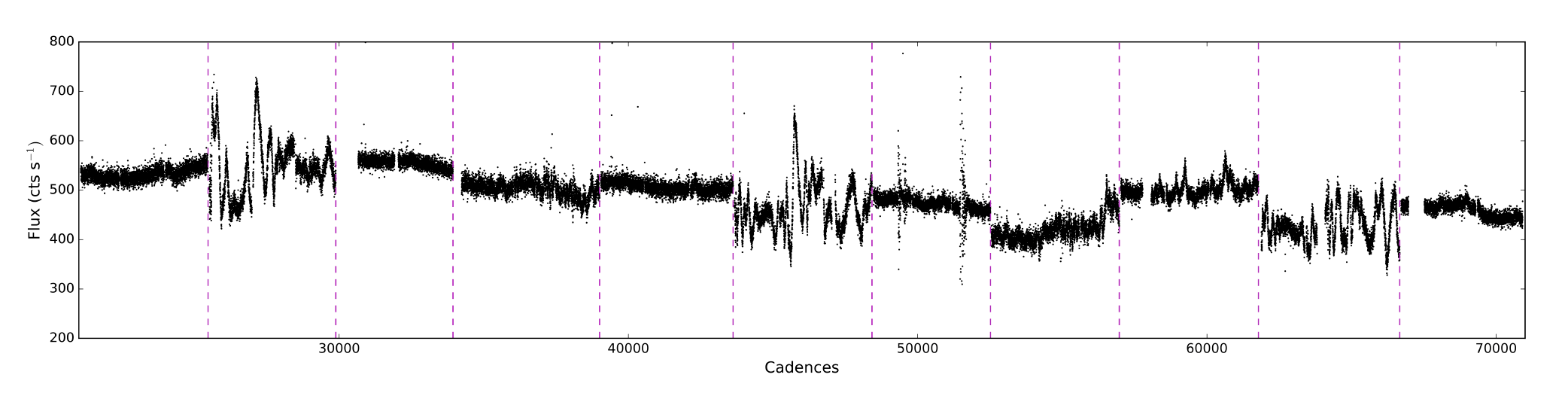}
    \caption{Light curve of KIC~10663134, an object badly affected by rolling band noise as shown in Figure~\ref{fig:mpd}.}
    \label{fig:badlc}
\end{figure*}

Electronic crosstalk between the science CCDs and the fine guidance sensor clocks produces an interference pattern known as rolling band noise or ``Moire pattern noise," which moves across the detector \citep{Kolodziejczak2010}. The Dynablack algorithm \citep{Van-Cleve2016}, a module in the \emph{Kepler} pipeline, is able to assess the level at which this pattern affects any given pixel. All TPFs from Data Release 25 (the most current at the time of this writing) include a column (RB\texttt{\_}LEVEL) with this information, given as a severity level in units of detection threshold calibrated to 20 ppm for a typical 12th magnitude star (see Section~A.1.1 in the data release notes for DR25). We have made plots of this severity level versus observing cadence for each object, and reject any object where the RB\texttt{\_}LEVEL severely affects a quarter. The detection threshold in our targets is considerably higher than for a 12th magnitude star, but the rolling band level rarely exceeds 2.0 in the majority of objects. We show three example cases in Figure~\ref{fig:mpd}. In cases such as KIC~10663134, the object was rejected due to serious rolling band contamination; the effect of the rolling band on the light curve is shown in Figure~\ref{fig:badlc}. Occasionally, the pattern is flagged to affect one to five individual 30-minute cadences, but immediately returns to undetectable levels. This can be seen in KIC~3347632. In these cases, we simply ignore the \emph{Kepler} flux for those cadences and linearly interpolate over them. 

\subsection{Long-term Systematics}
\label{longterm}

Large extraction masks still do not fully remove the spacecraft systematics. There are long-term trends in the data which are well known, especially differential velocity aberration (DVA).  For the long-timescale drifting, corrections can be obtained using the cotrending basis vectors (CBVs). A full description of their application can be found in the \emph{Kepler} Data Characteristics Handbook. In short, although stars can vary, their intrinsic variability should not be correlated with each other. The \emph{Kepler} software maintains a series of sixteen orthonormal functions that represent correlated features from a reference ensemble of stellar light curves. One can remove systematic trends from one's own light curve by fitting these CBVs to the data. The over-application of these vectors can easily result in over-fitting of the light curve, especially in AGN with intrinsic variability mimicking systematic trends. The ideal choice of number of applicable CBVs is therefore an optimization process between removing long-timescale systematics and overfitting genuine physical signal. 

The Data Characteristics Handbook points out that typically, eight CBVs is ideal for removing instrumental trends from stellar targets, flattening them enough to enable transit searches. Eight CBVs  always over-fit AGN in our trials, as spacecraft systematic features coincidentally overlapped with intrinsic behavior. We could see this by examining the CBVs themselves and nearby stars, noting that the degree to which a given CBV trend was actually present in the data was quite weak compared to the weight given to it in fitting a coincidentally-varying segment of an AGN light curve. In order to assess the optimal number of CBVs for fitting AGN light curves, we have incrementally increased the number from 1 to 8 while tabulating these effects. To be conservative, we have determined that two CBVs is the optimal number for correcting our large-aperture extracted light curves. After this point, legitimate variability begins to be mitigated by overfitting.  In the interest of reproducibility and consistency, we apply the same number of CBVs to each light curve. To illustrate the degree to which systematics likely still remain in the light curves, we apply two CBVs to three stars chosen from different locations on the \emph{Kepler} detectors. These stars are not included in any stellar variability surveys, and so are likely to be intrinsically quiet. We extracted their light curves across various sets of quarters in the same manner as our AGN (although this is not strictly necessary, since these stars are point sources without the faint extended emission that necessitates the approach in AGN), and applied the first two CBVs to each one using the PyKe task \texttt{kepcotrend}. Figure~\ref{fig:startest} shows the results. Long-timescale systematic trends are clearly removed, leaving the light curves mostly flat. Although some artificial variability surely remains, we have eliminated the trends most likely to affect our power spectral density analysis while preserving as much intrinsic AGN variability as possible. A thorough discussion of the effects of CBV application to each object and the individual likelihood of the CBVs removing intrinsic variability is given in Appendix~\ref{appendixb}.

\subsection{Interquarter Stitching and Interpolation}
\label{stitch}
The next obstacle is the inter-quarter discontinuities introduced by spacecraft roll. We have chosen to additively scale the light curves based on the average fluxes of the ten cadences before and after the discontinuity (from the light curves with the previously described extraction and CBV corrections already applied). Multiplicative scaling frequently resulted in artificial inflation of flare-type features, and so we consider it less physically valid. 

Finally, there are various data flags which were tabulated by \emph{Kepler} as the data collection proceeded, including those for attitude tweaks, reaction wheel zero crossings, intervals where the spacecraft briefly paused to transmit data to Earth, cosmic ray detections within the extraction aperture, etc. (see Table 2-3 in the \emph{Kepler} Archive Manual). We have excluded cadences with any of these data flags. In most cases, these exclusions consist of a single cadence, or 2-3 cadences grouped together. We have linearly interpolated over these gaps. In some cases, especially when the spacecraft is in Earth-point for data transmission, as many as $\sim$600 sequential cadences (12.5 days) can be flagged. To maintain the even sampling required for our method, we have also linearly interpolated over these gaps, inserting a point where each cadence would be (every 29.4 minutes). We have also interpolated over the gaps in between quarterly rolls (typically about 1 day). One might naturally wonder whether this interpolation method would affect the measured power spectral slopes or other properties. By interpolating linearly, we are not introducing any additional spectral power at a particular frequency, especially because the gaps are irregular and brief. We have performed both simple linear interpolation, as well as linear interpolation based on the LOWESS method of smoothing the existing data points and making calculations based on local values \citep{Cleveland1979}. The power spectral density slopes are nearly identical (to the 0.01 level) for both methods. In no case do the interpolated points consist of more than 10\% of the total light curve. Interpolated fluxes are used only in the power spectrum Fourier analysis, and are excluded from estimates of the light curve variance, rms-flux relation analysis, and flux distribution histograms. 

\subsection{Thermal Recovery Periods}
\label{thermal}

After the monthly data downlinks which require the spacecraft to change position for transmission to Earth, the photometer experiences thermal gradients. These result in focus changes, which eventually settle around 2 to 3 days after the downlink. The result is a transient-like flare or dip in the light curve. This period of focus settling is referred to as ``thermal recovery." These cadences are not flagged by the \emph{Kepler} pipeline, but the Earth-pointed downlinks are. In order to prevent these downlinks from affecting our analysis, we automatically replace the 150 cadences ($\sim$3 days) following an Earth-point flag with linearly interpolated data. The thermal recoveries can profoundly affect the power spectra, especially in high S/N light curves. 

\begin{figure*}
    \centering
    \includegraphics[width=0.9\textwidth]{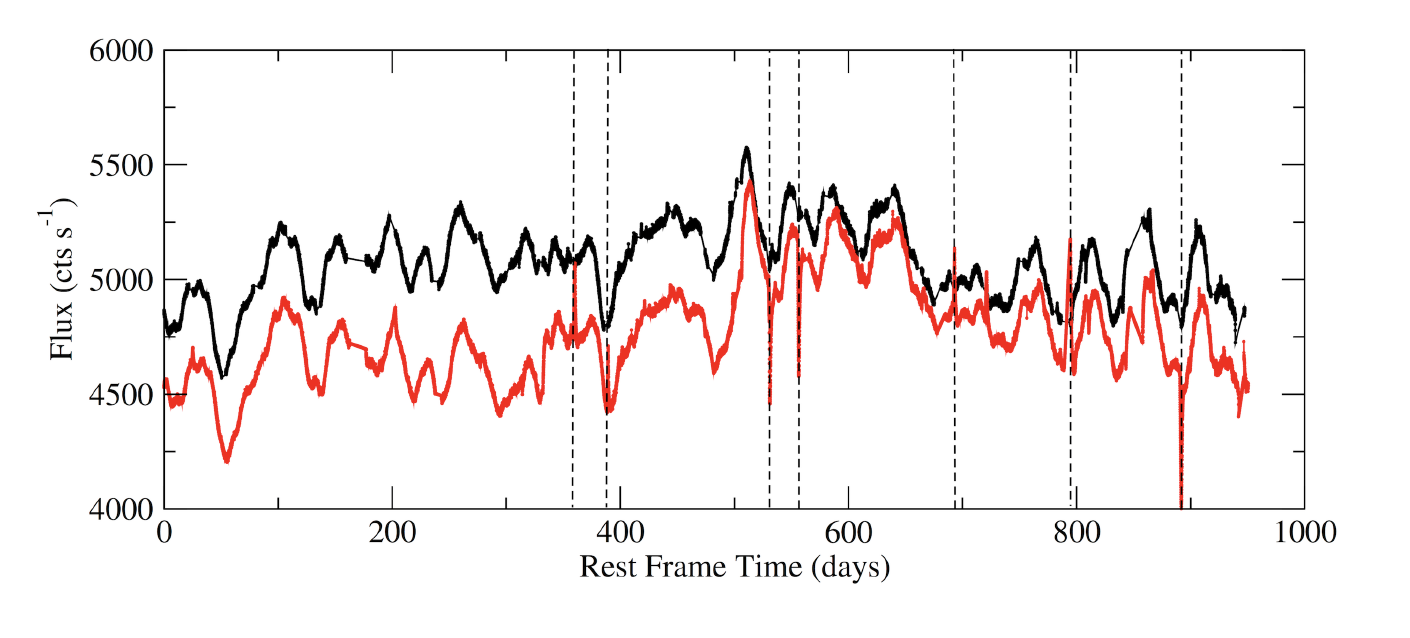}
    \caption{Light curve of KIC~9650712 before (red) and after (black) the correction of thermal recovery periods. The occurrences of some significant thermal events are shown by dotted lines.}
    \label{fig:thermlc}
\end{figure*}

\begin{figure}
    \centering
    \includegraphics[width=0.5\textwidth]{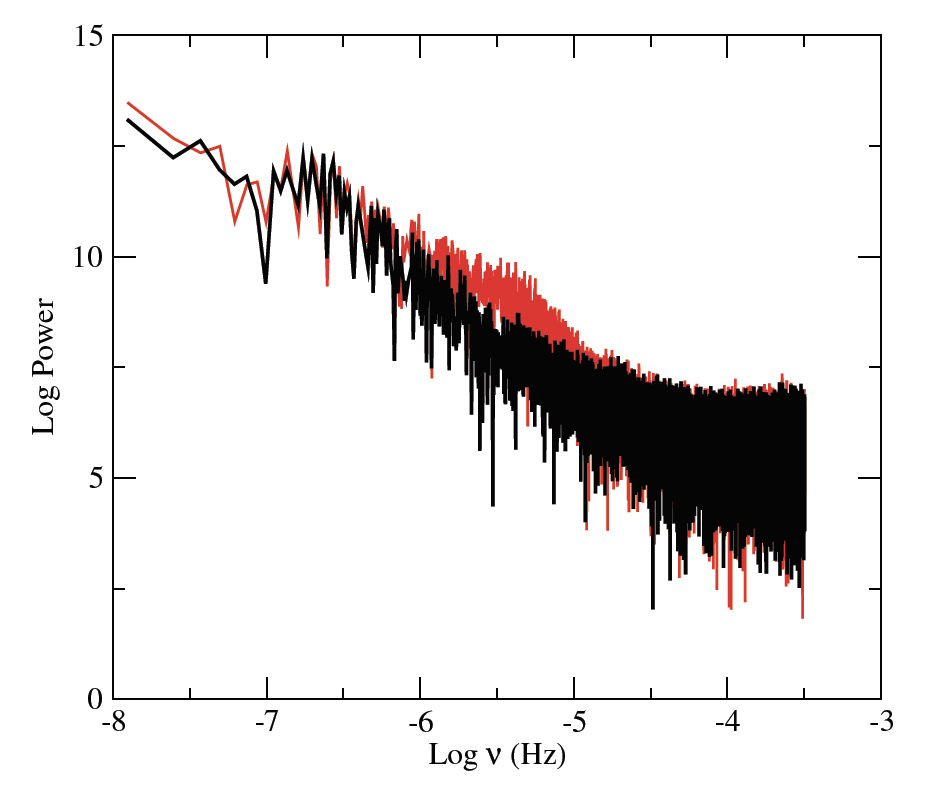}
    \caption{Power spectrum of KIC~9650712 before (red) and after (black) the correction of thermal recovery periods, generated from the light curves shown in Figure~\ref{fig:thermlc}.}
    \label{fig:thermps}
\end{figure}

As an example, in Figure~\ref{fig:thermlc} and Figure~\ref{fig:thermps} we show the result of flagging and removing the thermal recovery periods from the light curve of KIC~9650712. There is no other difference between the reduction methods of the light curves displayed in the graphs. The power spectrum is clearly affected, displaying a spurious characteristic timescale when generated by the uncorrected light curve. 


\begin{figure*}
    \centering
    \includegraphics[width=0.8\textwidth]{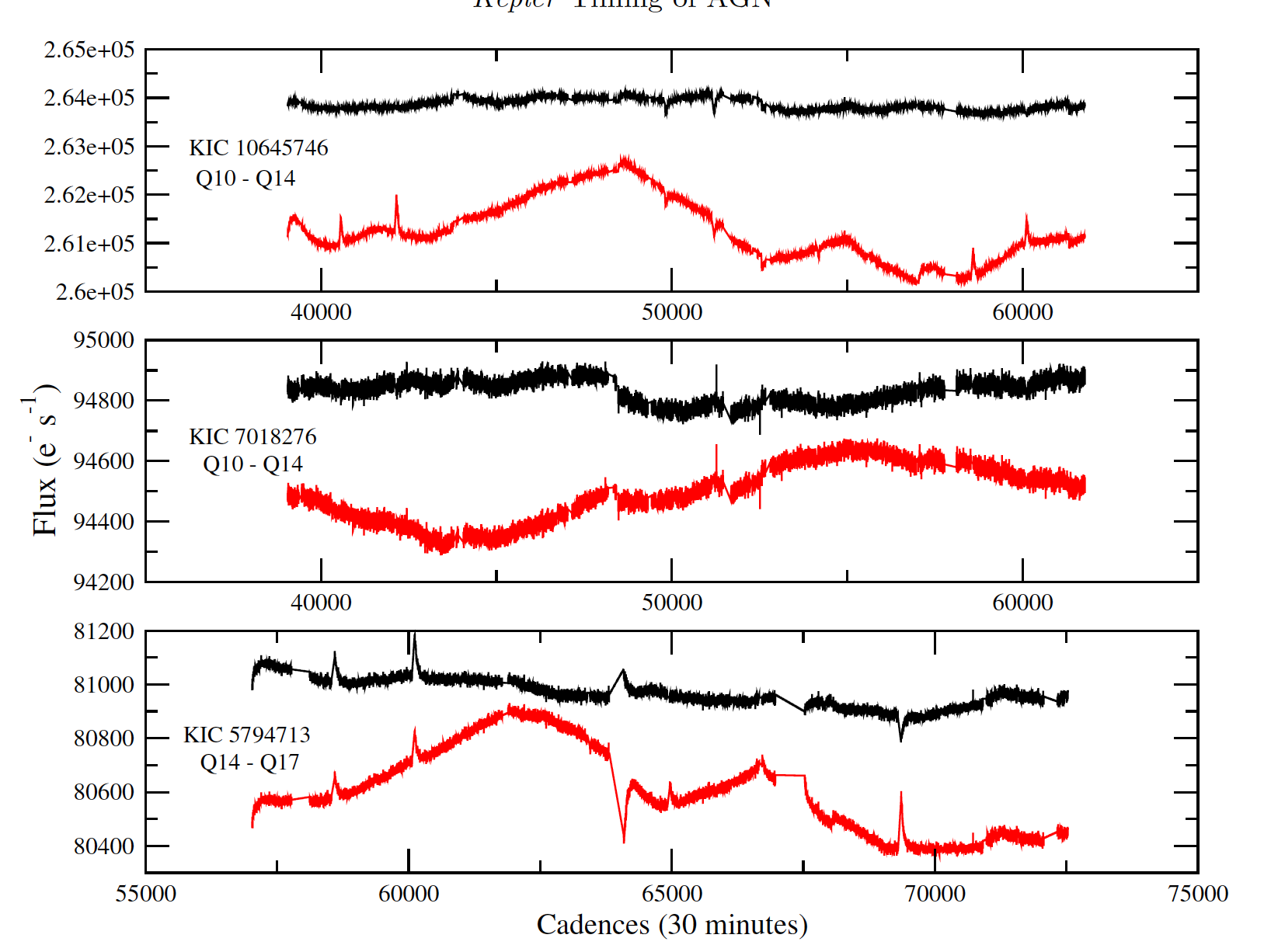}
    \caption{Light curves of three quiet stars before (red) and after (black) the application, fitting, and removal of the first two cotrending basis vectors.}
    \label{fig:startest}
\end{figure*}


\section{Light Curves}
\label{lightcurves}

The final reduced light curves of the sample are given in Appendix~\ref{appendixa}. The axes are scaled differently for each object so that interesting features can be seen, and time axes have been corrected to the galaxy's rest frame using the spectroscopic redshift. The only exception is KIC~7175757, a BL~Lac that has no spectroscopic features. Since a redshift could not be determined for this object, the light curve is shown in the observed frame. In order to better highlight the differences in behavior and monitoring baseline, the second set of figures in Appendix~\ref{appendixa} shows the same light curves, but on identical flux and time-baseline axes. The y-axis range is chosen to be 30\% of the mean flux (15\% in either direction). It is immediately obvious that most of the apparently-dramatic behavior in the light curves is well below the 10\% variability level, often occurring at approximately the 2-5\% level. Such variability would be very difficult to detect in ground-based variability surveys with current instruments. This is an important point: many AGN which are currently classified as non-variable are probably quite variable at these levels. All of the 24 spectroscopically-confirmed Sy~1 AGN in our sample exhibited variability. There are some cases, however, in which a shorter monitoring interval would have resulted in an object seeming non-variable. Take for instance the case of KIC~12208602: except for the event that begins at approximately 220 days in the rest frame, the light curve is statistically flat. This work exemplifies the importance of high-cadence, high-sensitivity, long-baseline monitoring for the classification of AGN as variable or otherwise based on optical light curves. We should refrain, for example, from drawing conclusions about obscuration and the unified model that rely on variability classifications from ground-based light curves.

\section{Power Spectra}
\label{powspec}
The shapes of the power spectral density functions (PSDs) are the product that is most immediately comparable to theoretical expectations. This function shows the relative power in the variability as a function of temporal frequency. The variability of AGN is a red noise process, meaning that successive samples are correlated in time. The power spectra of such processes are well described by a power law, where the spectral density $S$ varies with the temporal frequency as $S \propto f^{\alpha}$. Our investigation of the power spectral density functions (PSDs) has two main goals. The first is to determine whether any of our objects shows evidence for a characteristic variability timescale, which manifests as a ``break" in the power law, causing the PSD to be best fit by a steep power law at high frequencies and a shallower power law at low frequencies (e.g., a piecewise linear function in log-log space). The position of such a break,  $\nu_{\mathrm{char}}$, and its corresponding timescale,  $t_{\mathrm{char}}$ ~could conceivably be connected to a series of relevant physical timescales in the disk. Additionally, the MHD model of \citet{Reynolds2009} predicts characteristic frequencies in the power spectrum that correspond to local acoustic waves. Break timescales have been reported in a few \emph{Kepler} AGN, with inconsistencies in the only object studied by multiple authors: \citet{Carini2012} found that Zw~229-015 could be modeled as a break either at $\sim$95 or $\sim$43 days, while \citet{Edelson2014} reported an optical break timescale of $\sim$5 days.


\begin{figure}[t]
    \centering
    \includegraphics[width=0.5\textwidth]{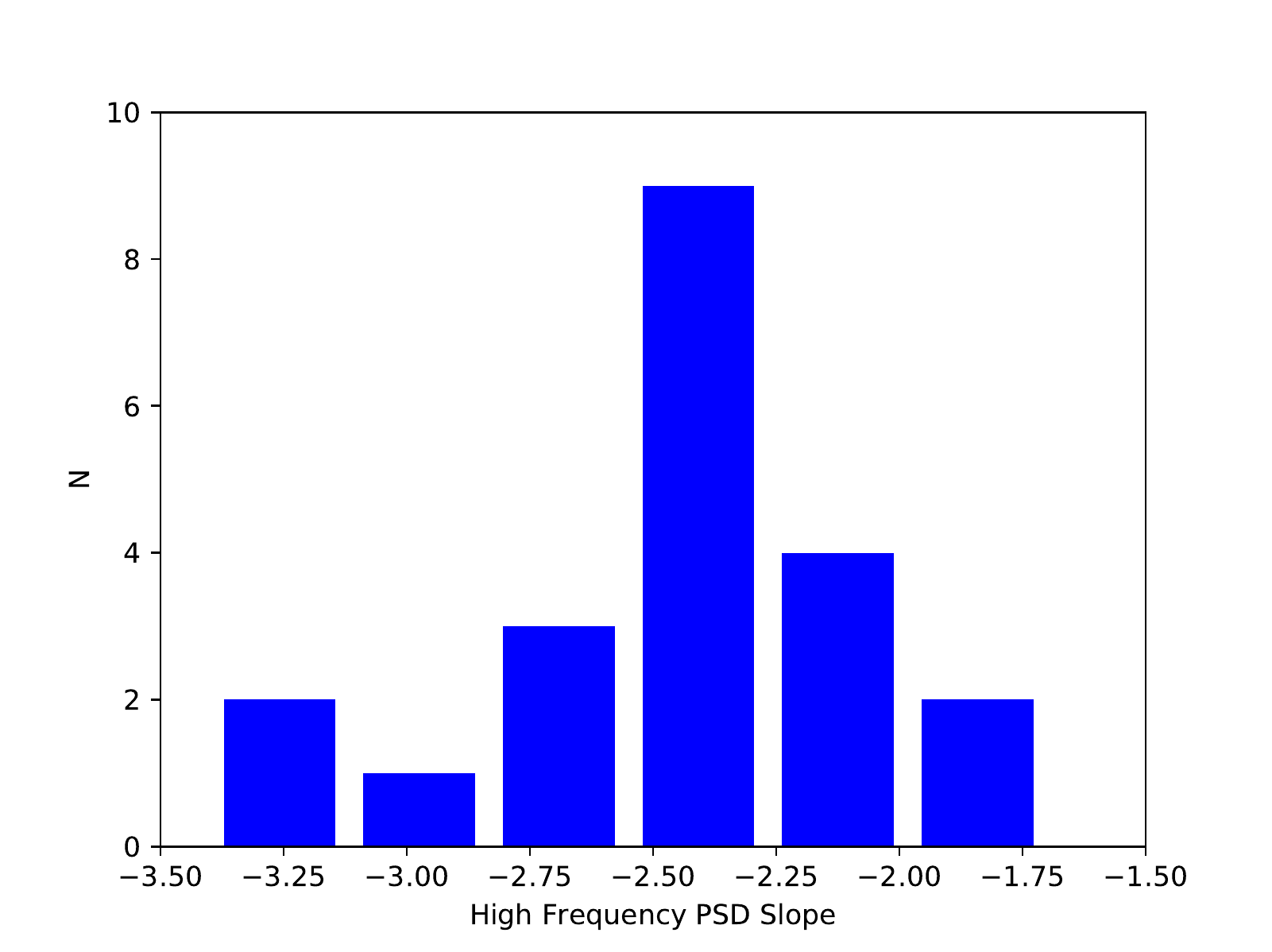}
    \caption{Histogram of the best-fitting high-frequency PSD slopes as measured by the simulations described in Section~\ref{powspec}.}
    \label{fig:psdslopes}
\end{figure}

\begin{figure*}
\begin{tabular}{ccc}

\includegraphics[width=6cm]{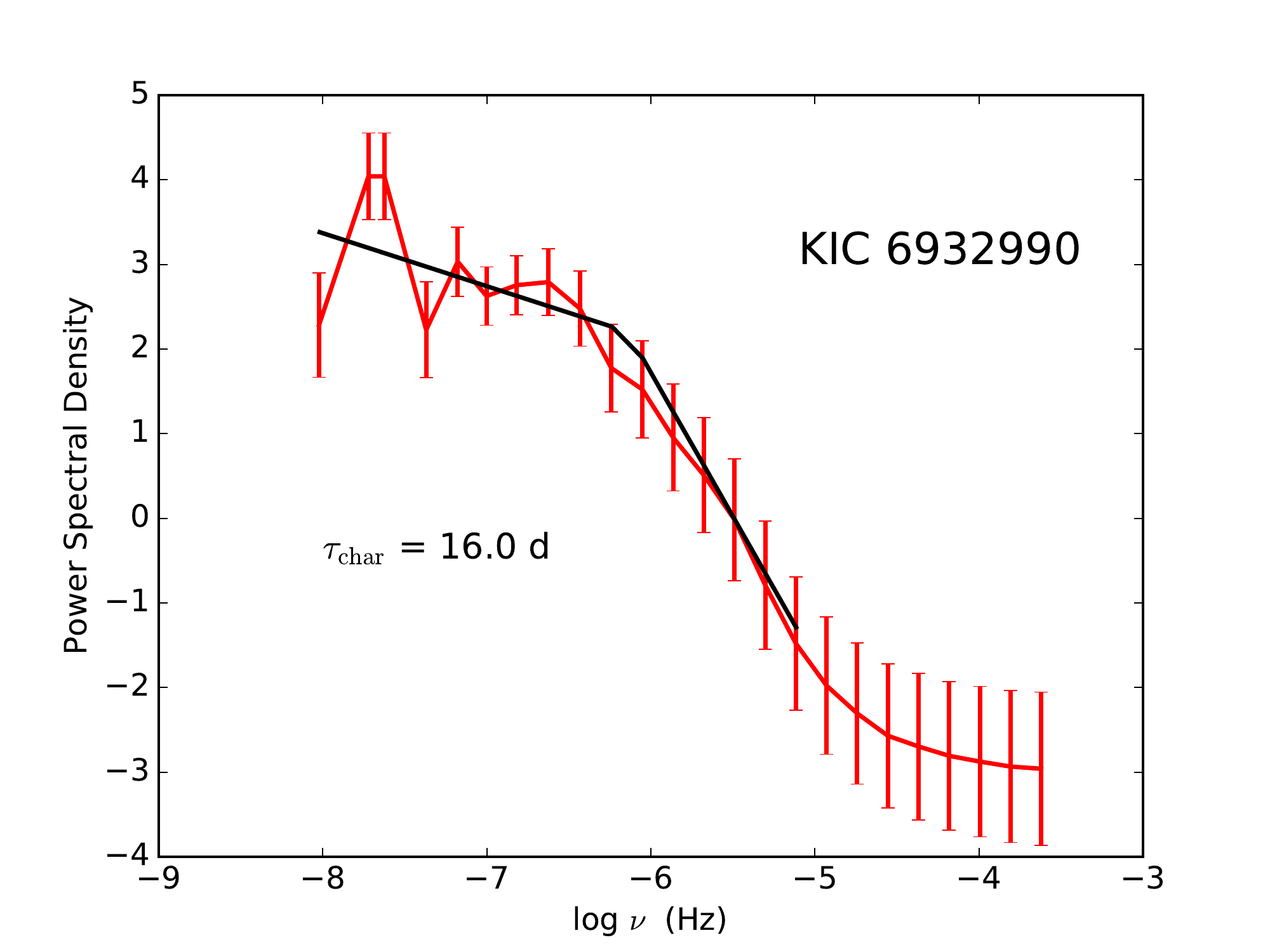} & \includegraphics[width=6cm]{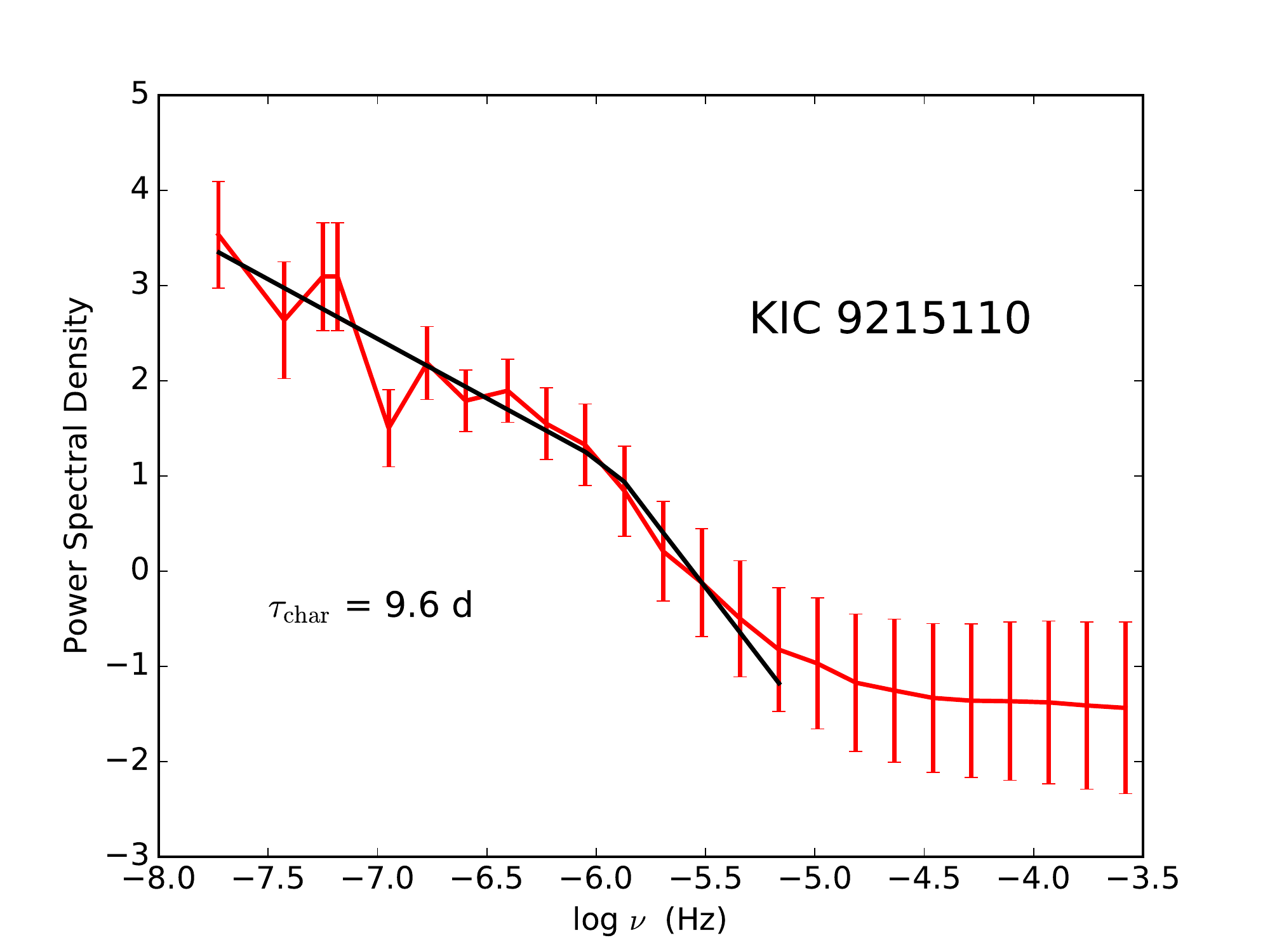} & \includegraphics[width=6cm]{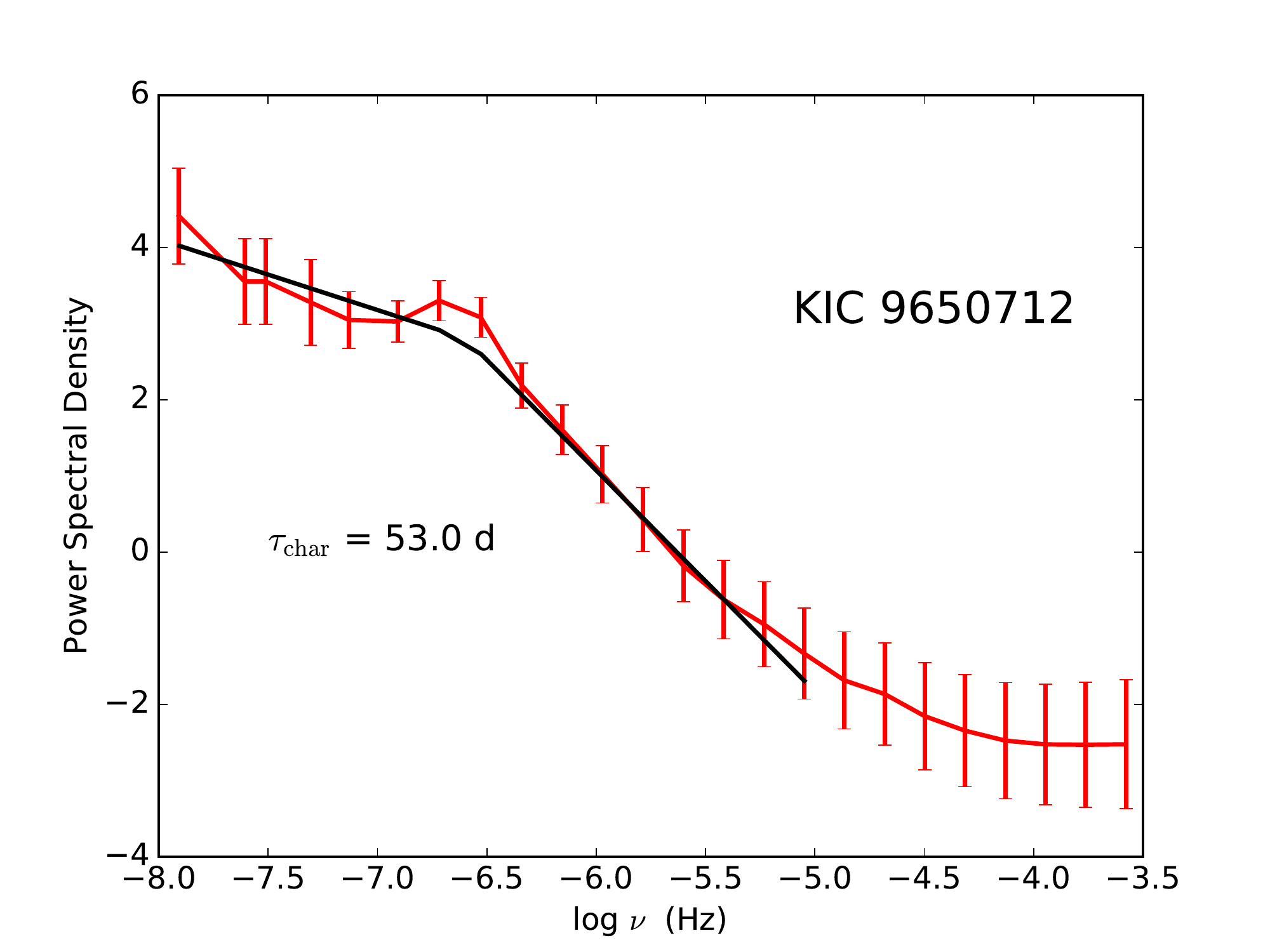} \\
\includegraphics[width=6cm]{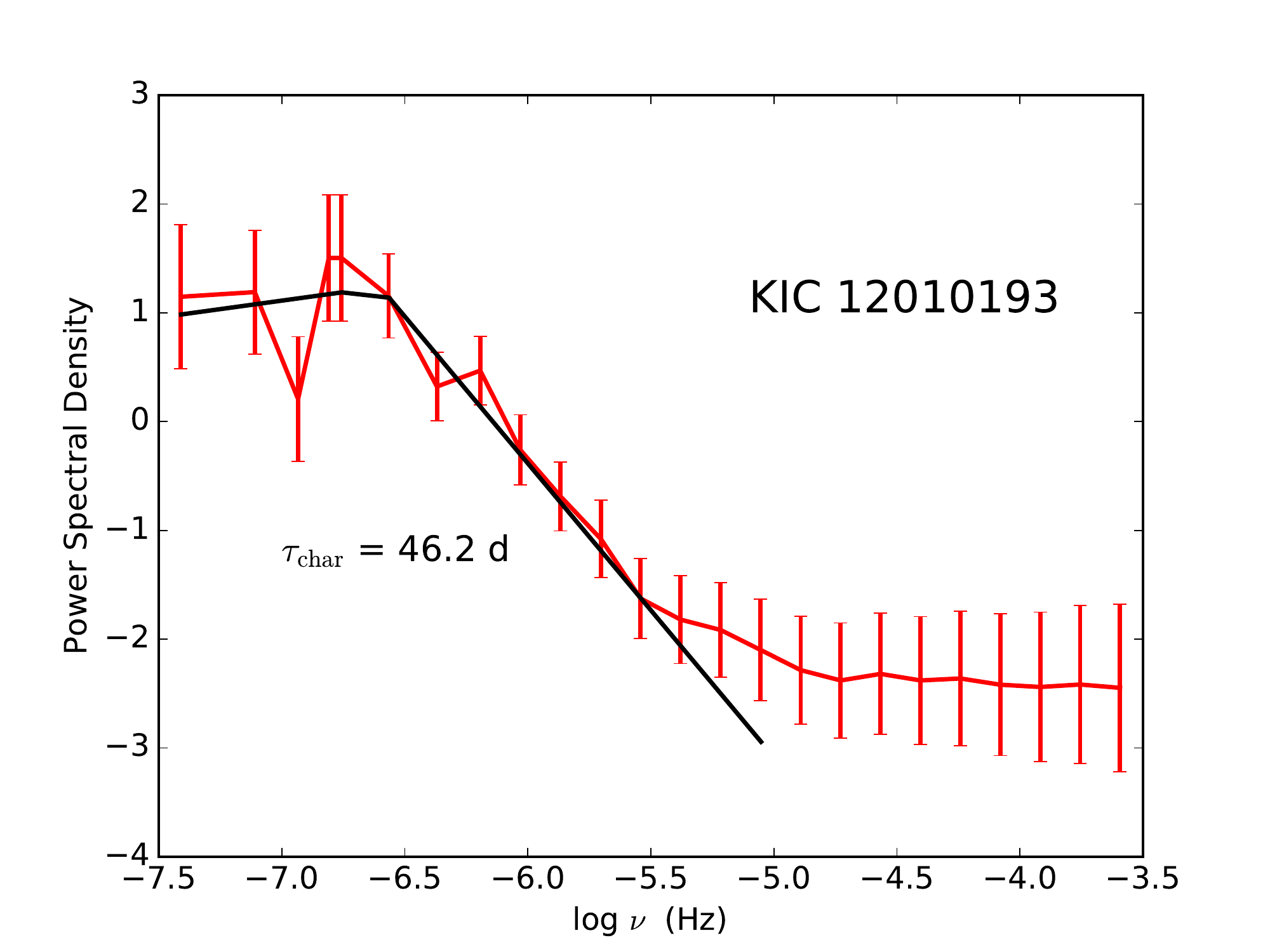} & \includegraphics[width=6cm]{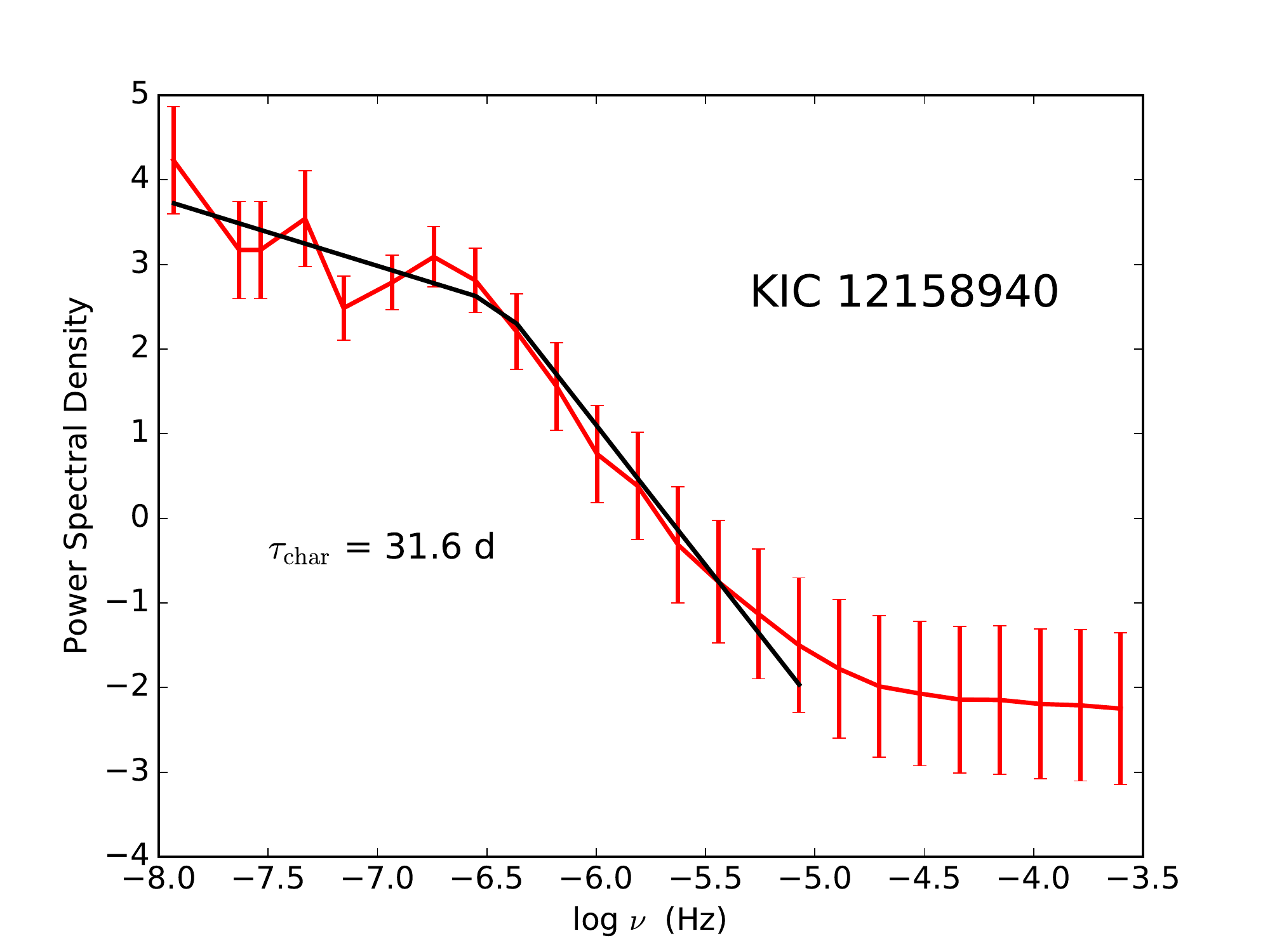} & \includegraphics[width=6cm]{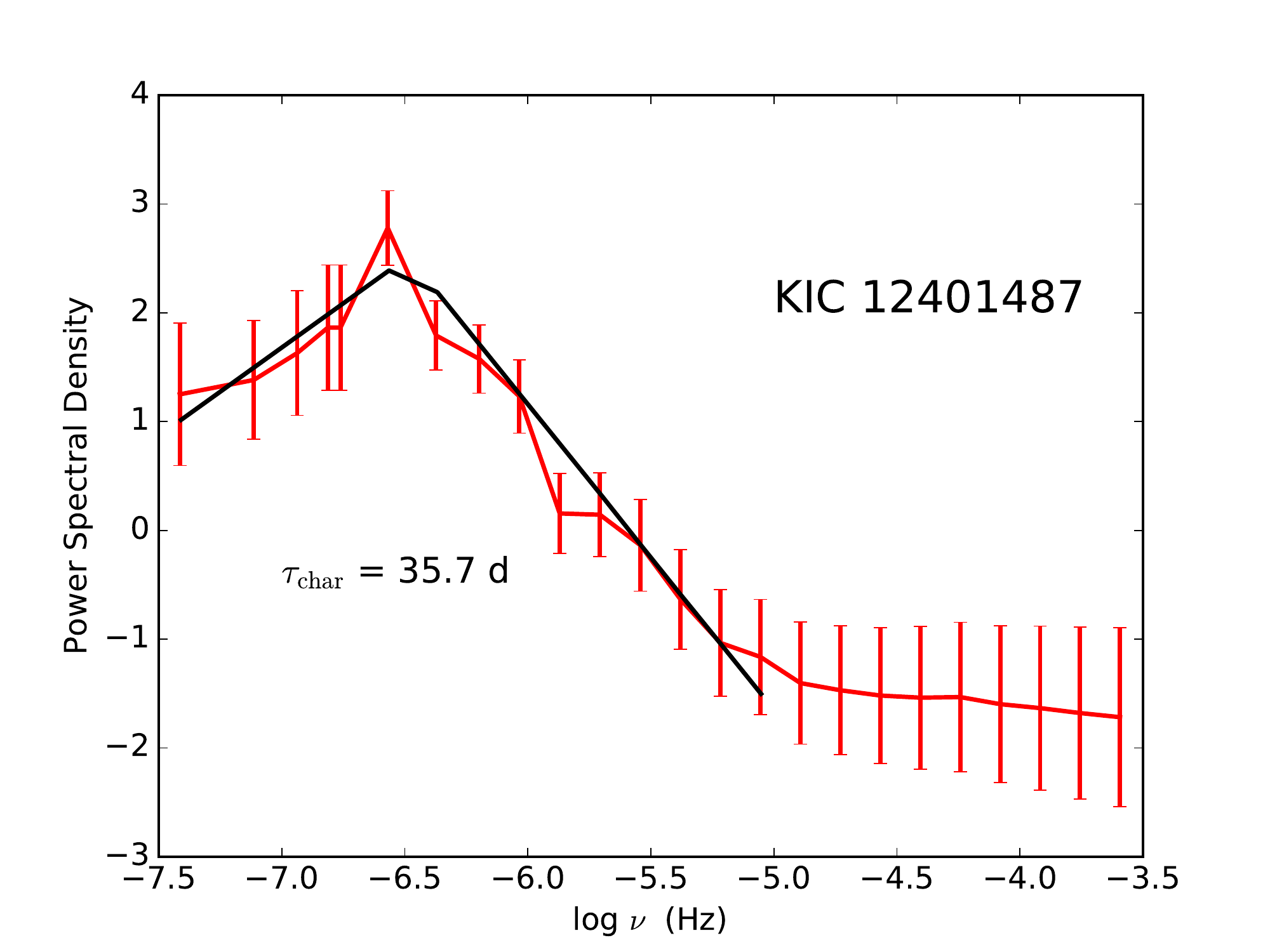} \\

\end{tabular}
\caption{Power spectra of the six AGN which require a broken power-law model for an acceptable fit. Error bars are derived from the Monte Carlo method of \citet{Uttley2002}. The best-fitting break timescale is shown in each plot. See Section~\ref{powspec} for details concerning the fitting.}
\label{fig:broken}
\end{figure*}



\begin{figure}[t]
    \centering
    \includegraphics[width=0.5\textwidth]{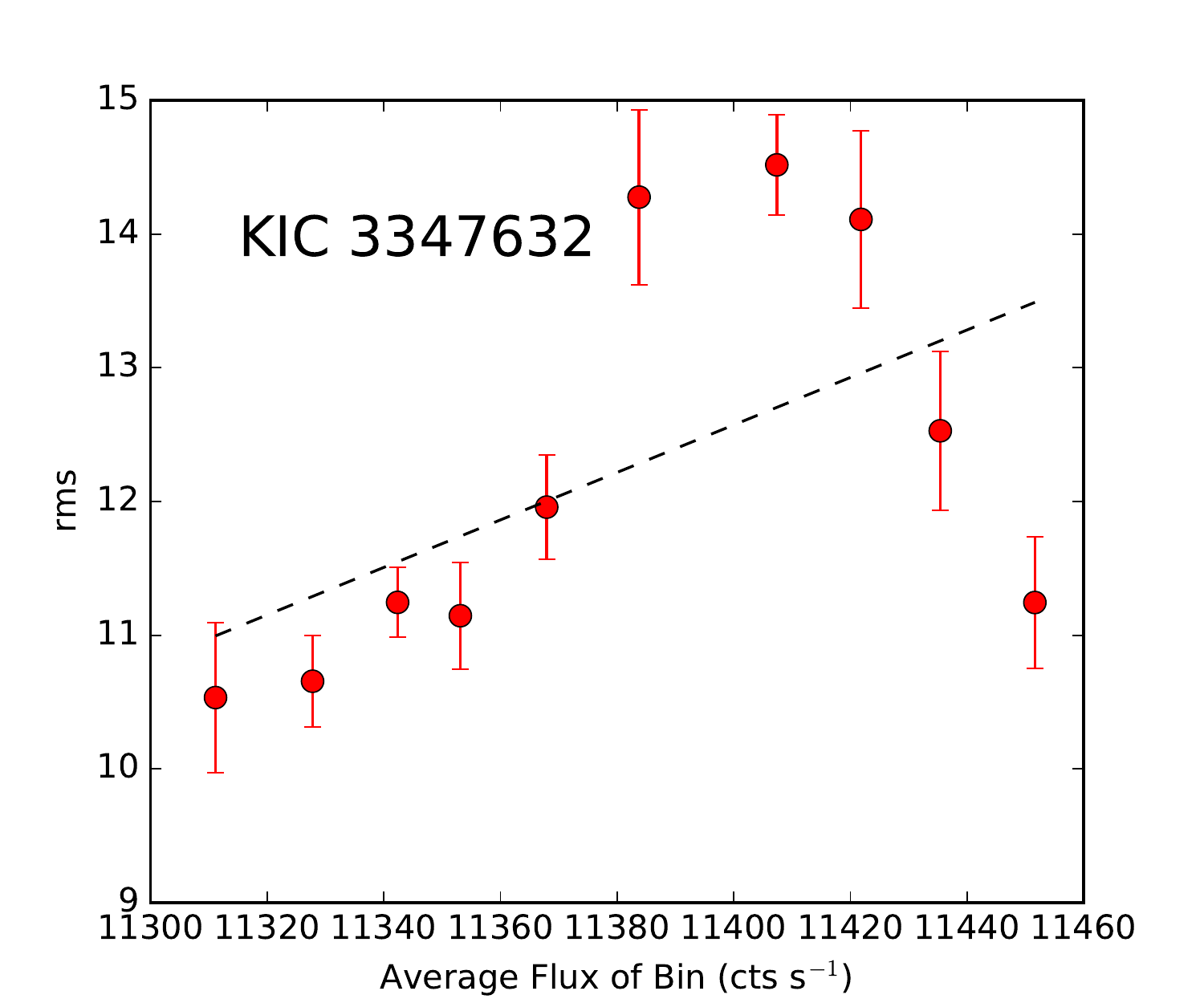}
    \caption{Typical rms-flux relation plot showing no correlation between the two quantities, with the best attempt at a linear fit.}
    \label{fig:exrms}
\end{figure}


The second goal is to measure the slope of the high-frequency portion of the PSD. The high-frequency slopes of optical PSDs have been found by many to be $\alpha\sim-2$ \citep[e.g.,][]{Czerny2003, Kozowski2010,Zu2013}. This value is appealing, because it is the same as the high-frequency slope observed in the X-ray literature and it is consistent with the very popular ``damped random walk" model for AGN variability proposed by \citet{Kelly2009}. However, earlier work using \emph{Kepler} has been in disagreement with these ground-based studies. \citet{Mushotzky2011} published early results for four \emph{Kepler}-monitored AGN, and found that their PSD slopes were inconsistent with predictions from the damped random walk model. The slopes reported varied from $\alpha = -2.6$ to $-3.3$.  This is considerably steeper than previous optical PSD measurements. This conclusion was also reached by \citet{Kasliwal2015}, who found that the damped random walk model was insufficient to capture the range of \emph{Kepler} AGN behavior. We find also that the high-frequency slope is unaffected by the CBV application in the majority of our objects, and that in the few exceptions it is likely that the CBV correction removed only spurious behavior. KIC~12158940 alone may have a shallower slope than reported here due to over-correction. These results are elaborated upon in Appendix~\ref{appendixb}.

To create our power spectra, we use the light curves produced as described in Section~\ref{sec:reduction}, including interpolation to enable Fourier methods. We then fit the entire light curve with a line, and remove this linear trend. This will remove the lowest-frequency component of the power spectrum. Although any linear rise or fall across the full baseline might indeed be real, we must use consistent methodology to study the variability in the same time regime for all objects. By removing this trend, all of the PSDs are now on equal footing for studying variability on timescales of 1 to 100s of days. The mean of these flattened light curves is subtracted, and a simple discrete Fourier transform is performed. We have normalized the power spectra by a constant $A_{\mathrm{rms}}^{2} = 2\Delta T_{\mathrm{samp}}/\bar{x}^{2}N$, where $\Delta T_{\mathrm{samp}}$ is the sampling interval, $\bar{x}$ is the mean count rate in cts~s$^{-1}$, and $N$ is the total number of data points. This normalization was defined by \citet{VanDerKlis1997} and is cited by \citet{Vaughan2003} as particularly useful for AGN, since the integrated periodogram yields the fractional variance of the light curve. In this scheme, the expected Poisson noise is given by Equation~A2 in \citet{Vaughan2003}, with the mean background count rate estimated from several background pixels nearby the source. We also resample the power spectra into 25 logarithmic frequency bins, for fitting purposes. The resulting PSDs are shown in Figure~\ref{fig:psds}, along with the best-fitting power law models that we describe next. 


\begin{figure*}[htb]
   \caption{Power spectra of the \emph{Kepler} AGN, continued on following page.}

   \includegraphics[width=\textwidth]{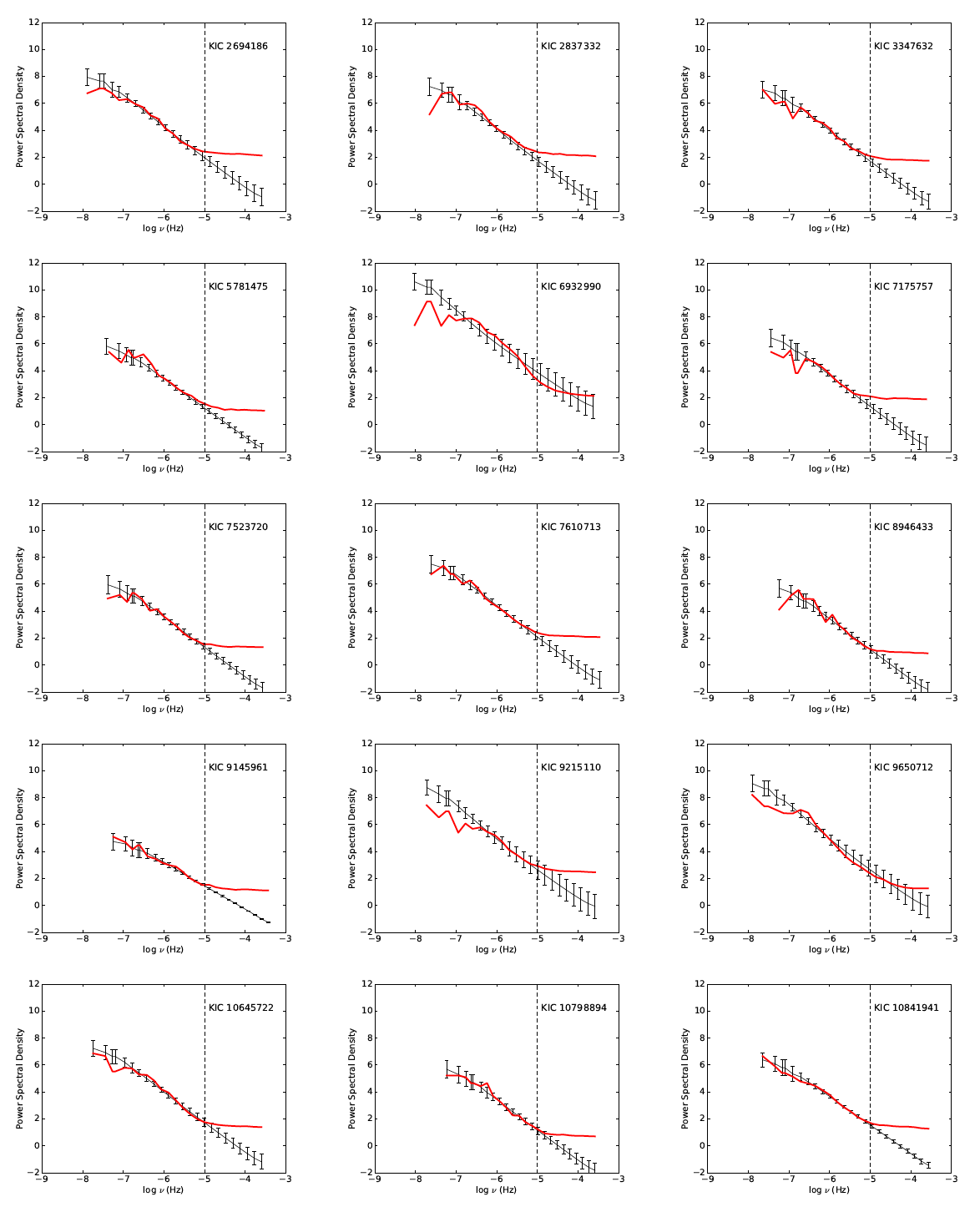}

\end{figure*}

\begin{figure*}[htb]\ContinuedFloat

   \includegraphics[width=\textwidth]{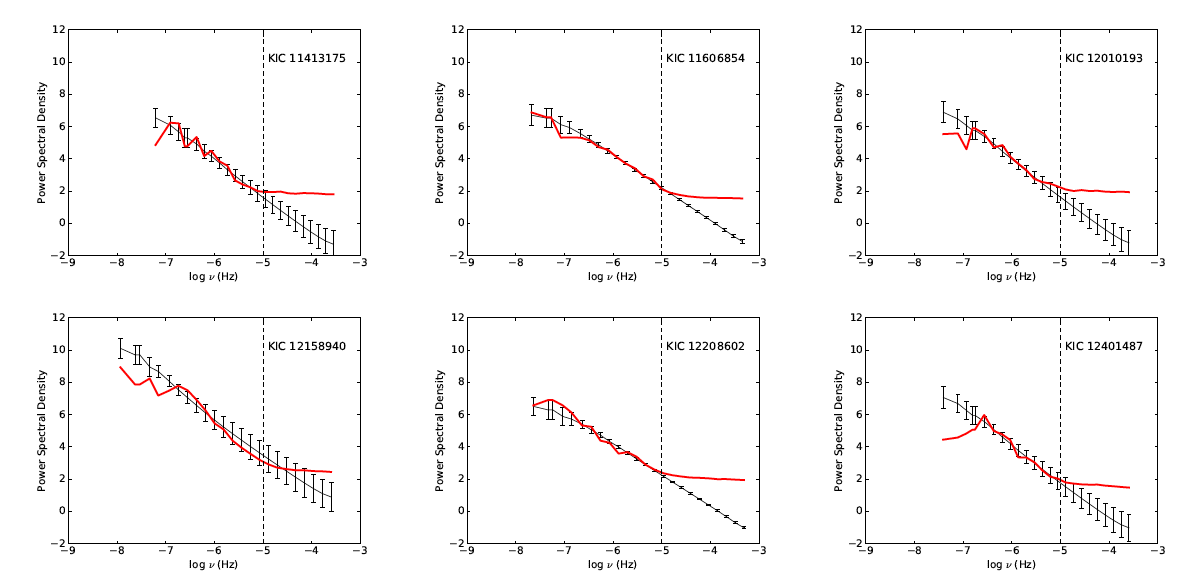}

\caption{Power spectra of the \emph{Kepler} AGN. Red denotes the observed binned power spectra, while black shows the best-fitting model from the \citet{Uttley2002} method with the error bars. Error bars in this case denote the spread in the simulated power spectra; see Section~\ref{powspec} for details. Fitting was only performed on frequencies lower than the dashed line, which denotes the point at which Poisson noise begins to dominate. Objects best-fit by broken power laws (e.g., KIC~12401487) are shown here for completeness. The broken power law fits can be seen in Figure~\ref{fig:broken}.  }
\label{fig:psds}
\end{figure*}


In order to determine the best-fitting high-frequency slope and the significance of any characteristic break frequencies, we employ the Monte Carlo method described in detail by \citet{Uttley2002}. Such methodology is necessary for measuring goodness-of-fit of various power law models, because it allows for an estimate of the error bars on the observed PSDs. We describe it briefly here. 

First, we use the \citet{Timmer1995} method to simulate a very long light curve with a given PSD slope. This light curve is made long enough that 500 light curves of the same length as the actual observations can be drawn from it without overlap, and with three-times finer resolution than the observed light curves. This mitigates the effects of both red noise leak and aliasing (see the discussion in Uttley et al. 2002). We rebin the simulated light curve to have the observed sampling (i.e., the bins have three measurements each). We then introduce gaps with the same location and duration as those in the observed light curve to each of the 500 simulacra, and treat them precisely as we treated the observed light curves, with the same flattening and interpolation techniques. This new simulated set is used to create 500 power spectra. The rms spread of the simulated power spectra about the mean determine the ``error bars" on the observed power spectra. We can then fit the data to the model, and calculate a $\chi^{2}$ value. We do not fit any data that is beyond the cutoff where Poisson noise takes over in the light curve, shown by the dashed lines in the figures in Appendix~\ref{appendixb}. Importantly, the $\chi^{2}$ value obtained from this fit cannot be compared to the standard $\chi^{2}$ distribution. To create a comparable distribution, we measure each of the 500 simulated power spectra against the model and create a sorted distribution of their $\chi^{2}$ values. The goodness-of-fit of the observed power spectrum to the model is then measured by calculating the percentile value above which the observed $\chi^{2}$ exceeds the simulated distribution. This percentage is the confidence with which we can reject the model. 

We do this for each object separately (so that each distribution of simulated power spectra have the same gaps and sampling as the observations), for PSD slopes of $-3.5 \leq \alpha \leq -1.5$, in steps of 0.1. The slope that can be rejected with the least confidence is the slope we consider to be the best fit for each object. These slopes are given in Table~\ref{t:tab1}, and are shown as a histogram in Figure~\ref{fig:psdslopes}. Most of our slopes agree with previous \emph{Kepler} results in that they are steeper than $\alpha=-2$, with some slopes steeper than $-3$ \citep{Mushotzky2011,Kasliwal2015}. Some objects have slopes consistent with $-2$. Our two radio-loud objects, KIC~12208602 and KIC~11606854, are well-fit by $\alpha=-1.9$ and $-2.0$ respectively. These relatively shallow slopes are the same as those measured for these objects by \citet{Wehrle2013} and \citet{Revalski2014}.

If all of the models in the set of $\alpha$ values can be rejected with more than 75\% confidence, we examine the PSD for possibilities of flattening at low-frequencies by testing whether or not a reduction in $\chi^2$~occurs with a broken power law model. If the broken model is a better fit, we calculate the turnover frequency by fitting the high-frequency data with the best-fitting slope, and iterating a model with break frequencies in the range of $-5.5<~$log~$\nu<-7.0$ in steps of 0.1. The break frequency resulting in the best reduced $\chi^{2}$ value is given in Table~\ref{t:tab1}. Of the 21 objects in the sample, 6 demonstrate significant low-frequency flattening in which a broken power-law model represents a reduction in $\chi^{2}$. The $\chi^{2}$ values are given in Table~\ref{t:broken} and their PSDs are shown in Figure~\ref{fig:broken}.  One of these, KIC~9650712, is poorly fit by all of our values of $\alpha$, and is better fit by a broken power law with $t_{\mathrm{char}} \sim 53$~days. However, it can also be well-modeled as a single power law with the addition of a quasi-periodic component. For this reason we use a different symbol for this break timescale in plots, so that readers may ignore it in this context. The possibility that this is a detection of an optical quasi-periodic oscillation (QPO) is discussed a paper being submitted concurrently with this one (Smith et al. 2017, submitted). We also point out that unfortunately, the H$\beta$ line in KIC~12010193 is too weak and noisy to permit a reliable black hole mass measurement, and so it is omitted from the plots of variability versus physical parameters in Section~\ref{mvar}. Finally, it is possible that applying the CBV corrections to the light curves as discussed in Section~\ref{longterm} could remove real low-frequency variability and artificially flatten the light curve. Most of the timescales in this project are likely safe, with the possible exception of KIC~12158940. Appendix~\ref{appendixb} provides an in-depth description of the CBV effects and a before-and-after look at the light curves and power spectra, and describes the reasoning for this exception.

\begin{table}
\caption{$\chi^{2}$ in Broken Power-Law AGN}
\centering
\footnotesize
{\renewcommand{\arraystretch}{0.5}
\begin{tabular}{lcccc}
\hline\hline
KIC \# & $\chi^{2}$ & $\chi^{2}$ &Red. $\chi^{2}$  & Red. $\chi^{2}$ \\
  & Single  & Broken  & Single  &  Broken   \\
\hline

6932990	&	76.05	&	13.44	&	6.91	&	1.22	\\
12401487	&	46.9	&	7.0	&	4.69	&	0.70		\\
12010193	&	18.18	&	10.9 &	1.81	&	1.09		\\
9215110	&	51.12	&	7.15	&	5.11	&	0.71		\\
12158940	&	35.46	&	73.4	&	3.22	&	0.56		\\
9650712	&	37.87	&	7.46	&	3.44	&	0.67	\\[1ex]

\hline
\end{tabular}}
\label{t:broken}
\\[10pt]
Values of $\chi^{2}$ and reduced $\chi^{2}$ for the objects with candidate characteristic timescales.
\end{table}

\section{Results}
\label{sec:results}

The goal of this project and future analyses is twofold: to determine whether any AGN physical parameters, or combination thereof, correlates with variability statistics in such a way as to clarify the physical mechanism of accretion and enable large-scale measurements of AGN parameters in future surveys, and to determine the extent to which the optical and X-ray variability is interrelated. We can attain the first goal using the parameters measured by our optical spectra, and checking for correlations across a comprehensive set of variability metrics and PSD properties based on the existing literature. We can attain the second goal by comparing the \emph{Kepler} timing behavior to well-known properties of X-ray variability in AGN. If the optical light curves exhibit similar properties, albeit on slower timescales, we can assume that a large fraction of the optical variability is the result of reprocessing. 

Before discussing our results, we would like to emphasize that the \emph{Kepler} light curves are probing an entirely new regime of variability, on timescales of hours with year-long baselines, and with amplitudes that would often be invisible from the ground. These results should be viewed as the properties of this new regime, not as contradictions or confirmations of ground-based optical timing studies.

\subsection{RMS-Flux Relationships and Flux Distribution Histograms}
\label{sec:rmshisto}

The traditional model for AGN accretion disks is the standard optically thin, radiatively thick $\alpha$-disk \citep{Shakura1973}. Although broadly consistent with spectral observations, this model is challenged by the fact that it should be thermally and viscously unstable, especially in the radiation-dominated inner regions where accretion is taking place. As \citet{Kelly2009} point out, variability is a natural probe of temperature fluctuations within the disk since they lead to an understanding of the stress-pressure coupling, and can provide a powerful test of the traditional $\alpha$-disk model.

Numerous theoretical models which deviate from or complicate the $\alpha$-disk have been proposed to explain the observed variability. The most intuitive explanation is fluctuations in the global mass accretion rate: radiation output increases and decreases as more or less matter is converted into energy. This model has enjoyed broad consistency with some observations of luminous quasars \citep{Gu2013}, but struggles to explain the short timescales of the variability: as matter is added to the disk, the rate at which this increases the total optical radiative output is approximately the viscous time. For typical black hole masses, this can be hundreds of years, while variability is seen on timescales of days and hours. 
A more favored idea is that the rapid optical variability is powered by reprocessing of fast variations in the X-ray and far-UV light very close to the black hole, which can happen on the observed timescales.

In the traditional model, $\alpha$ is a constant that embodies one prescription of the stress-pressure coupling which governs the outward transport of angular momentum and, thereby, the inward transport of material for accretion. More realistic prescriptions allow $\alpha$ to vary throughout the disk. These inhomogeneities can be generated by various instabilities, especially the magneto-rotational instability \citep[MRI; ][]{Balbus1991}. Local fluctuations then propagate inwards with the accreting material. 

Steady advances in the resolution and power of magnetohydrodynamic simulations have suggested several observable consequences of the propagating fluctuations model. One testable prediction is that histograms of the fluxes observed in a light curve will be log-normal, rather than gaussian. This has been seen in X-ray light curves of AGN, but not optical. They also predict that the rms varability of a given segment of the light curve will linearly correlate with the mean flux of that segment. This has also been seen in X-ray AGN variability studies, as well as X-ray light curves of stellar mass black holes and accreting neutron stars, but has not been confirmed in optical AGN light curves. If indeed the fluctuations originate further out in the disk where optical emission dominates, and then move inward to cause the observed X-ray variability, we should see these traits in optical AGN time series. Understanding the origin and interaction of the X-ray and optical emitting regions will enable tight constraints on important theoretical questions in AGN physics, especially on shape and location of the X-ray corona. 

We have rebinned each non-interpolated light curve into 2-day bins to overcome noise fluctuations, and then cut each light curve into 50-day segments. For each segment, we calculate the mean flux and the standard deviation of the flux values. None of our objects exhibits any correlation between these two quantities. A typical plot is shown in Figure~\ref{fig:exrms}. This is in stark contrast to X-ray light curves of AGN, which show remarkably tight linear correlations between flux and variability \citep[e.g.,][]{Edelson2002,Vaughan2003a,McHardy2004} that are also seen in stellar-mass black holes \citep{Uttley2001}. X-ray light curves typically probe much shorter timescales (on the order of seconds to minutes for AGN) than optical light curves, so this result may indicate that the origin of the variability on day to year timescales is not simply reprocessing of faster variability in the X-ray emitting region. 

We have also created histograms of the fluxes in each light curve. We alert the reader to the fact that flux histograms are the product most susceptible to incorrect CBV corrections and long-term systematics, as they necessarily include long timescale variability. Recall that the likely validity of CBV corrections for each object is explained in Appendix~\ref{appendixb}. These have a wide variety of shapes in our sample, and are shown in Figure~\ref{fig:histo}. Some are well-fit by log-normal distributions, as predicted by the propagating fluctuations model and frequently seen in X-ray AGN variability studies, but most are best represented by other shapes. Perhaps the most intriguing are those that are best-fit by a bimodal distribution of two gaussians. The light curves of these objects in Appendix~\ref{appendixa} are marked by red dotted lines that denote the peak values of each gaussian component. Most of the time, the bimodal behavior is due to the AGN seeming to transition to a lower flux state, and then return to the original state several hundred days later, as exemplified in KIC~3347632, KIC~9145961, KIC~9215110, and KIC~10841941. The fact that the light curve tends to return to the the same average original flux could indicate preferred accretion states; however, the fact that the variability properties do not seem to change when the object is in a low or high state is perhaps contrary to expectations for state transitions. Another possible explanation is the passing by of obscuring material. One can imagine a cloud of gas and dust passing in front of an unobscured varying continuum source, dimming the source without affecting the observed variability. When the cloud had passed, the average flux would return to the original state, as we see in these cases. 

\begin{figure*}
\begin{tabular}{cc}

\includegraphics[width=0.5\textwidth]{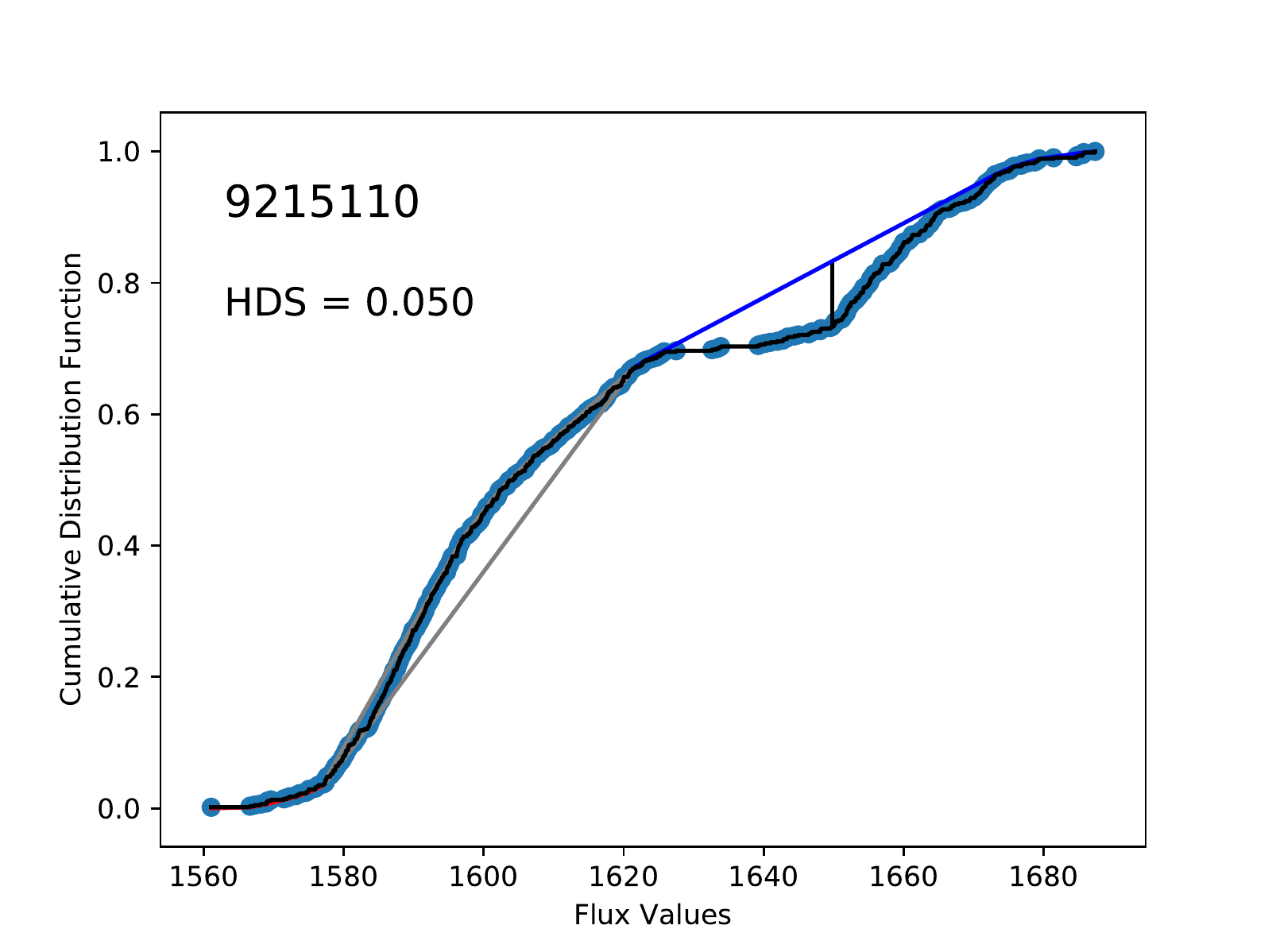} & \includegraphics[width=0.5\textwidth]{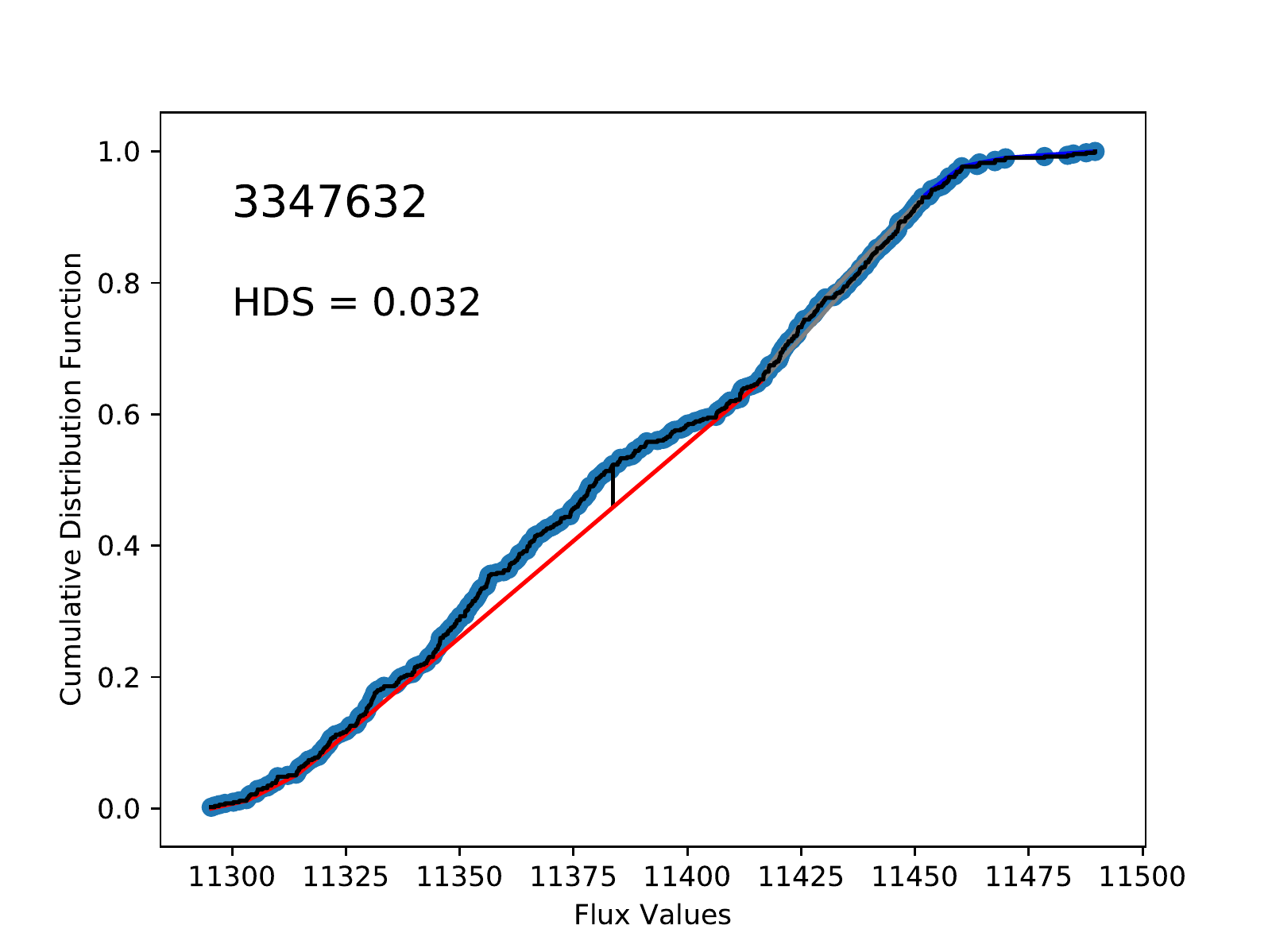} \\

\end{tabular}
\caption{Results of the Hartigan's Dip Test \citep{Hartigan1985} using 1-day binned cumulative distribution functions for the most significant (99.99\% in KIC~9215110, left) and least significant (6.0\%, KIC~7610713, right) bimodalities in our sample. The statistic determines the maximum difference, or ``dip," between the observed CDF and a unimodal one that minimizes the dip. In the case of total insignificance, as on the right, the difference between the empirical and unimodal CDF is not significant.}
\label{fig:hartigan}
\end{figure*}

To investigate the statistical reality of these bimodal states, we have conducted the Hartigan's Dip Test \citep{Hartigan1985} for all objects with bimodal histograms. The test compares the cumulative distribution function (CDF) of the data with a unimodal CDF that minimizes the maximum difference between itself and the empirical CDF. Figure~\ref{fig:hartigan} demonstrates the results of this test for KIC~9215110, which has the most significant ``dip" in the CDF, to KIC~7610713, the least significant\footnote{These plots were made using a Python implementation of the Hartigan Dip Test which can be found at https://github.com/tatome/dip\_test.}. Comparison of the Hartigan Dip Statistic (HDS) to its standard significance tables requires a substantially smaller sample size than the typical length of our light curves, so we bin the light curves into 4-day and 1-day bins to assess the dependence of the significance on bin size. Two objects, KIC~9125110 and KIC~10841941, are more than 99\% significant regardless of bin size. At 1-day binning, KIC~3347632 and  KIC~10798894 are also above 99\% significance, but are reduced to 78\% and 32\% in 4-day bin sizes respectively. The dramatic reduction for KIC~10798894 is due to its very short baseline: the 4-day bins reduces the total number of datapoints to $\sim50$. One of our objects, KIC~9145961, is only marginally significant, coming in at 97\% with 1-day bins and 55\% at 4-day bins, while KIC~7610713 is fully insignificant at both bin sizes (6\% and 16\% for 1- and 4-days). Indeed, KIC~7610713 is the only object with a bimodal histogram that does not exhibit the steady switching behavior just discussed; an examination of its light curve suggests that the bimodal histogram is simply due to red noise fluctuations. 

Finally, we note that if the variable AGN core does not outshine its host galaxy by very much, and if its own flux state occasionally drops below the constant flux of the galaxy, there would be a ``floor" in the light curve that might take the form of a sharp cutoff of low fluxes in the histograms. This may be the case,or example, in the light curve of KIC~11606854.

\begin{figure*}[htb]

\caption[Flux Histograms of the \emph{Kepler} AGN]{Flux distribution histograms, with either lognormal, gaussian, or bimodal gaussian fits (magenta curve), continued on next page.}
\label{fig:histo}

   \includegraphics[width=\textwidth]{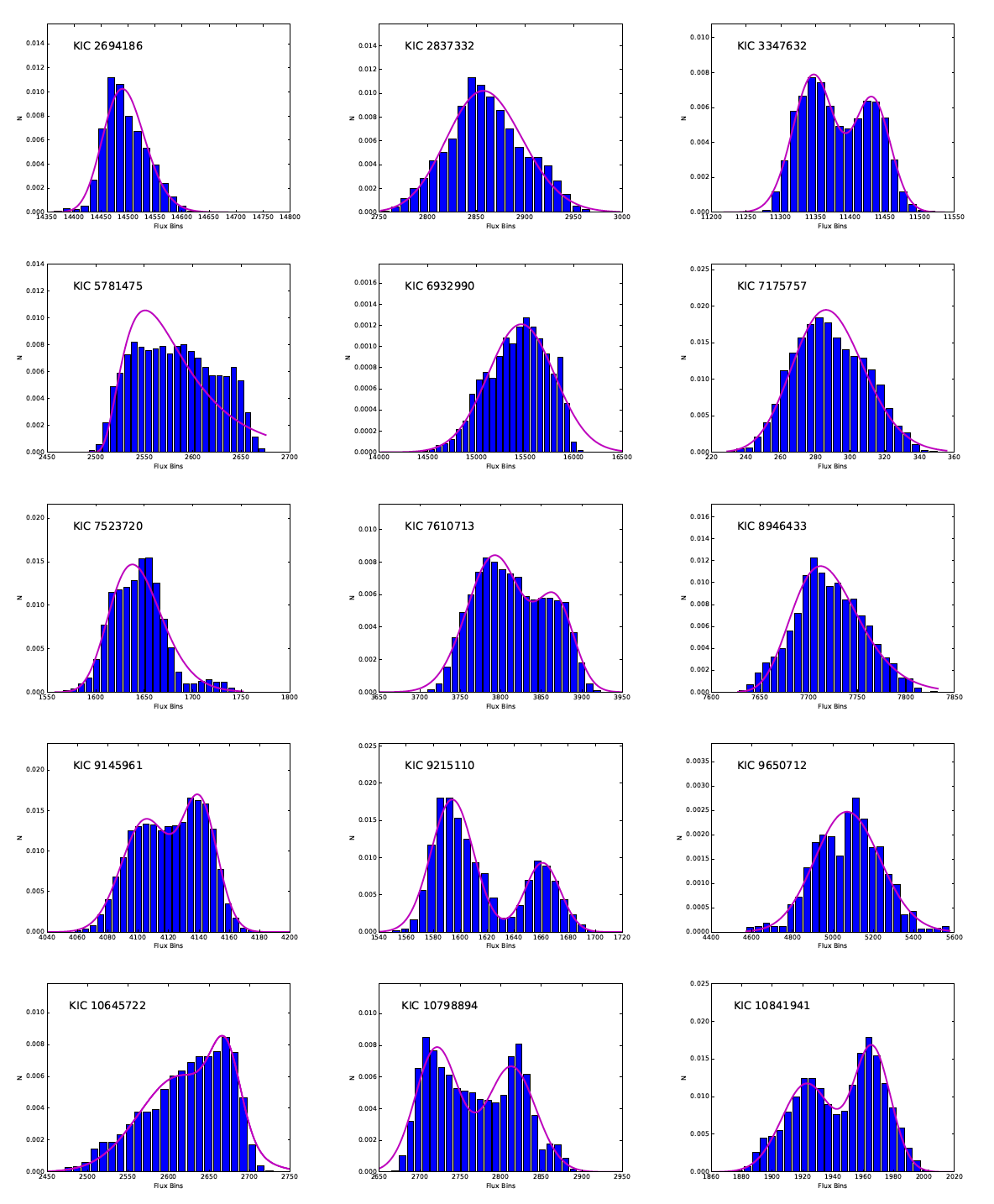}

\end{figure*}

\begin{figure*}[htb]\ContinuedFloat

   \includegraphics[width=\textwidth]{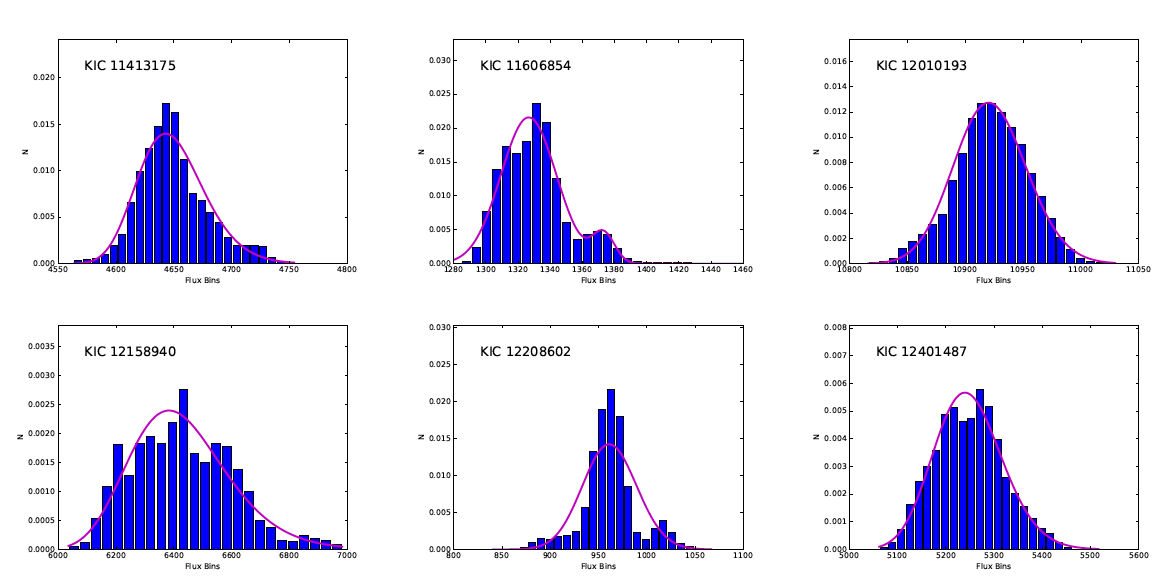}

\caption[Flux Histograms of the \emph{Kepler} AGN]{Flux distribution histograms, with either lognormal, gaussian, or bimodal gaussian fits (magenta curve).}
\label{fig:histo}
\end{figure*}

One might expect that objects with high X-ray/optical flux ratios would be more likely to experience significant reprocessing and therefore more likely to show the same properties as X-ray light curves. Of our 21 objects, 17 have archival X-ray information in the Second ROSAT All-Sky Survey \citep{Boller2016} or XMM Slew Survey \citep{Saxton2008}. We converted the X-ray count rates to fluxes assuming a power law with $\Gamma = 1.8$ as the X-ray spectral slope and galactic column densities from \citet{Kalberla2005}. For optical fluxes, we use the same 5100\AA~or 3000\AA~flux that was measured for bolometric luminosity calculations in Section~\ref{sec:spectra}. In the case of XMM data, we use only the flux in Band 6, 0.2-2 keV, which is the closest XMM band to ROSAT (0.1-2.4 keV). Although these are approximate measures, especially since they are disparate in time from the optical spectra, we require only a qualitative handle on the degree to which the X-ray flux may influence the optical behavior. The values range between $F_X/F_O = 0.03$~to 4.21, with a mean of 0.08. There is no variability-flux relationship even for the highest $F_X/F_O$ objects. We also do not see any tendency for higher $F_X/F_O$ objects to be more often well-fit by lognormal distributions.

\subsection{Variability Properties and Physical Parameters}
\label{correlations}

\subsubsection{A Brief Summary of Past Results}
\label{sec:briefsum}

Next, we test for possible correlations of various properties of the optical variability with the physical parameters of the AGN: black hole mass ($M_{\mathrm{BH}}$), bolometric luminosity ($L_{\mathrm{bol}}$), and Eddington ratio ($L / L_{\mathrm{Edd}}$). We have measured these properties from the optical spectra as described in Section~\ref{sec:spectra}. This is frequently done in variability literature, generally with two goals. The first is to gain physical insight into the geometry and causal relationships within the central engine. The second is to establish correlations that will enable measurements of vast samples of black hole masses and accretion rates from easily-observed variability statistics. If this can be accomplished, upcoming large timing surveys like LSST would enable unprecedented insights into the cosmological buildup of black holes in the universe, with implications for galaxy evolution and the growth of structure.

The literature on this subject is varied and often contradictory. This arises from two principal points. First, any given study must choose how to quantify the ``variableness" of a light curve. Many authors invent their own measures of variability; for example, \citet{Wold2007} use the standard deviation, mean, median and maximum of the distribution of values of $\Delta m_{ij} = m_i - m_j$, the difference in magnitude of subsequent photometric measurements. \citet{Kelly2013} use the square of a parameter of the Ornstein-Uhlenbeck process that describes the strength of the driving noise, $\varsigma^2$, which controls the amplitude of variability on timescales much shorter than the characteristic time. There are other such examples, and most of them indeed capture the ``strength" of variability across sources, but at the cost of complicating comparisons between different works. The second sticking point in the literature is the great variety of sources of light curves, with significant differences in sampling rate and uniformity, photometric accuracy, and the properties of the AGN sample. The relatively small literature ranges from studies of 13 Sy1 galaxies from AGN Watch \citep{Collier2001} to ensemble studies of tens of thousands of quasars from the Palomar-QUEST survey \citep{Bauer2009}. Depending on the properties of the light curves, authors choose to use structure functions, periodograms, or autocorrelation analyses. In addition to the different analysis techniques, the intrinsic variability properties of relatively low-luminosity Seyferts may be quite different from luminous quasars, and there is even evidence that radio loudness might be correlated with variability \citep[most recently, by ][]{Rakshit2017} - and so, the selection effects that plague the rest of AGN literature are also at work here.

Space-based light curves like those from \emph{Kepler}, K2 and the upcoming Transiting Exoplanet Survey Satellite (TESS) have the potential to address some of these problems, in that their sampling rate, uniformity, and photometric precision is far better than ground-based surveys. Such light curves can be downsampled to resemble any ground-based survey, and the timing results from a given study could be compared to the ``true" behavior (the space-based measurements being, presumably, as close as we can get).

The \emph{Kepler} sample of AGN is small, due to the unexpectedly short lifetime of the spacecraft preventing our obtaining long-baseline light curves of a much larger sample (see Section~\ref{sec:selection}). The photometry allows us to measure variations as small as 0.1\%, and the thirty-minute sampling rate over several years is not comparable to any ground-based studies. Therefore, results from this study should not be seen as contradicting any of the ground-based results, as the behavior being measured is quite different. Indeed, much of the variability captured by the \emph{Kepler} light curves would be very challenging to detect from the ground. The sample is too small to allow for binning by any parameter, which is a strong limitation since the variable behavior in any given light curve, and when we happened to observe it, could dilute any correlations. 

There are a few conclusions that seem to be reasonably consistent across the optical variability literature so far. First, variability has been seen to correlate with redshift and hence anticorrelate with rest wavelength, which is best understood as the tendency of UV light to be more variable than optical \citep[e.g.,][]{Cristiani1997}.  An anticorrelation between luminosity and variability was first discovered by \citet{Angione1972}, and has been confirmed many times since. This relationship was seen by \citet{Hook1994} in a sample of 300 optically-selected quasars, by \citet{Cristiani1997} in a sample of 149 optically-selected QSOs, by \citet{Giveon1999} in a sample of 42 Palomar-Green quasars, by \citet{Kelly2013} in a sample of 39 AGN, by \citet{Simm2016} in 90 X-ray selected AGN across a broad redshift range, in ensemble studies of thousands of quasars binned by luminosity and black hole mass \citep{Bauer2009,Wilhite2008,Zuo2012}, and all the way into the most recent studies of Seyferts in the Catalina survey \citep{Rakshit2017}. In contrast, \citet{Wold2007} do not see any trend between variability and luminosity in a matched sample between the Quasar Equatorial Survey Team Phase 1 (QUEST1) survey and the SDSS; however, they believe this may be due to the strong trend between redshift and luminosity in their data.

While the relationship with luminosity seems fairly well established, the relationship of various properties with black hole mass and Eddington ratio is much less certain. \citet{Wold2007} found a scattered, positive correlation between variability and $M_{\mathrm{BH}}$. This was also seen by the ensemble study of \citet{Bauer2009}, and \citet{Giveon1999} reported a positive correlation between variability and H$\beta$~equivalent width. In contrast, \citet{Kelly2013} found a scattered anticorrelation between their $\varsigma^2$ variability measure and $M_{\mathrm{BH}}$, as well as a weak anticorrelation of variability and Eddington ratio. \citet{Simm2016} found no relationship between variability amplitude and $M_{\mathrm{BH}}$, and an anticorrelation of the excess variance and variability amplitude with Eddington ratio. Finally, \citet{Zuo2012} found that whether or not variability and $M_{\mathrm{BH}}$ were correlated depended on the Eddington ratio and luminosity. 

Having summarized the current state of confusion, we will now present our results, which as stated before should be viewed as the properties of an entirely different sort of variability than those reported on before.

\subsubsection{Correlations with Physical Parameters}
\label{mvar}

We have chosen to measure the variability of our light curves via two metrics. Before calculating either of these quantities, we rebin our light curves into 2-day time bins, to mitigate gaps and overcome the effects of noise. 

\begin{figure}[t]
    \centering
    \includegraphics[width=0.5\textwidth]{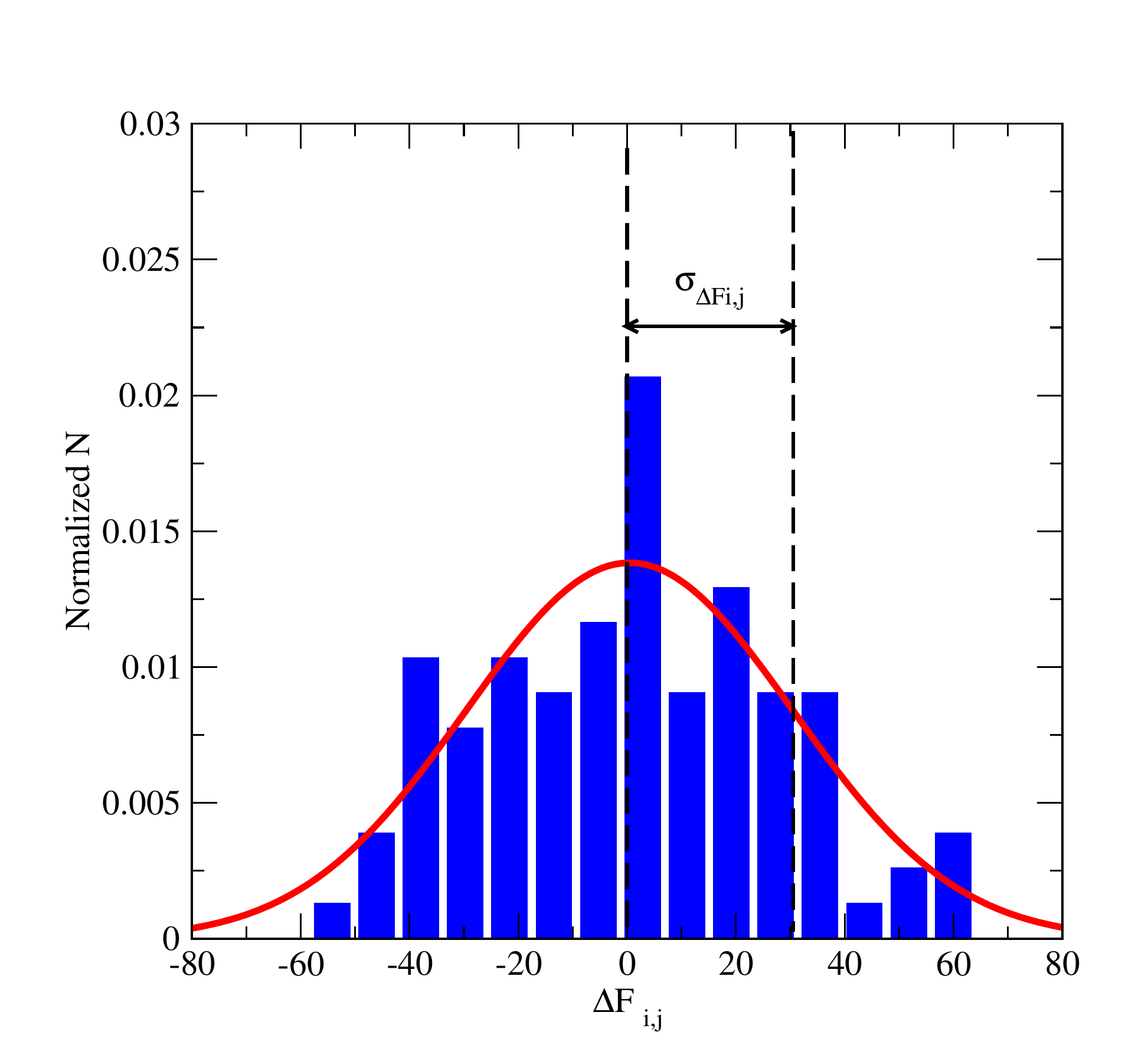}
    \caption{Example of the distribution of the difference between subsequent flux measurements $\Delta F_{ij}$ used to measure the variability of our light curves; the metric used is the width of the gaussian fit (red). This is the distribution for KIC~10645722.}
    \label{fig:woldex}
\end{figure}

\begin{figure}[t]
    \centering
    \includegraphics[width=0.5\textwidth]{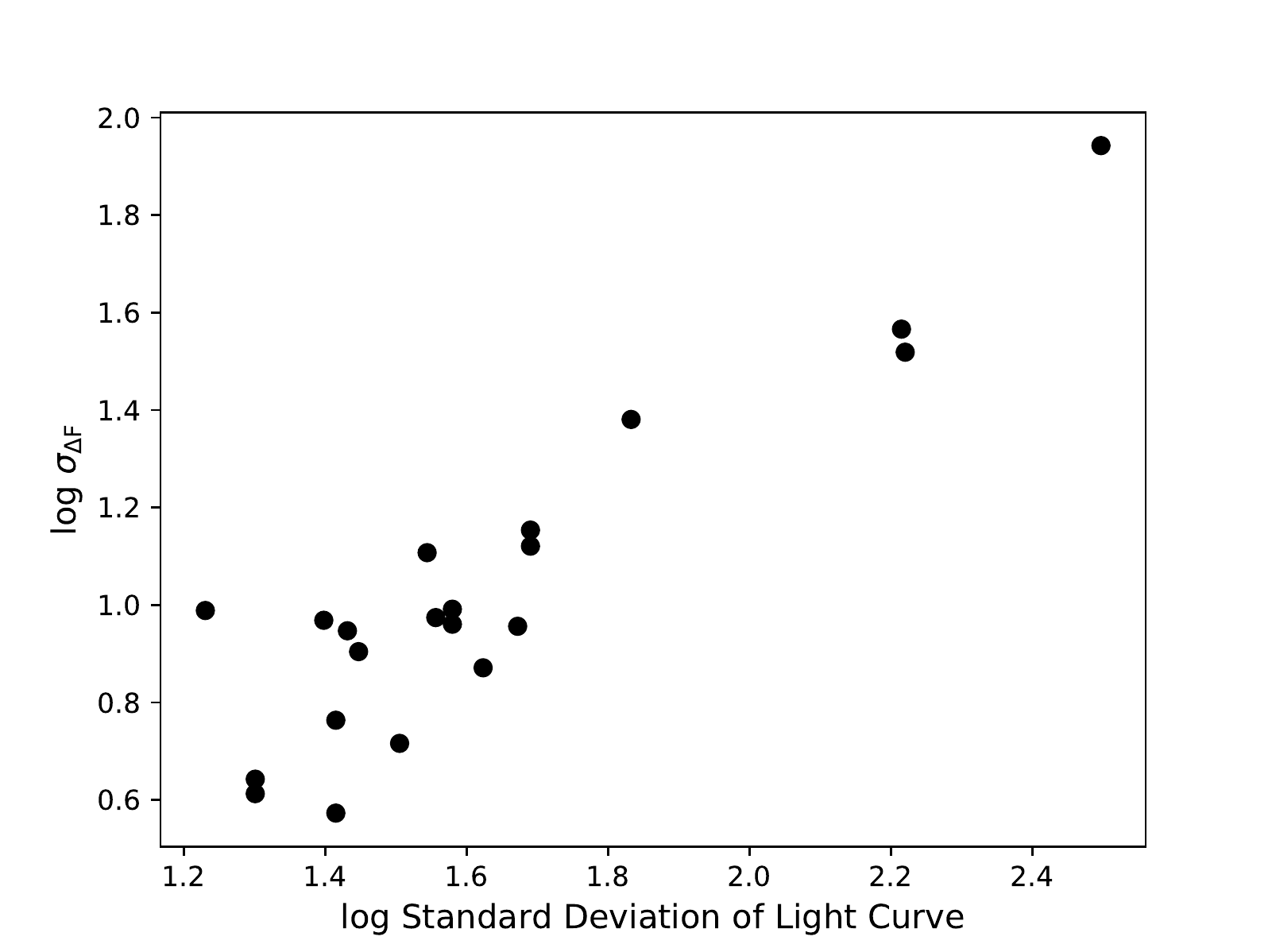}
    \caption{Variability metric $\sigma_{\Delta\mathrm{F}}$ versus the standard deviation of the light curve. Both are shown for light curves with 2-day binning.}
    \label{fig:varmetrics}
\end{figure}

The first measure of variability is the standard deviation of the distribution of values of the difference between subsequent flux measurements. We calculate the distribution of $\Delta F_{ij} = F_i - F_j$, and fit this distribution with a gaussian, the width of which is our variability criterion $\sigma_{\Delta\mathrm{F}}$. As an example, Figure~\ref{fig:woldex} shows this distribution for KIC~10645722. Our second measure of variability is simply the standard deviation of the binned light curves. Figure~\ref{fig:varmetrics} compares these two metrics for the objects in our sample. Both capture variability and quantify it generally the same way, but there is scatter at low variability. We believe there is value in exploring correlations with both metrics, since there is such a wide variety of homegrown metrics used in the literature. The values of $\sigma_{\Delta\mathrm{F}}$ and the standard deviation are given in Table~\ref{t:tab2}. In our very shortest light curves, the 15-day binning reduced the sampling too far for the distribution of $\sigma_{\Delta\mathrm{F}}$ to be gaussian. These three objects are not included in the table or the figures.

Appendix~\ref{appendixb} discusses the effects of CBV correction on these metrics. In summary, the $\sigma_{\Delta\mathrm{F}}$ values are not largely affected, since they measure only short-term variations. The standard deviation is generally inflated in the uncorrected light curves, but is also definitely incorrect for some uncorrected objects where spacecraft systematics are certainly present before CBV application. In these plots only the values for the corrected light curves are shown.


\begin{table*}
\caption{The Variability Properties of the \emph{Kepler} AGN}
\centering
\footnotesize
{\renewcommand{\arraystretch}{0.5}
\begin{tabular}{lccccc}
\hline\hline
KIC \# & $\sigma_{\Delta\mathrm{F}}$ & $\sigma_{\Delta\mathrm{F}}$ & $\sigma_{\Delta\mathrm{F}}$ &$\sigma_{\Delta\mathrm{F}}$ & Std. Dev. \\
  & 2 days  &  5 days &  10 days & 15 days & 2 days    \\
\hline
\\
10841941	&	5.8	&	10.4	&	11.05	&	12.5	&	26	\\
10645722	&	14.24	&	26.24	&	39.9	&	45.2	&	49	\\
7175757	&	9.74	&	7.97	&	12.78	&		&	17	\\
2694186	&	9.8	&	18.04	&	25.2	&	30.6	&	38	\\
6932990	&	87.58	&	145.89	&	241	&	284	&	314	\\
2837332	&	9.42	&	18.05	&	29.3	&	46.24	&	36	\\
9145961	&	4.39	&	8.12	&	9.09	&	7.76	&	20	\\
12401487	&	24.02	&	24.02	&	43.7	&	169.02	&	68	\\
5781475	&	9.13	&	24.3	&	53	&	42	&	38	\\
8946433	&	12.8	&	17.7	&	5.2	&		&	35	\\
11606854	&	4.1	&	5.1	&	8.06	&	4.9	&	20	\\
12010193	&	8.02	&	21.1	&	39.7	&		&	28	\\
9215110	&	5.2	&	9.1	&	12.6	&	10.9	&	32	\\
7523720	&	9.3	&	18.8	&	21.04	&	26.13	&	25	\\
12158940	&	33	&	73.4	&	104	&	115	&	166	\\
12208602	&	3.74	&	3.62	&	5.48	&	4.9	&	26	\\
9650712	&	36.8	&	73.67	&	134.74	&	170.9	&	164	\\
10798894	&	13.2	&	21.54	&	18.69	&	24.32	&	49	\\
7610713	&	7.43	&	16.1	&	30.4	&	31.1	&	42	\\
3347632	&	9.04	&	18.9	&	22.8	&	35.2	&	47	\\
11413175	&	8.85	&		&		&		&	27	\\
[1ex]

\hline
\end{tabular}}
\label{t:tab2}
\\[10pt]
Our two variability metrics given for each object. The first, $\sigma_{\Delta\mathrm{F}}$, is the subsequent flux difference distribution width as shown in Figure~\ref{fig:woldex}, for bins of 2, 10, 15, and 20 days. The second is the standard deviation of the light curve with 2-day bins.
\end{table*}


\begin{figure}[t]
    \centering
    \includegraphics[width=0.5\textwidth]{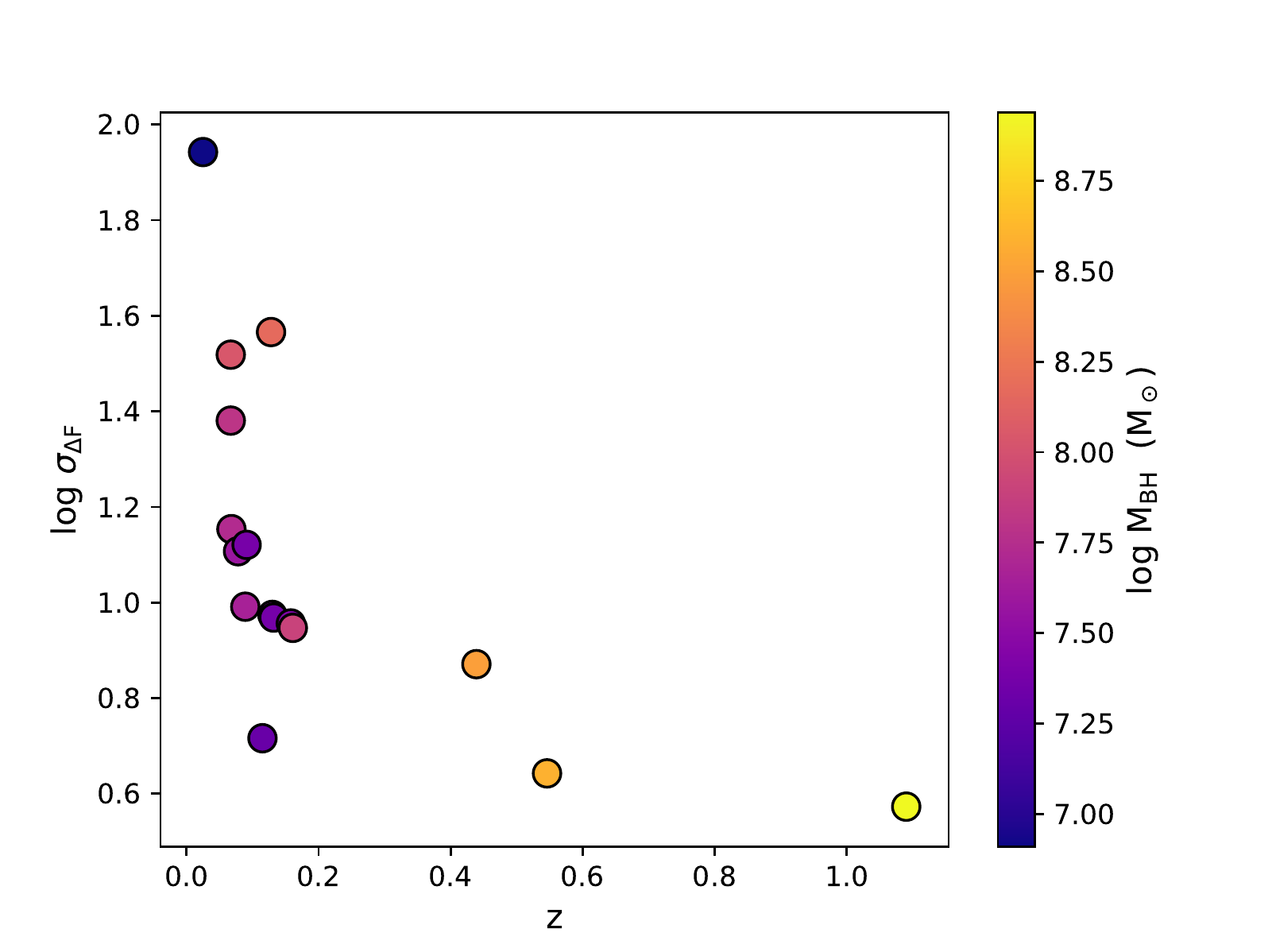}
    \caption{Variability metric $\sigma_{\Delta\mathrm{F}}$ (2-day bins) versus redshift, with log~M$_{\mathrm{BH}}$ as a color gradient.}
    \label{fig:zstd}
\end{figure}


We first note that our sample, unlike most others, does not show higher variability at higher redshifts. In fact, the opposite is true. This is most likely because our very few high-redshift objects are also our most massive black holes, which tend to vary less or at least more slowly than the other objects in our sample. This relationship is shown in Figure~\ref{fig:zstd}. 

\begin{figure*}
\begin{tabular}{cc}

\includegraphics[width=0.5\textwidth]{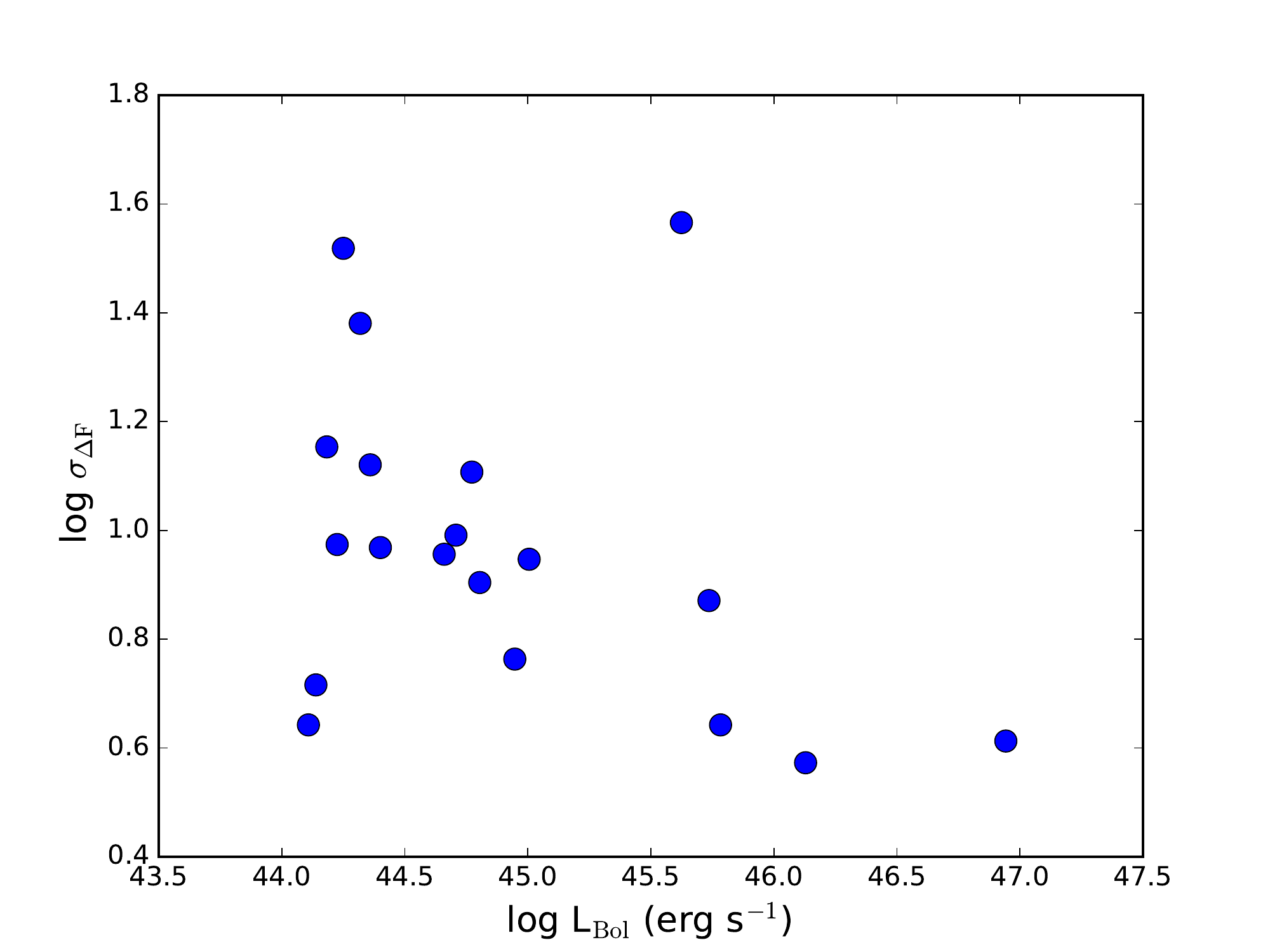} & \includegraphics[width=0.5\textwidth]{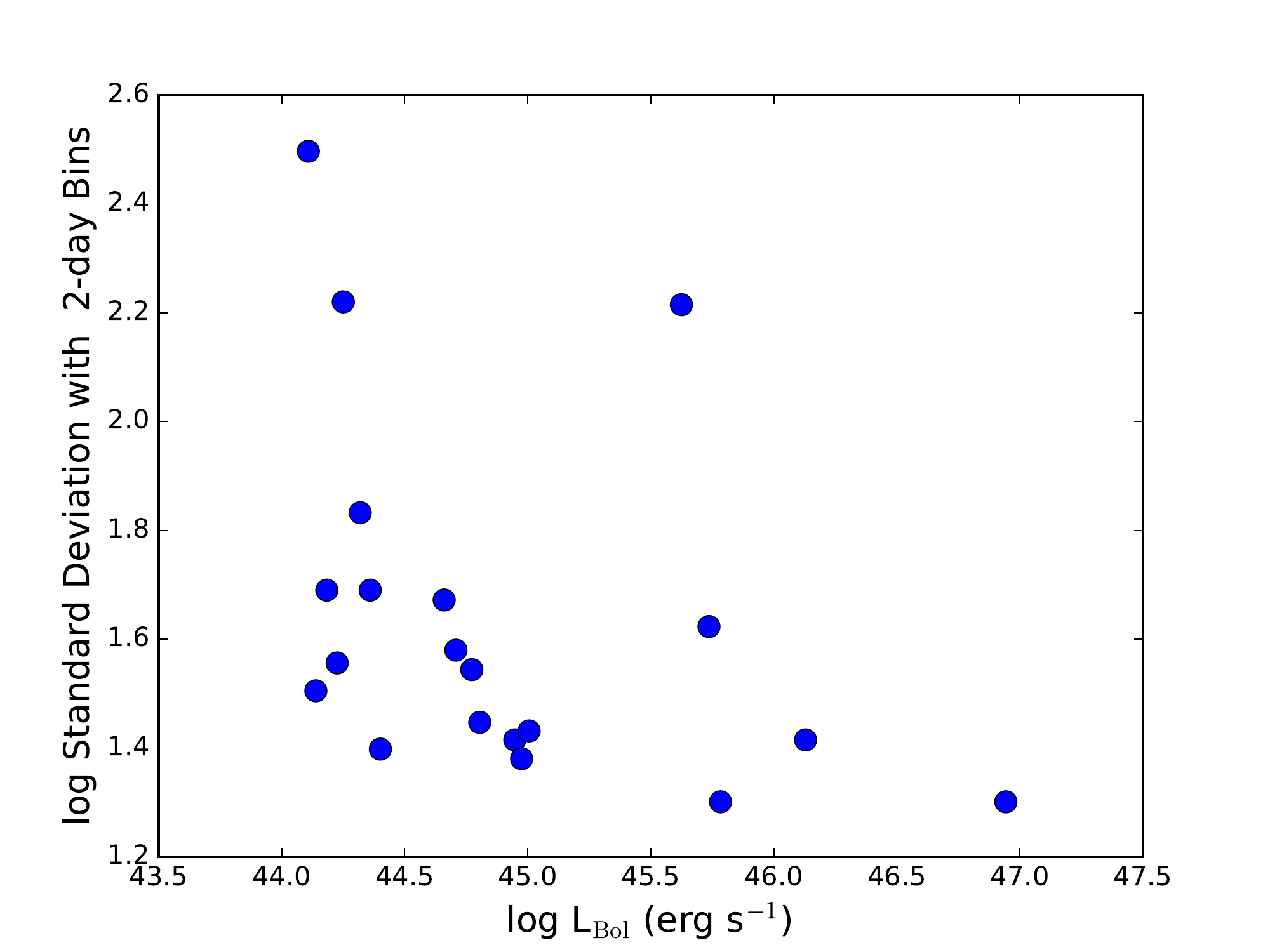} \\

\end{tabular}
\caption{Relationship of bolometric luminosity and our two measures of variability: the width of the distribution of $\Delta F_{ij} = F_i - F_j$ between subsequent flux measurements (for 2-day bins; left), and the standard deviation of the binned light curve (right). We generally duplicate the long-known tendency of brighter objects to be less variable.}
\label{fig:lbolvar}
\end{figure*}

\begin{figure}[t]
    \centering
    \includegraphics[width=0.5\textwidth]{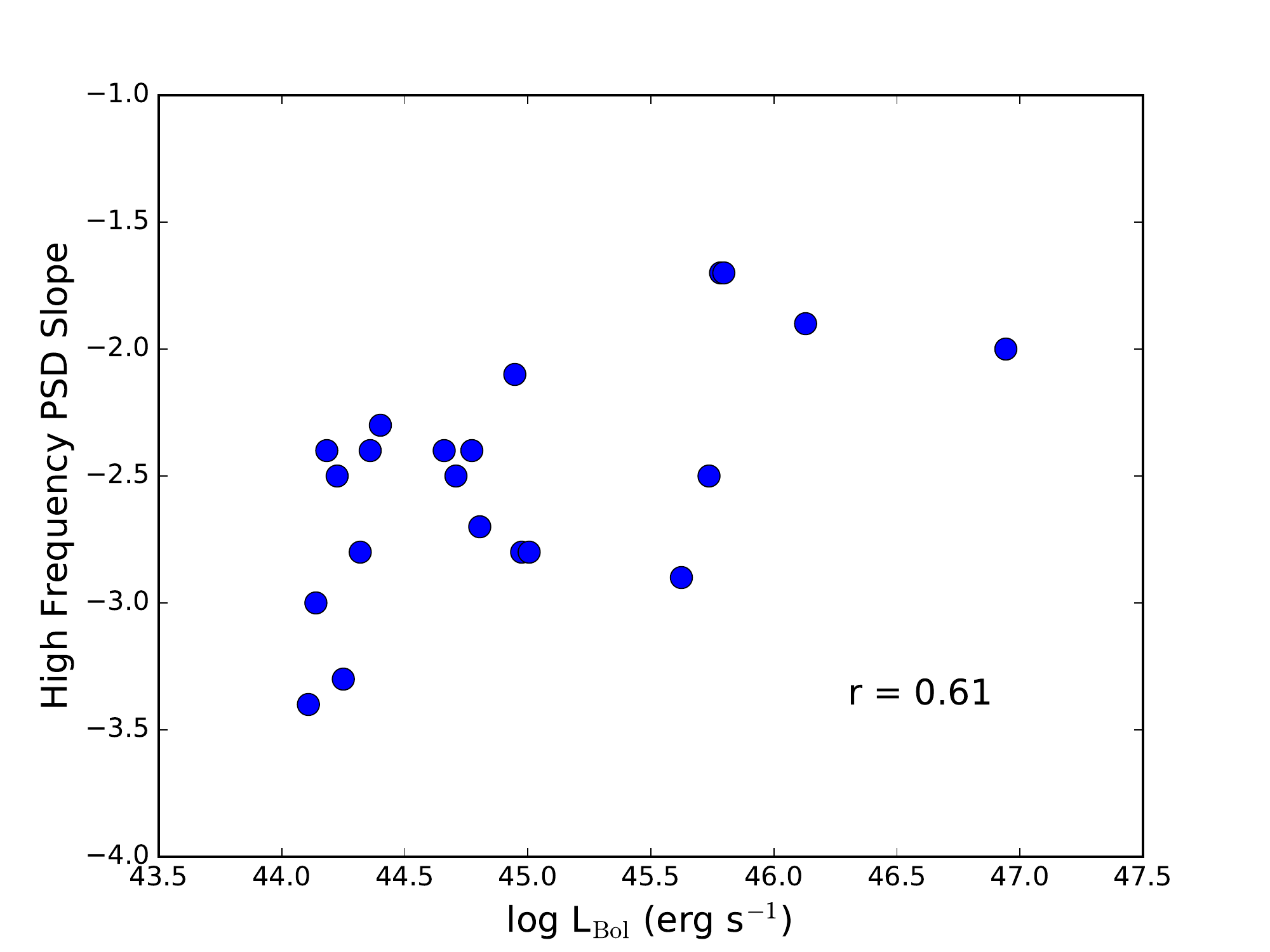}
    \caption{Bolometric luminosity versus best-fitting high-frequency PSD slope. The Pearson correlation coefficient is shown in the bottom right.}
    \label{fig:lbolslope}
\end{figure}

We also see that the \emph{Kepler} data conform to the long-known anticorrelation of luminosity and variability. Although not a linear correlation, the brighter objects in our sample also tend to be the least variable by both of our metrics, as can be seen in Figure~\ref{fig:lbolvar}. We also find a weak anticorrelation between bolometric luminosity and the best-fitting high-frequency PSD slope with a Pearson correlation coefficient of $r = 0.61$, shown in Figure~\ref{fig:lbolslope}, but no relationship between the slope and the Eddington ratio.

\begin{figure}[t]
    \centering
    \includegraphics[width=0.5\textwidth]{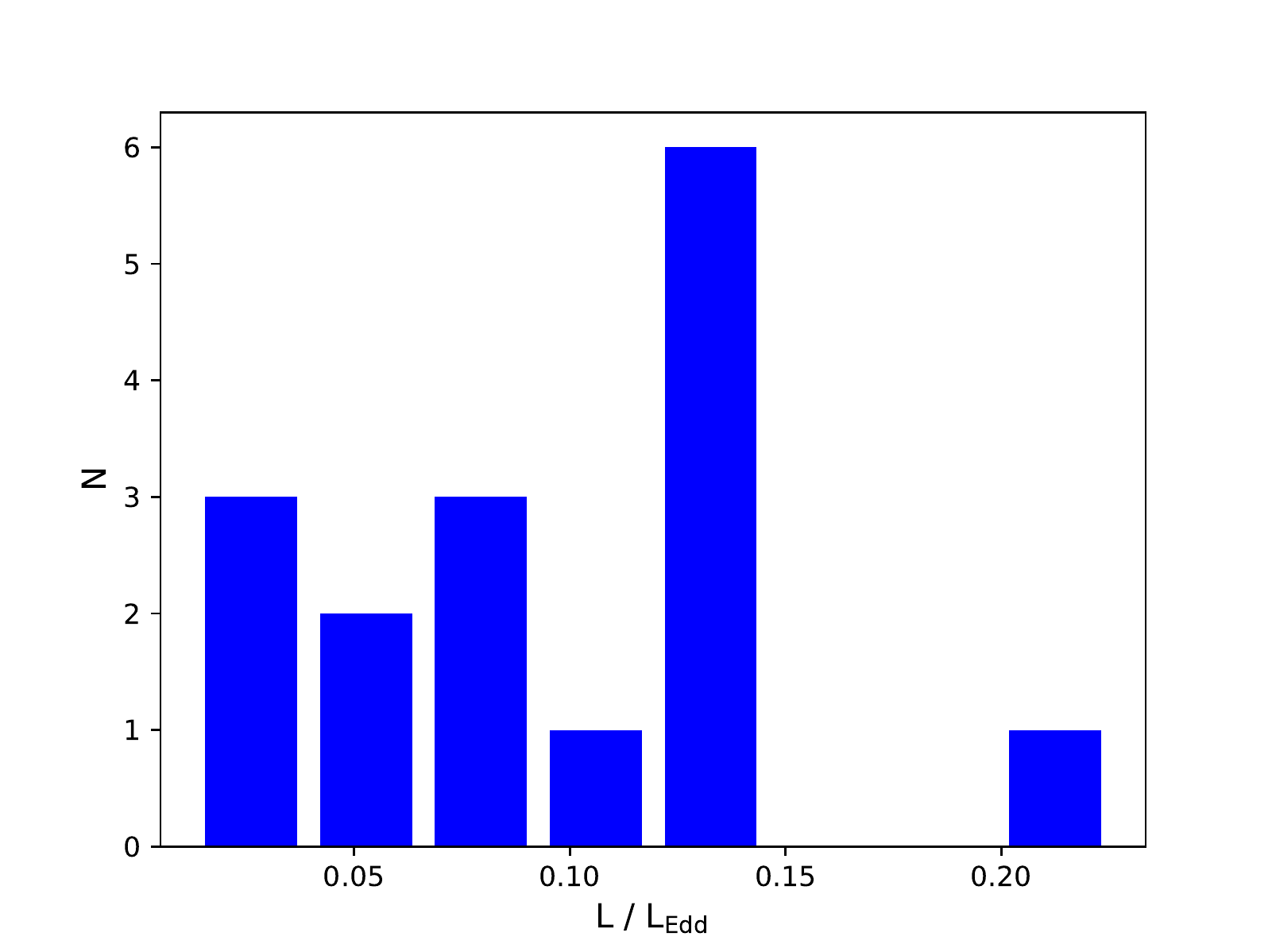}
    \caption{Histogram of values of Eddington ratio for \emph{Kepler} AGN with optical spectra.}
    \label{fig:eddhisto}
\end{figure}

Because of the hints in the literature that accretion rate and mass may both be related to the variability, we split our sample into low- and high-Eddington ratio subsets, with $L / L_{\mathrm{Edd}} = 0.1$~ as the dividing value. This choice was motivated by the possibility that accretion flows with log $L / L_{\mathrm{Edd}} \sim -2$ may be more like advection-dominated flows (ADAFs) than standard thin disks, and because it was a natural value based on the distribution of Eddington ratios in our sample, shown in Figure~\ref{fig:eddhisto}. 

\begin{figure*}
    \centering
    \includegraphics[width=0.9\textwidth]{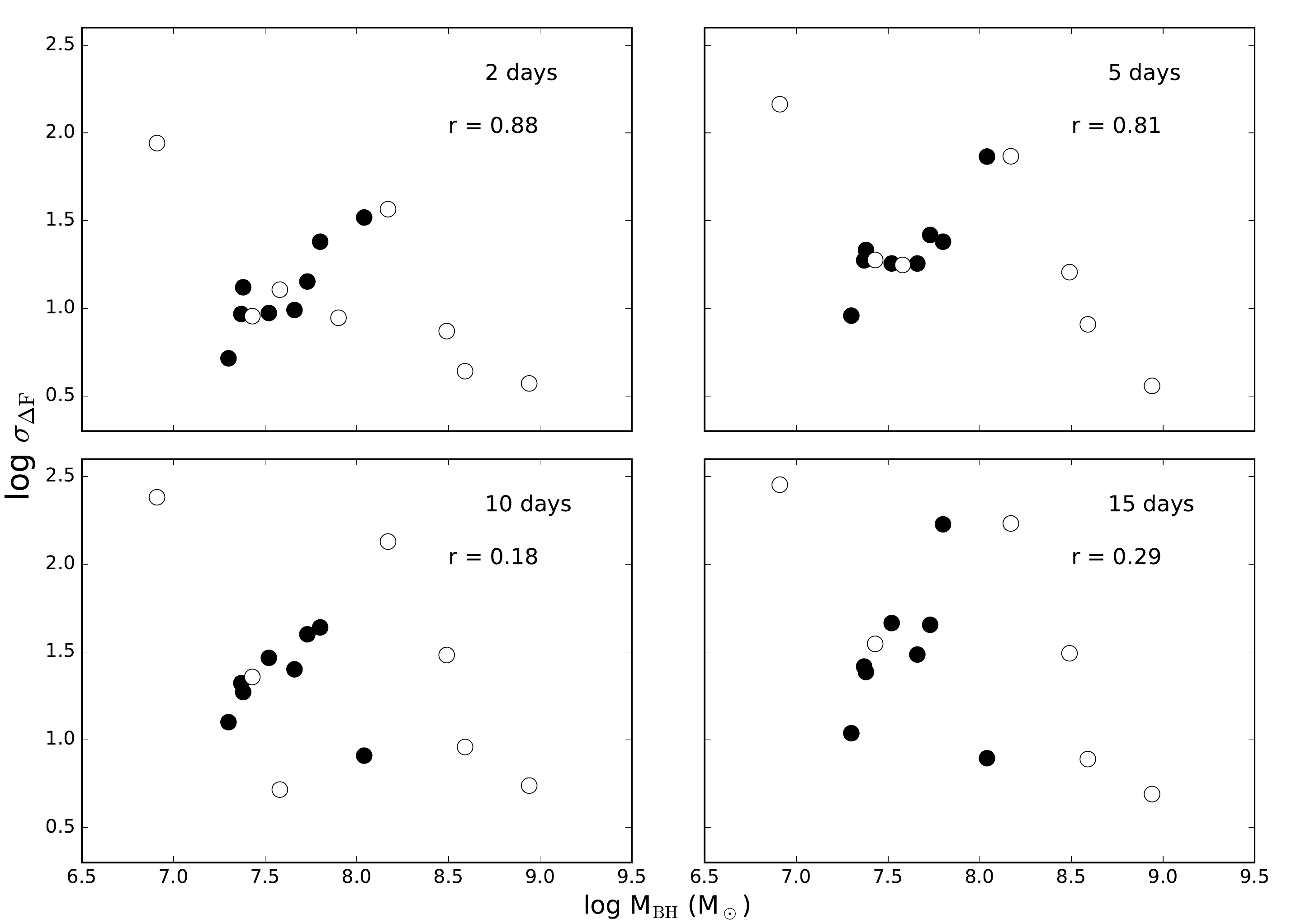}
    \caption{The width of the distribution of $\Delta F_{ij} = F_i - F_j$ versus black hole mass for four increasing light curve bin sizes. In each panel, solid circles represent objects with  $L/L_{\mathrm{Edd}} < 0.1$, and hollow circles represent objects with $L/L_{\mathrm{Edd}} > 0.1$. The Pearson correlation coefficient is shown for the trend seen in low Eddington ratio objects.}
    \label{fig:mbhstdgrid}
\end{figure*}

We find no correlation between $M_{\mathrm{BH}}$~and variability for the full sample. Although the numbers are small once the sample is divided by Eddington ratio, we do see hints that both measures of variability correlate positively with $M_{\mathrm{BH}}$ for low values of Eddington ratio ($L/L_{\mathrm{Edd}} < 0.1$). The strength of the correlation of $M_{\mathrm{BH}}$ with $\sigma_{\Delta\mathrm{F}}$ depends on the width of the time bins as shown in Figure~\ref{fig:mbhstdgrid}, where the Pearson correlation coefficients are also given. Any difference in behavior by Eddington ratio would be consistent with the idea in \citet{Zuo2012} that the correlation with $M_{\mathrm{BH}}$ depends on accretion rate. This effect may disappear with larger samples that permit stricter and more physically-motivated Eddington ratio binning, but any such difference would be an important insight to the dependence of disk structure on accretion rate.


Finally, we find that $M_{\mathrm{BH}}$ is correlated with the break timescale, $\tau_{\mathrm{char}}$. The sample is obviously too small to allow physical conclusions, but is consistent with previous results in \citet{Collier2001}. This relationship is shown in Figure~\ref{fig:mbhtchar}. 
\\

%


%


\subsection{An Unusual AGN Flare}
\label{sec:flare}

\begin{figure}[t]
    \centering
    \includegraphics[width=0.5\textwidth]{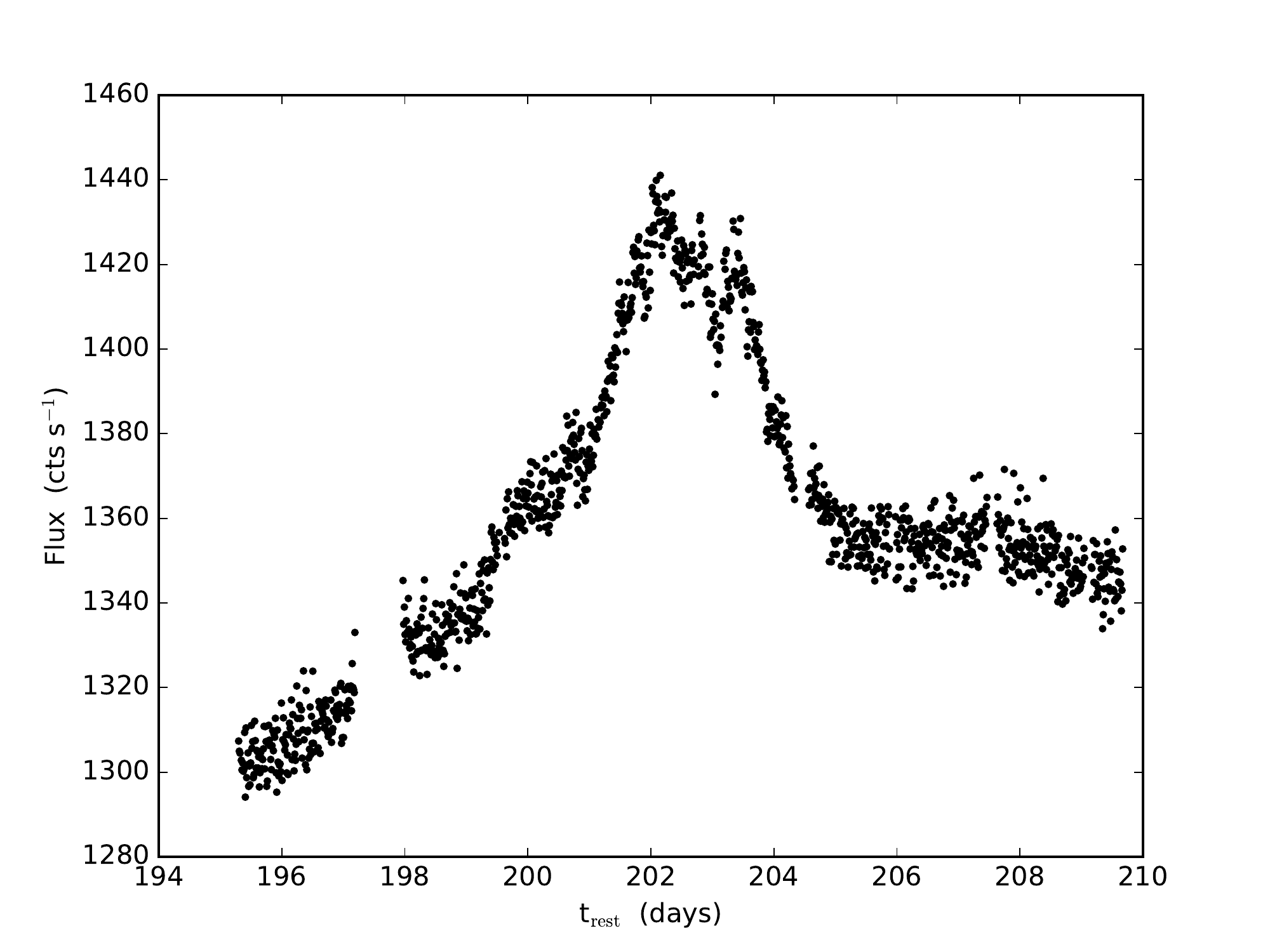}
    \caption{An excerpt of the light curve of KIC~11606852 showing the flare-like event.}
    \label{fig:flarezoom}
\end{figure}

\begin{figure}[t]
    \centering
    \includegraphics[width=0.5\textwidth]{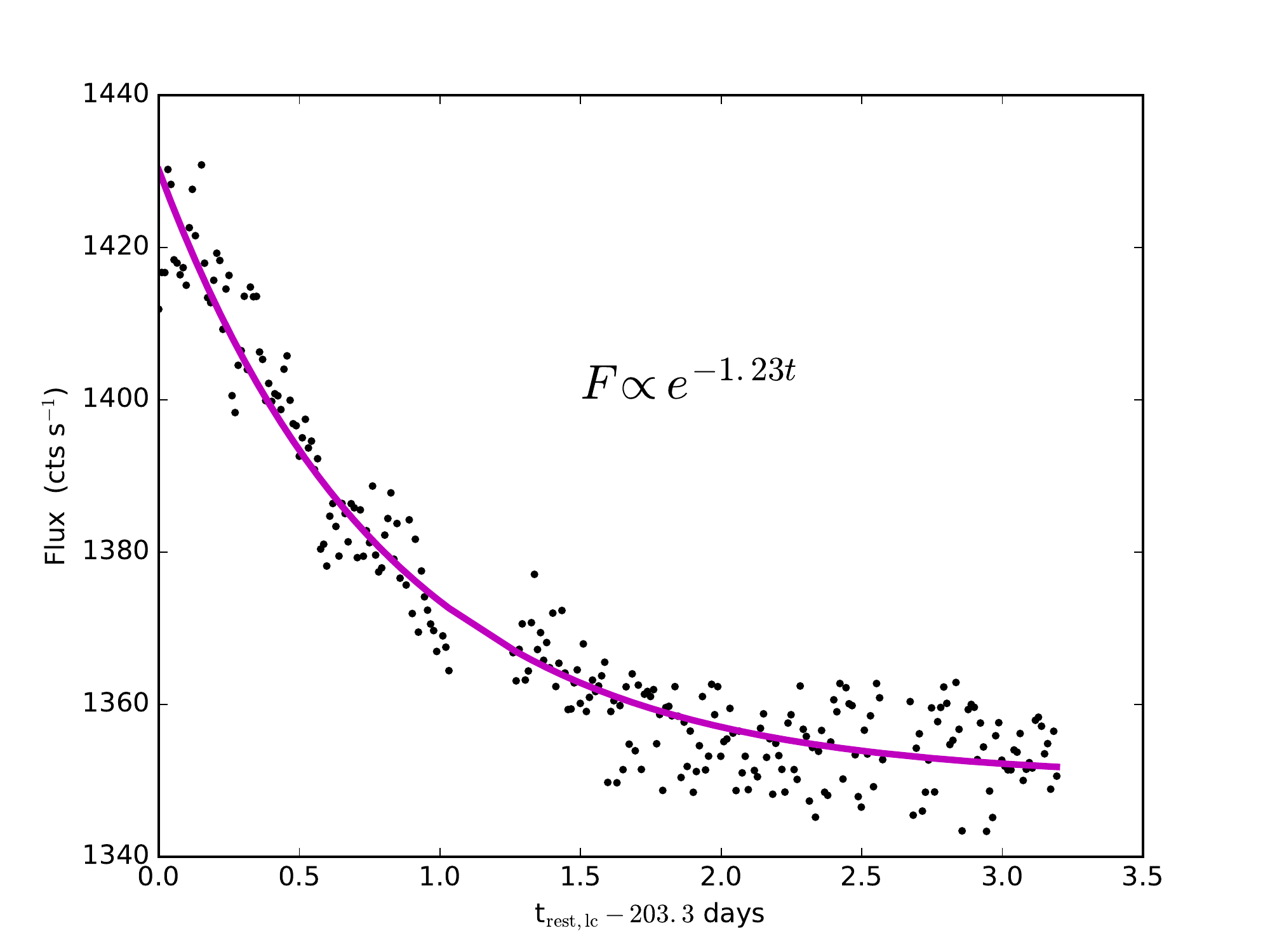}
    \caption{The best-fitting model to the decay of the flare-like event in KIC~11606852.}
    \label{fig:flarefit}
\end{figure}

The light curve of the relatively high-redshift AGN KIC~11606852 exhibits a large flare-type outburst at around 200 rest-frame days. This feature is not seen in the light curves of nearby KIC objects, which means it is not likely to be instrumental. Although it is possible that this event is taking place in a star that is convolved with the PSF of our galaxy, it would have to be quite close to the AGN, as there are no objects in the DSS image near enough to have been included in our extraction aperture. As with the QPO candidate, there is also no contaminating object seen in the J-band UKIRT image.
Figure~\ref{fig:flarezoom} shows the flaring portion of the light curve. The flare only lasts for a few days, too short to be a supernova afterglow. Such a feature could indicate a tidal disruption event (TDE) within the accretion disk \citep[e.g., ][]{McKernan2011} or the result of grazing stellar collisions produced by an extreme mass ratio inspiral (EMRI) pair orbiting the central supermassive black hole \citep{Metzger2017}. We could not get a good fit with the traditional $t^{-5/3}$ profile of a TDE \citep{Komossa2015}. Instead, the feature is best fit by an exponential decay, which we show in Figure~\ref{fig:flarefit}. Exponential decays are possible with the EMRI model, but the odd behavior at the flare's peak may be inconsistent with this approach. This remains a mysterious phenomenon.

\section{Physical Implications}
\label{sec:implications}
\subsection{The Damped Random Walk Model}
\label{sec:drw}

As we have stated before, the origin of optical variability in AGN is not known, and many physical models exist. Quantitative, testable predictions are scarce. The damped random walk model predicts that the high-frequency portion of the power spectrum should be well-fit by a slope of $\alpha=-2$. Like previous work using this sample \citep{Mushotzky2011,Kasliwal2015}, we have found that the PSD slopes are in general steeper than allowed by the damped random walk. Two of our sources have slopes shallower than $-2$, one is best-fit by precisely $-2$, and the remainder exceed this value, sometimes reaching values as steep as $-3.4$. We agree, then, that this model is at the very least not sufficient to capture the variability of AGN as seen by \emph{Kepler}, which we again stress is a new regime of variability. It is indeed possible that this model is correct for ground-based quasar light curves, as maintained by \citet{Kelly2009}, \citet{MacLeod2010} and \citet{Zu2013}; however, some ground-based studies have also found slopes that are steeper than $-2$ \citep{Simm2016,Kozowski2016,Caplar2017}.


\begin{figure}[t]
    \centering
    \includegraphics[width=0.5\textwidth]{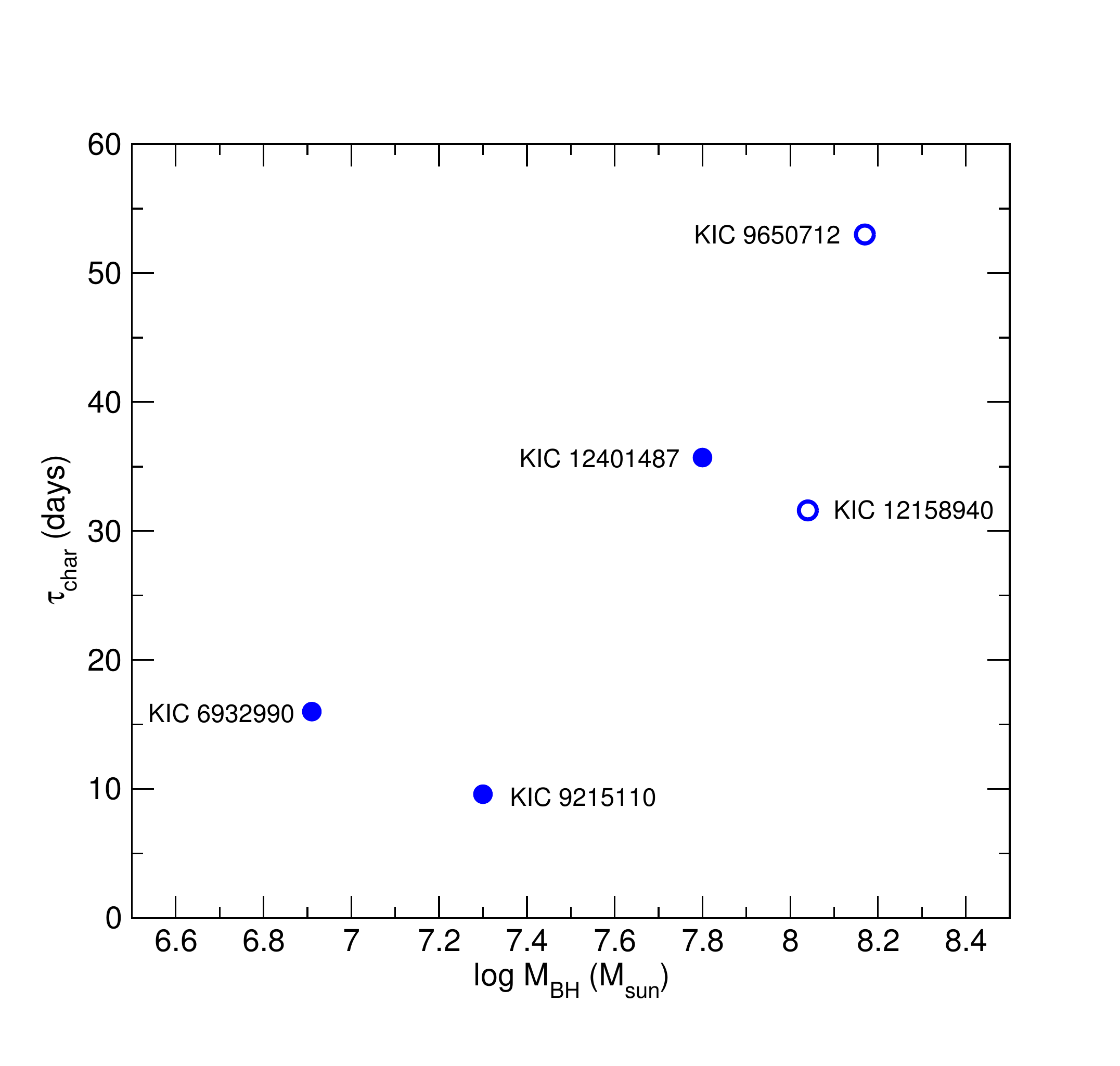}
    \caption{Characteristic timescale from the objects with PSDs best-fit by a broken power-law (see Section~\ref{powspec}) versus black hole mass. The hollow circles represent timescales that are somewhat uncertain as they could be interpreted as a QPO (KIC~9650712) or spuriously added in the CBV-correction (KIC~12158940).}
    \vspace{1cm}
    \label{fig:mbhtchar}
\end{figure}


Much of what we have found aligns with an idea put forth by \citet{Caplar2017}. What if $\tau_{\mathrm{char}}$ was the timescale at which the behavior switches from a steep, red variability to a damped random walk or similar? The timescale would need to be correlated with $M_{\mathrm{BH}}$ to explain what they see in their ensemble study of $\sim$28000 Palomar Transient Factory AGN: bins with higher masses tend to have a higher fraction of steep slopes. They observe steepening in their more massive objects, where they are sampling the pre-break timescale, but do not observe steep slopes in their lower-mass objects, where the putative transition frequency is below their sampling rate. Our sample is too small to allow for such binning - we have only one object that would fall in their highest-mass bin, and it has a shallow slope (as well as the numerous other caveats that we have put forth regarding direct comparison of \emph{Kepler} and ground-based light curves). We realized, however, that we may be able to test an aspect of this model.

Consider that perhaps the PSD of an AGN becomes redder and redder due to a cyclic accretion-disk duty cycle akin to that in X-ray binaries, building towards a critical moment when the behavior switches to a damped random walk. The timescale on which this switch occurs depends on the black hole mass, with longer timescales for larger masses (i.e., larger disks in general). In this case, those objects with observed PSD breaks would have the reddest high-frequency slopes compared to the rest of the sample, which has been caught somewhere in the middle of its reddening phase. This is indeed true in our broken-PSD objects. Five of the six broken-PSD objects have the five steepest slopes in our sample (the exception is best-fit by a slope of $-2.7$). The mean value for our broken-PSD objects is $<\alpha> = -3.0$, compared to $<\alpha> = -2.3$ for the unbroken sample. \citet{Gonzalez-Martin2012} also find steeper high-frequency slopes in objects fit with broken power laws. In larger upcoming surveys with appropriate cadences (sampling at least as often as 3-5 days), this hypothesis could be better tested.
\\

\subsection{X-ray Reprocessing}
Next, we note that the paucity of lognormal flux distributions and the total lack of a correlation between the variability and average flux of a given light curve segment places limits on how much of the 0.1\%$-10$\% level optical variability can be due to X-ray reprocessing. If reprocessing were an important source of optical variability, we would at least expect objects with high X-ray/optical flux ratios to show these traits. As we stated in Section~\ref{sec:rmshisto}, we do not see an rms-flux relationship in any of our light curves, and while some of the flux histograms are well-fit by a lognormal distribution, this does not seem to relate to $F_X/F_O$ in any way. We conclude that the variability probed by \emph{Kepler} is not dominated by X-ray reprocessing, and is more likely to be due to properties inherent to the optically-emitting disk. 

\subsection{Bolometric Luminosity}
A physical cause for the anticorrelation of bolometric luminosity and high-frequency PSD slope is difficult to conjure in the absence of any relationship with Eddington ratio. The bolometric luminosity by itself does not necessarily predict the structure of the disk. It is possible that the phase of a steepening PSD could be accompanied by a general dimming, but it is too early to speculate on why this may happen. It appears that the PSD slope is most consistent with the $\alpha=-2$~ prediction from the damped random walk model in luminous objects. Indeed, many of the ground-based studies cited earlier that found consistency with the damped random walk were performed on samples of luminous quasars. It will be important to look for this trend in future large-scale AGN timing studies of Seyferts and quasars.

\subsection{Characteristic Disk Timescales}
We now turn to the characteristic timescales. There are a number of physical timescales in analytical accretion disk theory. The light-crossing time is too short to be of interest for current optical timing studies and viscous timescales are too long; so we focus here on the orbital, thermal, and freefall timescales. From \citet{Edelson1999}, the orbital timescale is given by

\begin{equation}
t_{\mathrm{orb}} = 0.33 \Big( \frac{M}{10^7 M_\odot} \Big) \Big( \frac{R}{10R_S} \Big)^{3/2} \mathrm{days},
\end{equation}
the thermal timescale is given by

\begin{equation}
t_{\mathrm{th}} = 5.3 \Big( \frac{\gamma}{0.01} \Big)^{-1} \Big( \frac{M}{10^7 M_\odot} \Big) \Big( \frac{R}{10R_S} \Big)^{3/2} \mathrm{days}, 
\end{equation}
where $\gamma$ is the viscosity parameter, and the ADAF accretion timescale is essentially the freefall time \citep{Manmoto1996}:

\begin{equation}
t_{\mathrm{ff}} = 4.62\times10^{-5}  \Big( \frac{M}{10M_\odot} \Big)  \Big( \frac{R}{1000R_S} \Big)^{3/2} \mathrm{days}.
\end{equation}

 All of these are related to the black hole mass. This makes a correlation of characteristic timescales with mass a natural expectation, which is why its elusiveness is particularly vexing. One complication that afflicts all studies, including and perhaps especially this one, is that the light being monitored comes from a significant range of radii within the disk, diluting the correlation. \citet{Collier2001} noted that in their structure function analysis of 13 Sy1 galaxies, higher mass black holes tended to have longer characteristic timescales, but stopped short of calling it a correlation. The relationship they found between M$_{\mathrm{BH}}$~ and $\tau_{\mathrm{char}}$~ is closest to the theoretical expectation for orbital timescales, but with much scatter. They speculate that timescales on the order of a few to tens of days may have a different physical origin than previously-observed characteristic timescales of hundreds of days. \citet{Simm2016} reported break timescales of hundreds of days in all 90 of the Pan-STARRS light curves of XMM-COSMOS AGN, but saw no correlation of these timescales with any physical parameters including black hole mass.

Many arguments related to characteristic timescale involve the question of whether or not AGN are scaled-up models of accreting stellar-mass black holes; i.e., whether accretion is a universal process across a huge range of masses and relativistic geometries. \citet{McHardy2006} found that after accounting for accretion rate differences, the characteristic timescales in X-ray light curves of stellar mass black holes and AGN are very well correlated with mass, implying a universal accretion scenario. Indeed, the X-ray literature has been far more successful in finding correlations with physical parameters than optical studies. \citet{McHardy2004} found that X-ray break timescales correlated with black hole mass all the way from Cygnus~X-1 to several Seyfert galaxies, although within the Seyfert galaxies there was no correlation. With a larger sample, \citet{Gonzalez-Martin2012} found that out of 104 XMM power spectra of AGN, 15 were best-fit by broken power laws with characteristic timescales that correlated well with black hole mass. \citet{Kelly2013} found that the amplitude of the high-frequency X-ray power spectrum showed a significant anticorrelation with mass, and could predict masses that agreed well with reverberation mapping values for the same objects. The fact that these relationships have been much less obvious in the optical variability may again speak to the presumably low degree to which optical variations are driven by reprocessing of X-ray variations, or may be due to the fact that the existing body of optical light curves and timing surveys is much more varied than the space-based X-ray light curves of only a few satellites.

Recently, \citet{Scaringi2015} has proposed that the best predictor of characteristic timescales is a combination of the mass, accretion rate, and size of the accreting object. They show that a relationship with the form $\mathrm{log}~\nu_b = A~\mathrm{log}R + B~\mathrm{log} M + C~\mathrm{log} \dot M + D$ is able to correctly predict $\nu_b$ for accretors ranging from young stellar objects (YSOs) and white dwarfs to AGN (using $R_\mathrm{ISCO}$ with a spin parameter $a=0.8$ as the ``size" of the black holes). According to the results of their grid search, the size is actually the most important parameter. Motivated by this and the desire to determine which of the physical disk timescales best matches our observed break timescales, we have calculated an effective size of the portion of the accretion disk probed by the \emph{Kepler} light curves, and compared the relationship of our observed timescales with these radii to those expected for the physical timescales described above. Because calculating a characteristic radius for the very broad bandpass of \emph{Kepler} is somewhat uncertain, we proceed in two ways. 

First, we treat the disk locally as a blackbody, and use Wien's Law $\lambda = b / T$ to calculate the temperature at the disk radius probed by \emph{Kepler}, using $\lambda = 6600$\AA, the central wavelength of the bandpass. We then use the following equation relating disk temperature and radius from \citet{Peterson1997}:

\begin{equation}
T(r) \approx 6.3\times10^5 \Big(\frac{\dot M}{\dot M_\mathrm{Edd}} \Big)^{1/4} \Big(\frac{M}{10^8M_\odot}\Big)^{-1/4} \Big(\frac{R}{R_S} \Big)^{-3/4} \mathrm{K}
\end{equation}

An alternative way to calculate a characteristic radius is used by \citet{Mushotzky2011}, based on the expression for the effective size of the region emitting light at a given frequency from \citet{Baganoff1995}:

\begin{equation}
\frac{R_\nu}{R_S} = 7.5\times10^{23} \epsilon^{-1/3} \nu^{-4/3} \Big(\frac{M}{M_\odot}\Big)^{-1/3} \Big(\frac{L}{L_\mathrm{Edd}}\Big)^{1/3},
\end{equation}

where $\epsilon$~is the accretion efficiency. Using this relation will predict shorter timescales for the same radius and black hole mass, because it takes into account light from the inner, hotter regions of the disk contributing to the \emph{Kepler} bandpass. However, we note that both the known effect of atmospheric scattering, which requires a color correction that reddens the emission, and the observational evidence (i.e., the 1000\AA~break seen in quasar spectra) for truncated disks in which hot inner regions do not contribute much to the optical light, would move the effective radius closer to our original, more naive estimate.


\begin{figure}[t]
    \centering
    \includegraphics[width=0.5\textwidth]{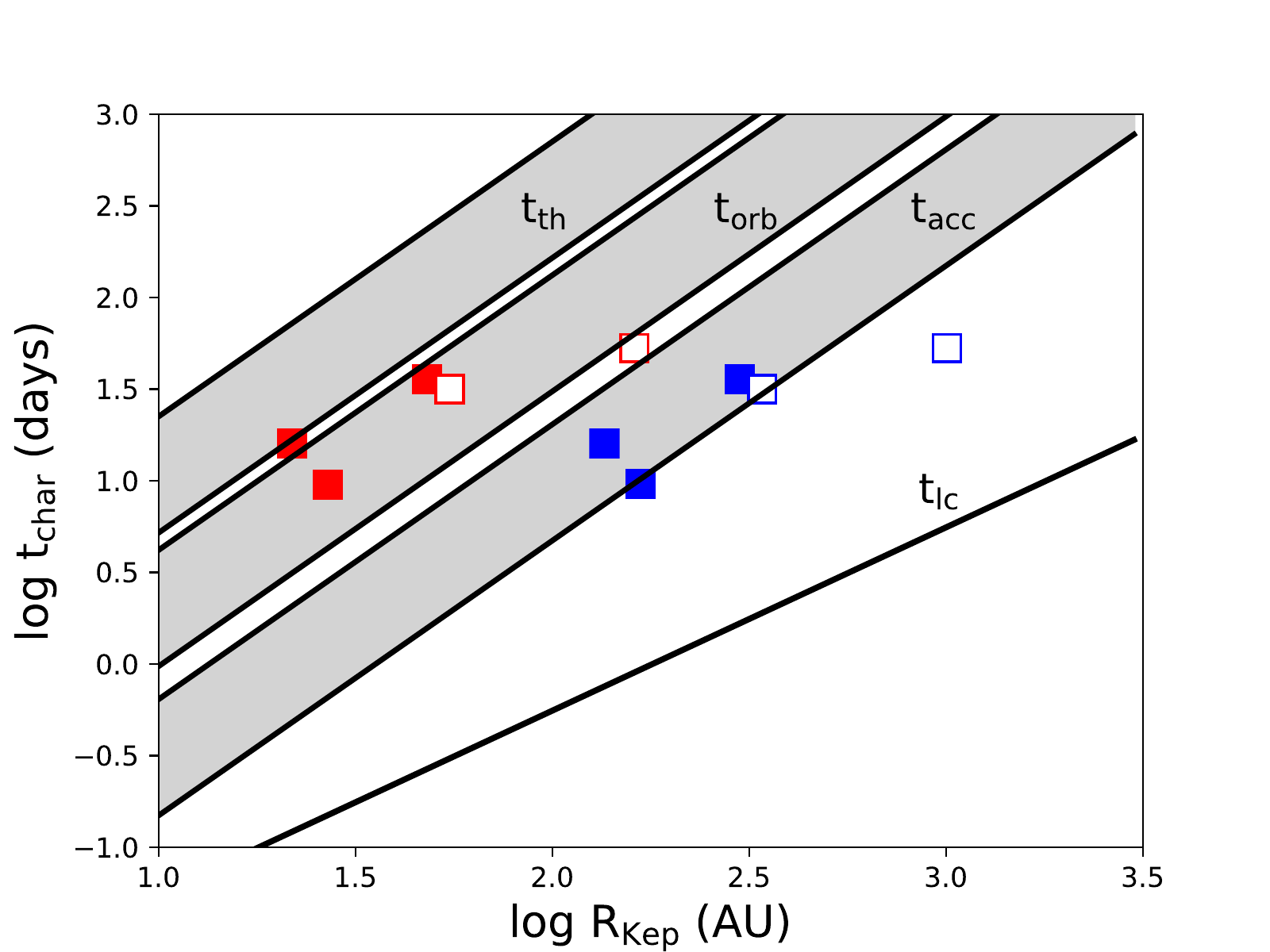}
    \caption{The four break timescales (solid squares) and the break timescales for KIC~9650712 and KIC~12158940 (hollow squares) in the plane of physical timescale and effective radius of the disk probed by the \emph{Kepler} bandpass. Blue symbols are for radii calculated using Equation~4, and red squares for Equation~5. Hollow symbols are used for the objects for which the break may either be interpreted as a QPO instead, or the break is in doubt due to the CBV application, as discussed in the text.}
    \label{fig:tcharrkep}
\end{figure}

In Figure~\ref{fig:tcharrkep}, we show the regions in the $\tau_\mathrm{char} -$radius plane described by each physical timescale for the range of masses in our broken-PSD sample. If calculated using our blackbody estimation and Equation~4, the timescales are most consistent with the freefall or ADAF accretion timescale. If they are calculated using the effective emitting region size with Equation~5, they are more consistent with orbital timescales. This would be in agreement with the orbital timescale consistency found by \citet{Collier2001}. 

We conclude by noting that, as \citet{Scaringi2015} points out, accreting white dwarfs and stellar-mass black holes have shown identical break timescales in their optical and X-ray PSDs. This has not been observed in AGN, but this dataset would be a good candidate for X-ray timing follow up. Many X-ray break timescales in the AGN literature are on the same order as our optical timescales, such as those reported by \citet{McHardy2006} and \citet{Gonzalez-Martin2012}. However, some of the \citet{Gonzalez-Martin2012} timescales are on the order of minutes to hours, far shorter than we could have seen with these data (white noise begins to dominate our PSDs at around 1~day timescales). Whether or not the X-ray and optical characteristic timescales are identical in a given object would do much towards identifying and disentangling the various sources of variability.


\section{Summary}
\label{sec:summary}

We have analyzed a sample of 21 \emph{Kepler} light curves of Type~1 AGN using a customized pipeline and Fourier techniques. Our results are as follows.

1. We have found that the high-frequency PSD slopes are largely inconsistent with the value of $\alpha=-2$ predicted by damped random walk models, in agreement with other studies.\\
2. Despite the possibility that the \emph{Kepler} space-based optical light curves would be more consistent with X-ray timing studies than optical ground-based surveys, we do not see the lognormal flux distributions or linear rms-flux relationships that characterize X-ray AGN light curves. This holds true even for the highest $F_X/F_O$ objects in our sample, indicating that X-ray reprocessing is unlikely to be a large contributing factor to the 0.1-10\% optical variability of AGN.\\
3. Some of our light curves exhibit bimodal flux distributions, transitioning between what appear to be fixed flux levels. This is possibly the signature of preferred accretion states, but may also be an indication of obscuring material passing along the line of sight. \\
4. The \emph{Kepler} light curves uphold the general anticorrelation between bolometric luminosity and variability, but do not show an increased variability with redshift. This latter is probably due to the fact that we do not have enough high-redshift AGN to test this properly.\\
5. Bolometric luminosity is weakly correlated with the shallowness of the high-frequency PSD slope.\\
6. Black hole mass is positively correlated with variability for low Eddington-ratio objects, but there is no correlation with high Eddington objects or overall. The low Eddington correlation weakens with increasing light curve bin size (i.e., probing variability on longer timescales).\\
7. Six of our AGN show statistically significant PSD flattening, with characteristic break timescales ranging from 9 to 53 days. The black hole mass roughly correlates with these timescales, and objects with break timescales also have the steepest PSD slopes.\\
8. The characteristic timescales are most consistent with orbital or freefall / ADAF accretion timescales, depending on how the characteristic radius of the disk at the observed wavelength is calculated. \\

The results of \emph{Kepler} timing studies were hampered by the premature end of the primary mission. Ongoing space-based optical timing missions like K2 \citep{Howell2014} and TESS \citep{Ricker2016} will have similar cadences to \emph{Kepler} but with significantly shorter monitoring baselines. The data shown in this study can be used to understand how much the statistical properties of variability in K2 and TESS AGN light curves is affected by the shorter baseline, and potentially act as a bridge between K2/TESS and upcoming ground-based surveys like LSST. At the very least, studies of the \emph{Kepler} and upcoming space-based AGN light curves show us that AGN vary at an astounding diversity of timescales and amplitudes, and that with the confluence of all types of surveys in the optical and X-ray, accretion physics will soon be accessible and testable in an entirely unprecedented way.

\acknowledgments

KLS is grateful for support from the NASA Earth and Space Sciences Fellowship (NESSF), which enabled the majority of this work. Support for this work was also provided by the National Aeronautics and Space Administration through Einstein Postdoctoral Fellowship Award Number PF7-180168, issued by the Chandra X-ray Observatory Center, which is operated by the Smithsonian Astrophysical Observatory for and on behalf of the National Aeronautics Space Administration under contract NAS8-03060. KLS also acknowledges Tod Strohmayer and Nathan Roth for helpful discussions regarding QPOs and AGN flares.

\small
\bibliographystyle{apj}
\bibliography{biblio3}
\newpage

\appendix
\section{Appendix A: The Light Curves }
\label{appendixa}
We present here the full sample of AGN light curves. We display them two ways: first, on different flux axes that are best for each object, to allow distinct features to be easily seen. Second, we show each light curve on identical axes: the range of the ordinate is always 30\% of the mean flux ($\pm$15\% in each direction), and the abscissa is 1200 rest-frame days, which is the length of our longest light curve. This allows the amount of variability and the baselines to be compared at-a-glance. Each plot displays flux in units of counts sec$^{-1}$ versus time in rest-frame days.

\begin{figure}[htb]
\caption{Light Curves of the \emph{Kepler} AGN}
\label{fig:lightcurves}
\centering

   \includegraphics[width=0.9\textwidth]{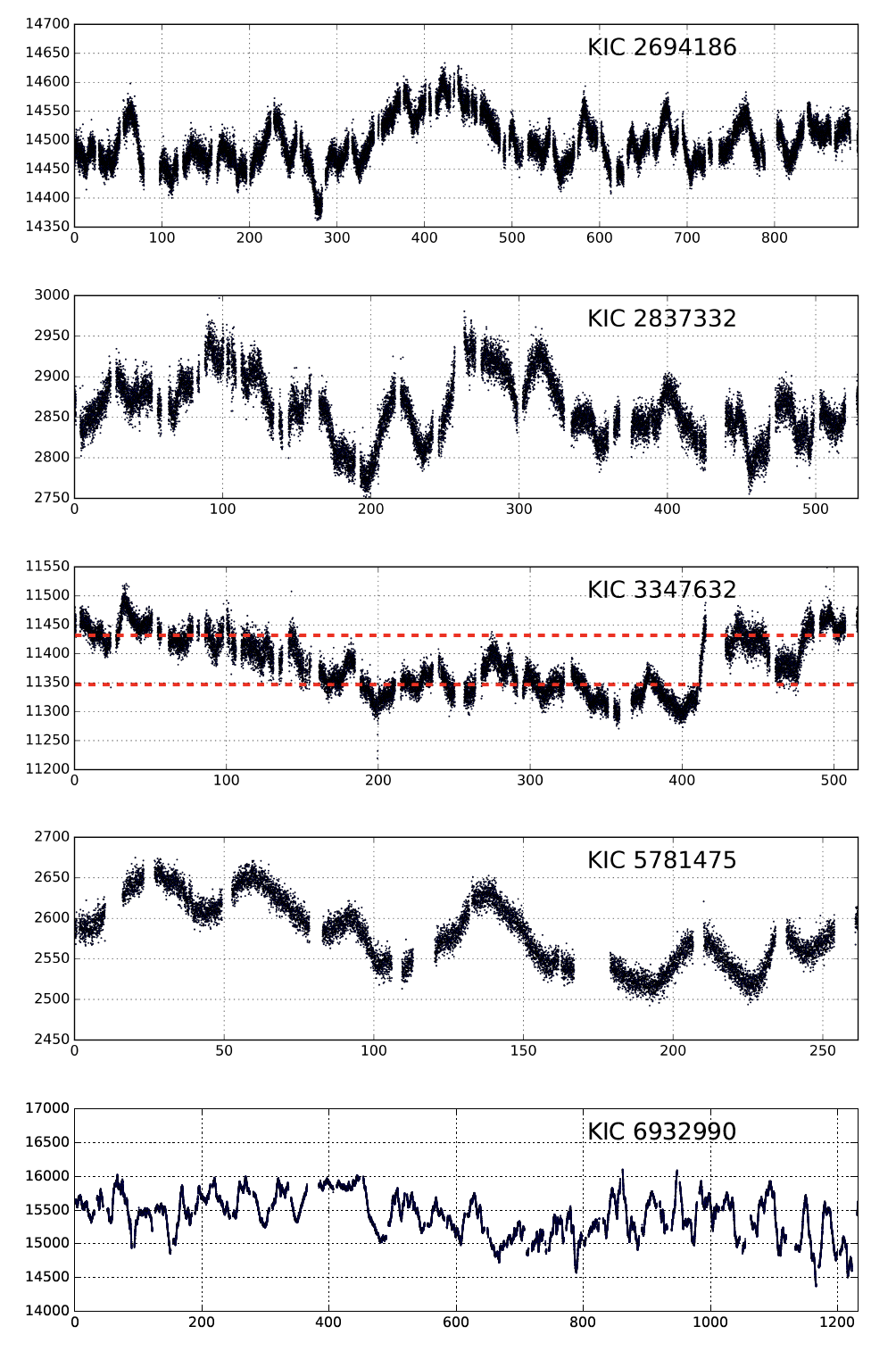}

\end{figure}

\begin{figure}[htb]\ContinuedFloat
\centering

   \includegraphics[width=0.9\textwidth]{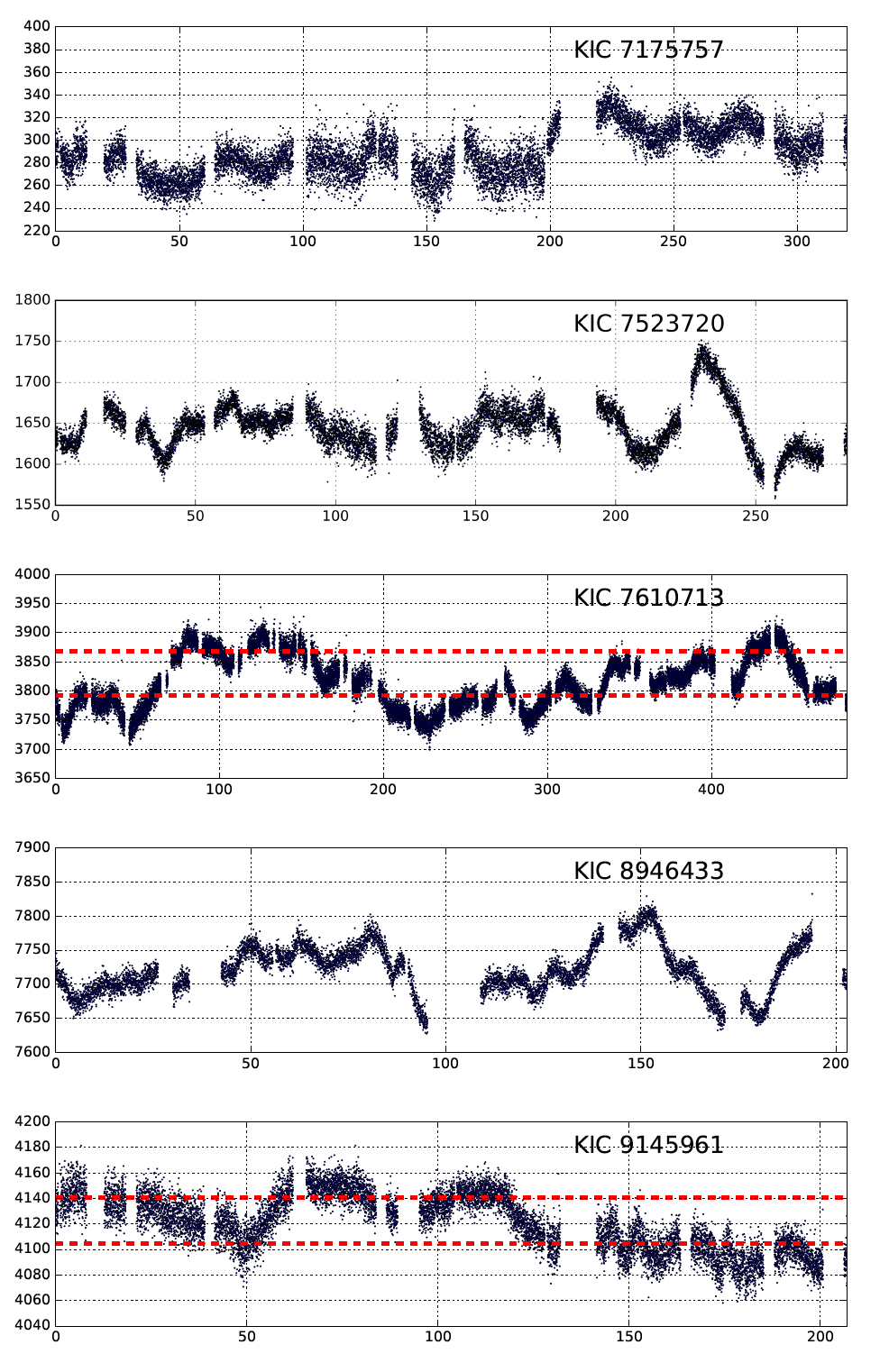}

\end{figure}

\begin{figure}[htb]\ContinuedFloat
	\centering	
 
   \includegraphics[width=0.9\textwidth]{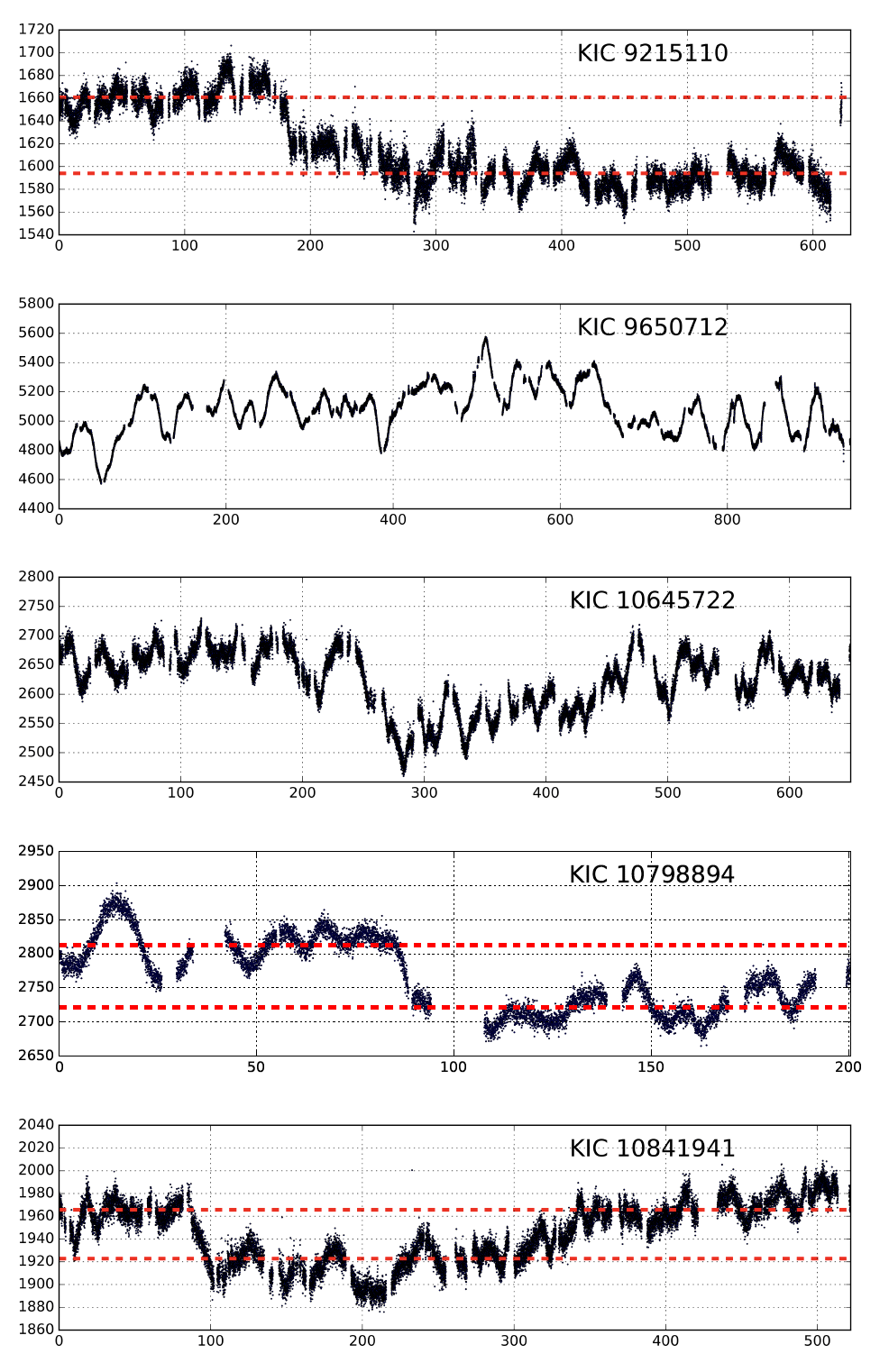}

\end{figure}

\begin{figure}[htb]\ContinuedFloat
	\centering

   \includegraphics[width=0.9\textwidth]{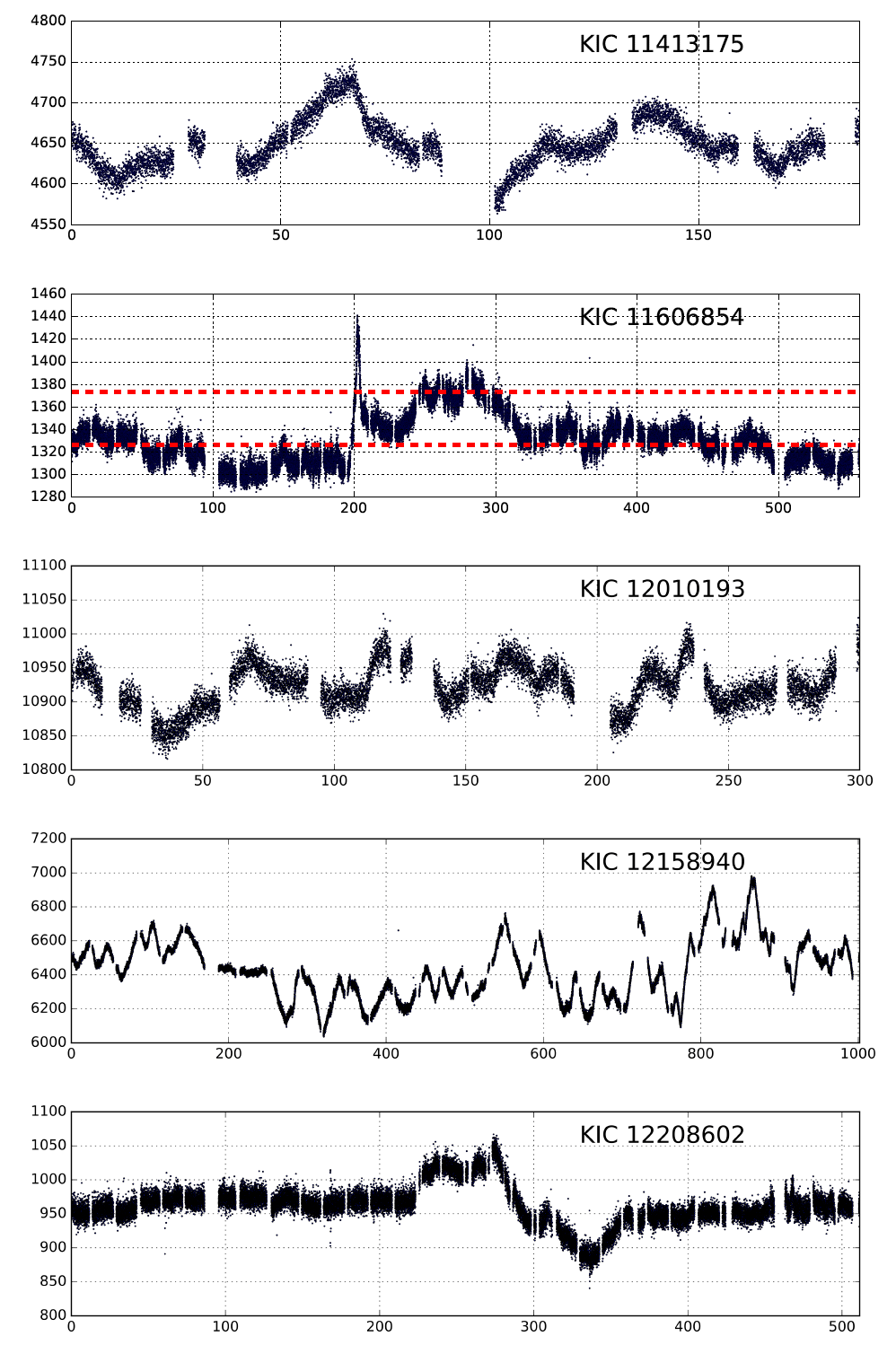}

\end{figure}

\begin{figure}[htb]\ContinuedFloat
	\centering

   \includegraphics[width=0.9\textwidth]{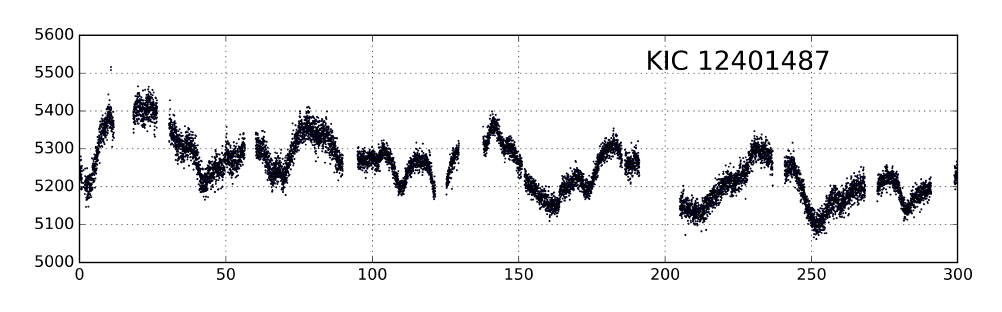}

\end{figure}

\begin{figure}[htb]
\caption{Light Curves of the \emph{Kepler} AGN on Identical Time and Flux Axes}
\label{fig:1200d}
\centering

   \includegraphics[width=0.9\textwidth]{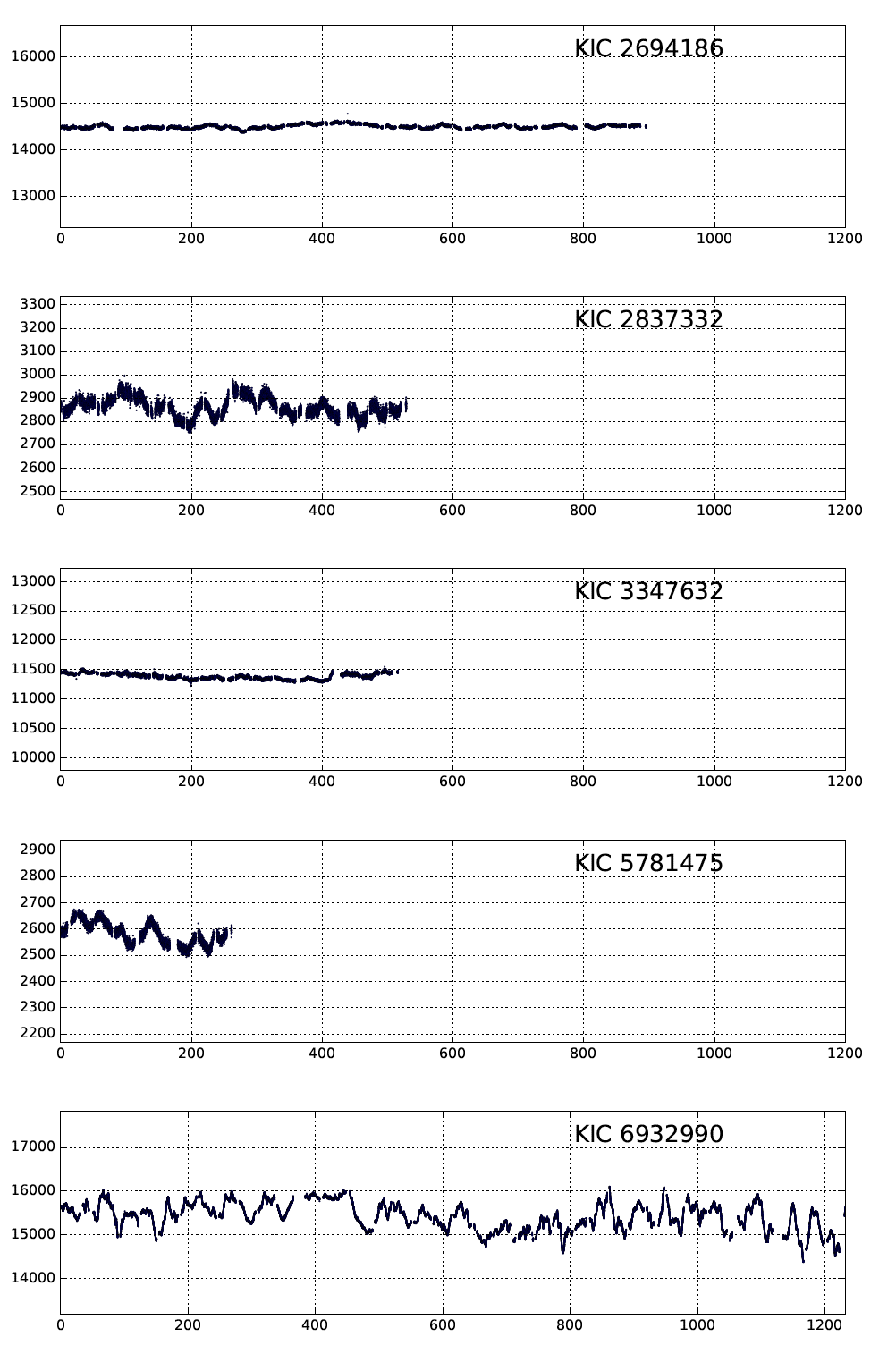}

\end{figure}

\begin{figure}[htb]\ContinuedFloat
	\centering

   \includegraphics[width=0.9\textwidth]{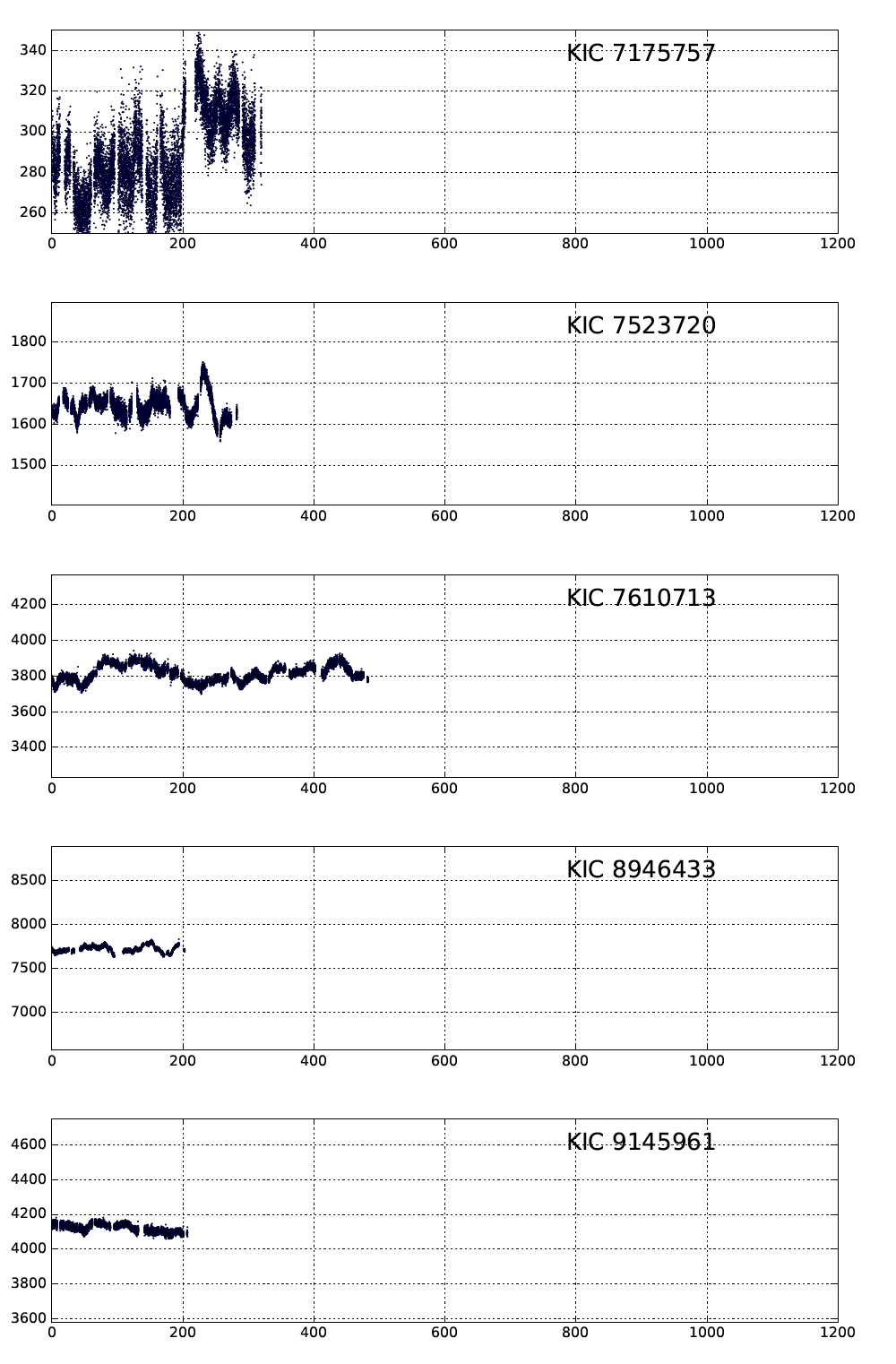}

\end{figure}

\begin{figure}[htb]\ContinuedFloat
	\centering	
	
   \includegraphics[width=0.9\textwidth]{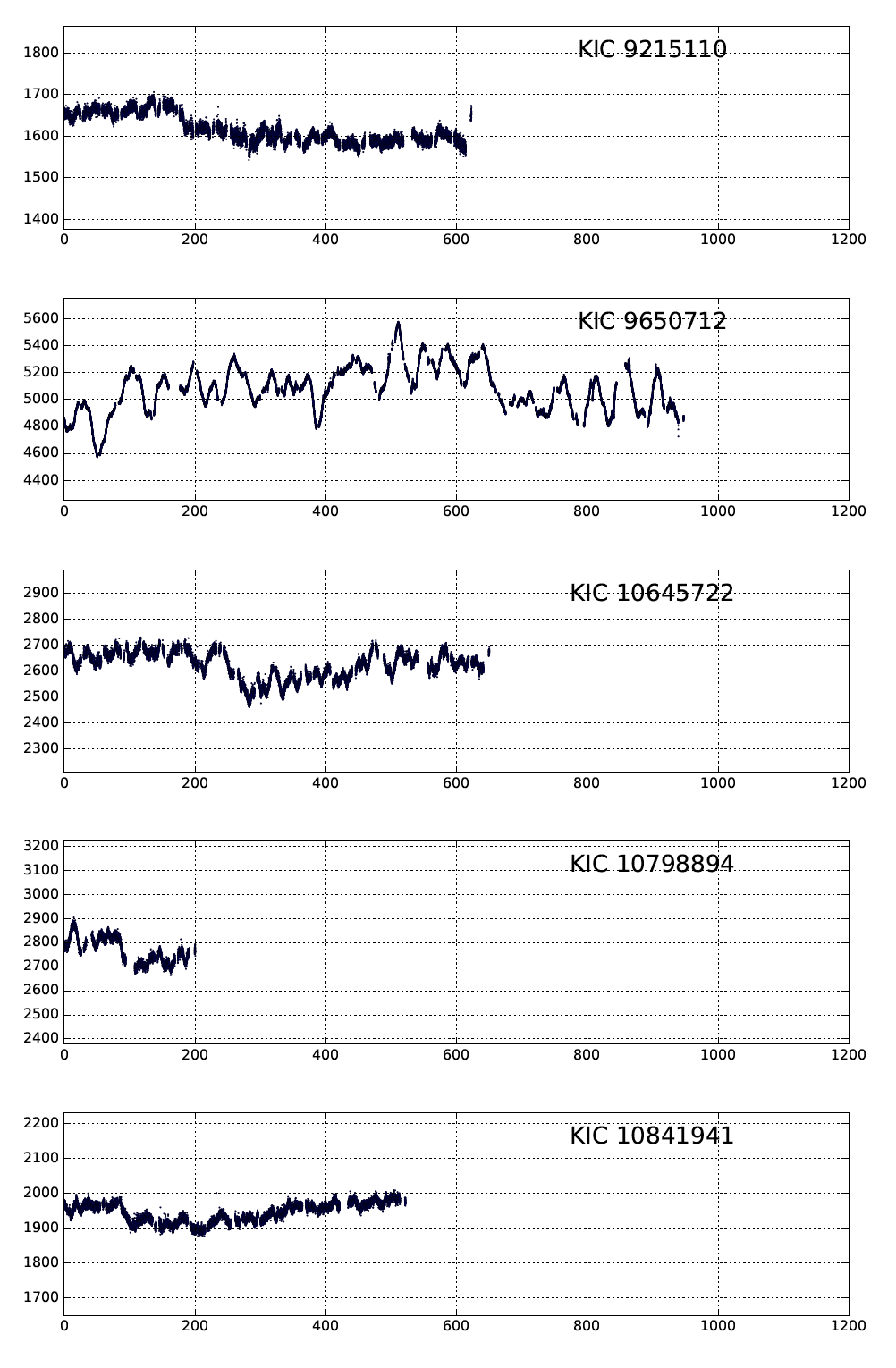}

\end{figure}

\begin{figure}[htb]\ContinuedFloat
	\centering
	
   \includegraphics[width=0.9\textwidth]{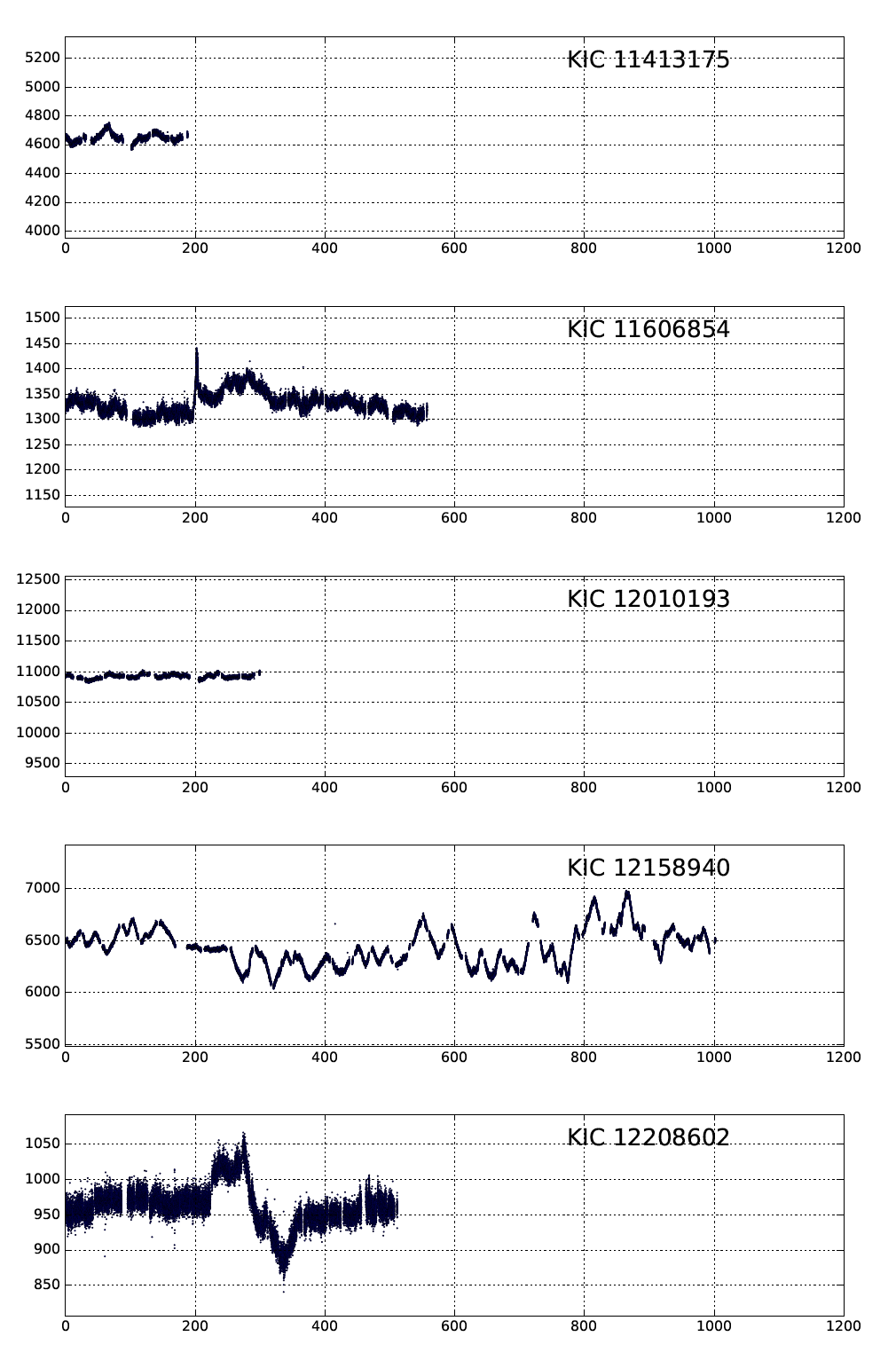}

\end{figure}

\begin{figure}[htb]\ContinuedFloat
	\centering

   \includegraphics[width=0.9\textwidth]{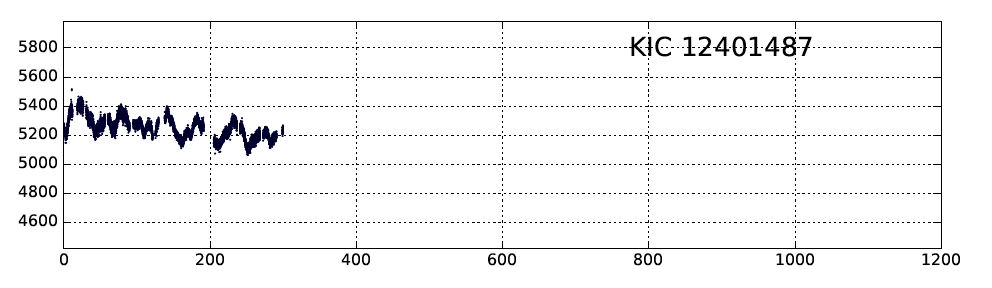}

\caption{Light curves of the \emph{Kepler} AGN with identical flux and time axes. Each plot displays flux in units of counts sec$^{-1}$ and time in rest-frame days.}
\end{figure}
\clearpage
\newpage
\section{Appendix B: Effects of Detrending using Cotrending Basis Vectors}
\label{appendixb}

As mentioned in the main text, spacecraft systematics in the \emph{Kepler} data make analyzing stochastically-variable sources very challenging. Long-term trends occurring on timescales of $\sim$~hundreds of days are clearly extant and nonphysical (see, for example, Figure~\ref{fig:startest}.) Such trends exist in almost every star, and so there must be some attempt to remove these trends before light curves can be used for astrophysics. In an ideal situation, simultaneous ground-based photometry would serve as a calibration source, verifying whether or not a long-term rise or fall in flux is intrinsic or due to spacecraft drift. However, since the majority of these AGN were not known before the launch of \emph{Kepler}, Zw~229-015 (KIC~6932990) is unique in this sample for its simultaneous ground-based monitoring (from the Lick AGN Monitoring Project; \cite{Barth2011}), which allows us to determine whether or not the application of CBVs has overlapped with and removed real signal. 

Applying CBV corrections to AGN is an admittedly dangerous enterprise. With the \emph{Kepler} data, however, we are unfortunately trapped into the conundrum of either abandoning the data entirely for the 20 other AGN, or trying to walk the line between systematics removal and affecting the science.

The only hope for simultaneous monitoring for the other 20 AGN was the public data release of the Palomar Transient Factory \citep[PTF;][]{Law2009}. Only two (one of which is Zw~229-015) have PTF light curves with more than one epoch overlapping the \emph{Kepler} data. These can serve as anchor points to determine whether true behavior was removed by the CBV application or not. In the case of Zw~229-015, the PTF light curve and the light curve obtained by the Robotically-Controlled Telescope (RCT) monitoring shown in \citet{Williams2015} match better the pre-CBV light curve. However, in KIC~9650172, the only other case with simultaneous monitoring, the CBV treatment clearly removed a dramatic flux increase due to spacecraft systematics. Both cases are shown in Figure~\ref{fig:ptfkep}. So, the application of CBVs is necessary and not always deleterious. One can see for oneself that the ``bumps and wiggles" on short timescales are almost always preserved, while long-term trends are mitigated. However, as we will see shortly, long-term trends are not always mitigated by CBV removal; i.e., intrinsic long-term behavior can survive the CBV process if it does not have the misfortune (as it evidently did in Zw~229-015) to coincide with the spacecraft behavior.

In the absence of ground-based simultaneous monitoring, the only way forward to is to investigate the effect of CBV removal on each object, determining the robustness of the results to the detrending process, and, if a result is changed after detrending, considering whether or not the result was spurious or is a physical result only detectable after systematics have been removed.
\begin{figure*}[h!]

   \includegraphics[width=\textwidth]{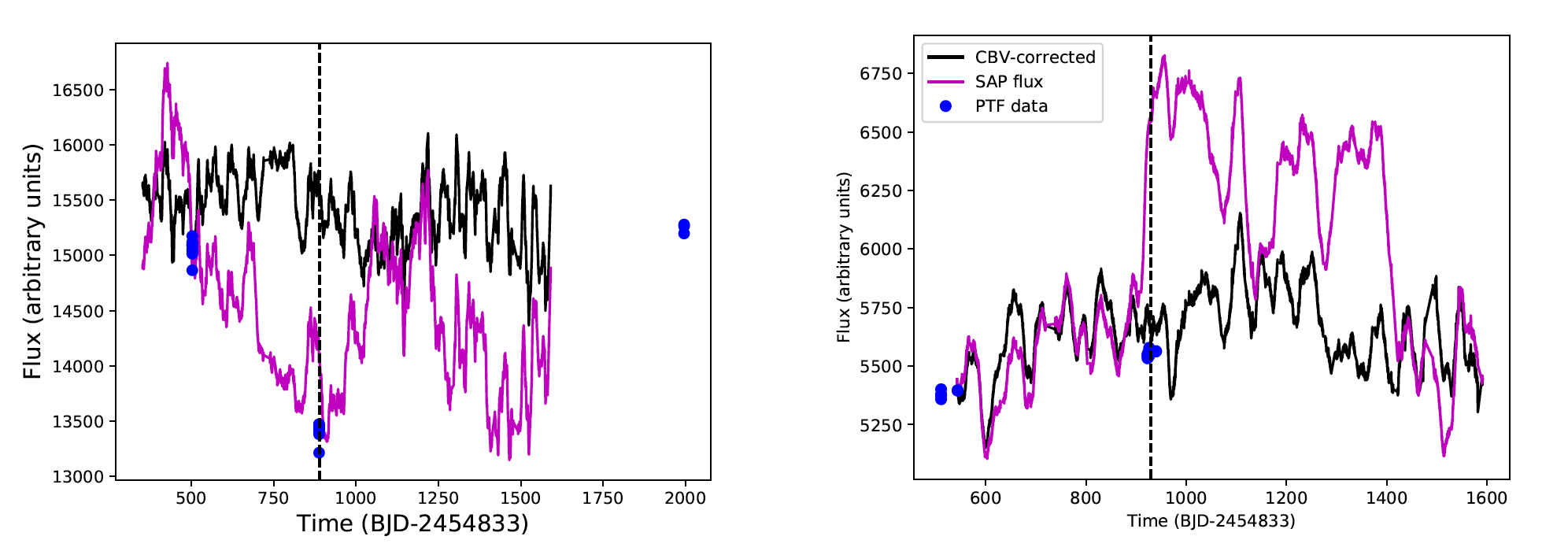}

\caption{Light curves of Zw229-015 (KIC~6932990) and KIC~9650712 with CBV corrections applied (black) and without (magenta). The Palomar Transient Factory light curves are shown in blue. Note that other pipeline steps, such as large aperture extraction and thermal corrections, have been applied in both cases. Black dashed lines show the median location of the second batch of PTF data points. The light curves have been arbitrarily normalized to match at the mean flux of the first set of PTF points.}
\label{fig:ptfkep}
\end{figure*}


To this end, we display in Figure~\ref{fig:cbvcomp} the light curves and power spectra before and after CBV removal for all of our objects. Because of the abundance of large figures elsewhere, the sizes are condensed, but the purpose of this figure is merely to indicate that sometimes there are massive, clear systematics in the pre-CBV light curves which are removed (e.g., KIC~2694186), sometimes there is behavior that may or may not be intrinsic that is inadvertently removed (e.g., 9215110), and sometimes the CBV treatment has almost no effect at all (e.g, KIC~8946433). This last situation presumably occurs because there are no major systematic effects on the relevant portion of the CCD during the monitoring baseline.

\begin{figure*}

   \includegraphics[width=\textwidth]{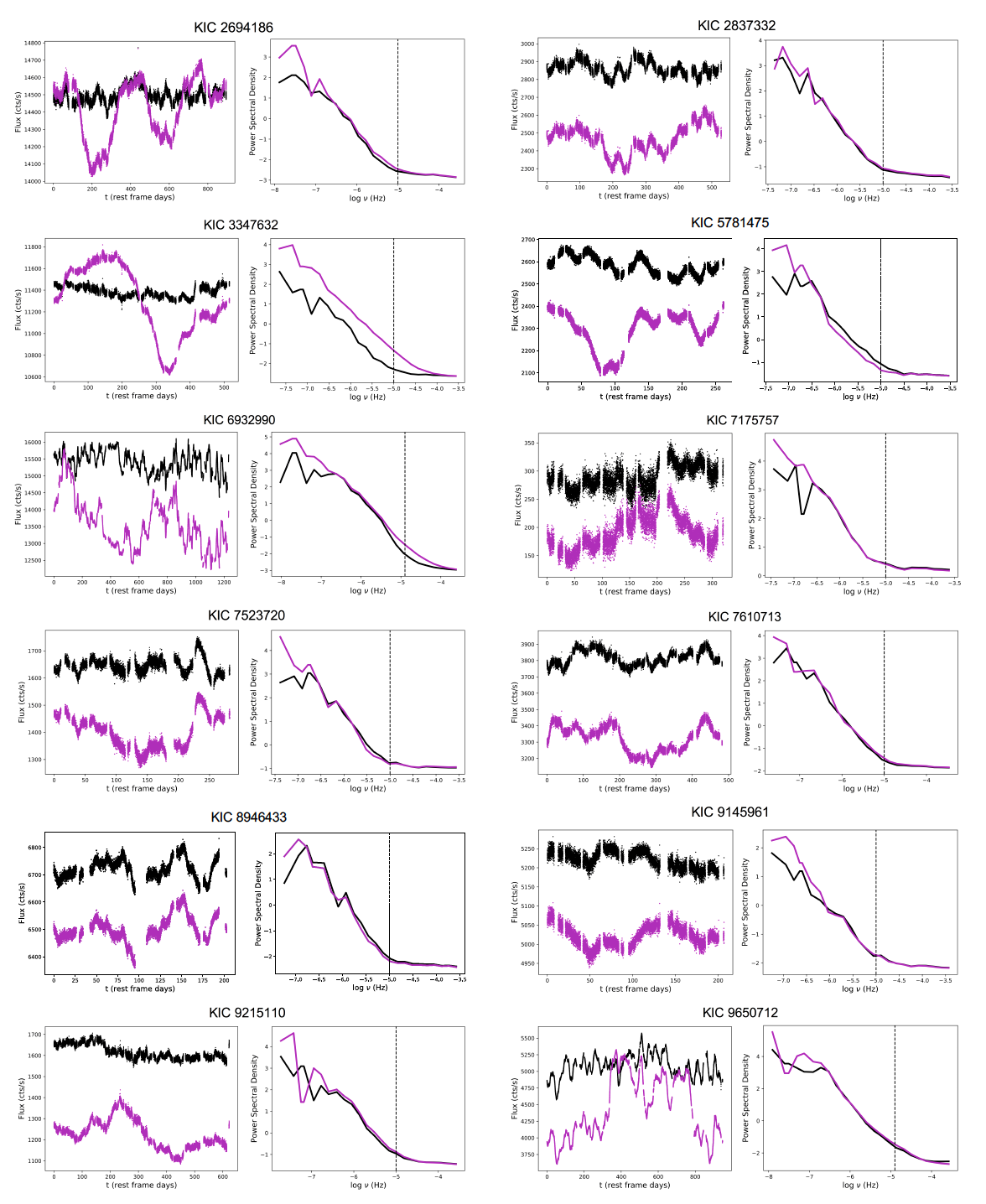}

\caption{Continued on next page...}
\label{fig:cbvcomp}
\end{figure*}

\begin{figure*}[htb]\ContinuedFloat

   \includegraphics[width=\textwidth]{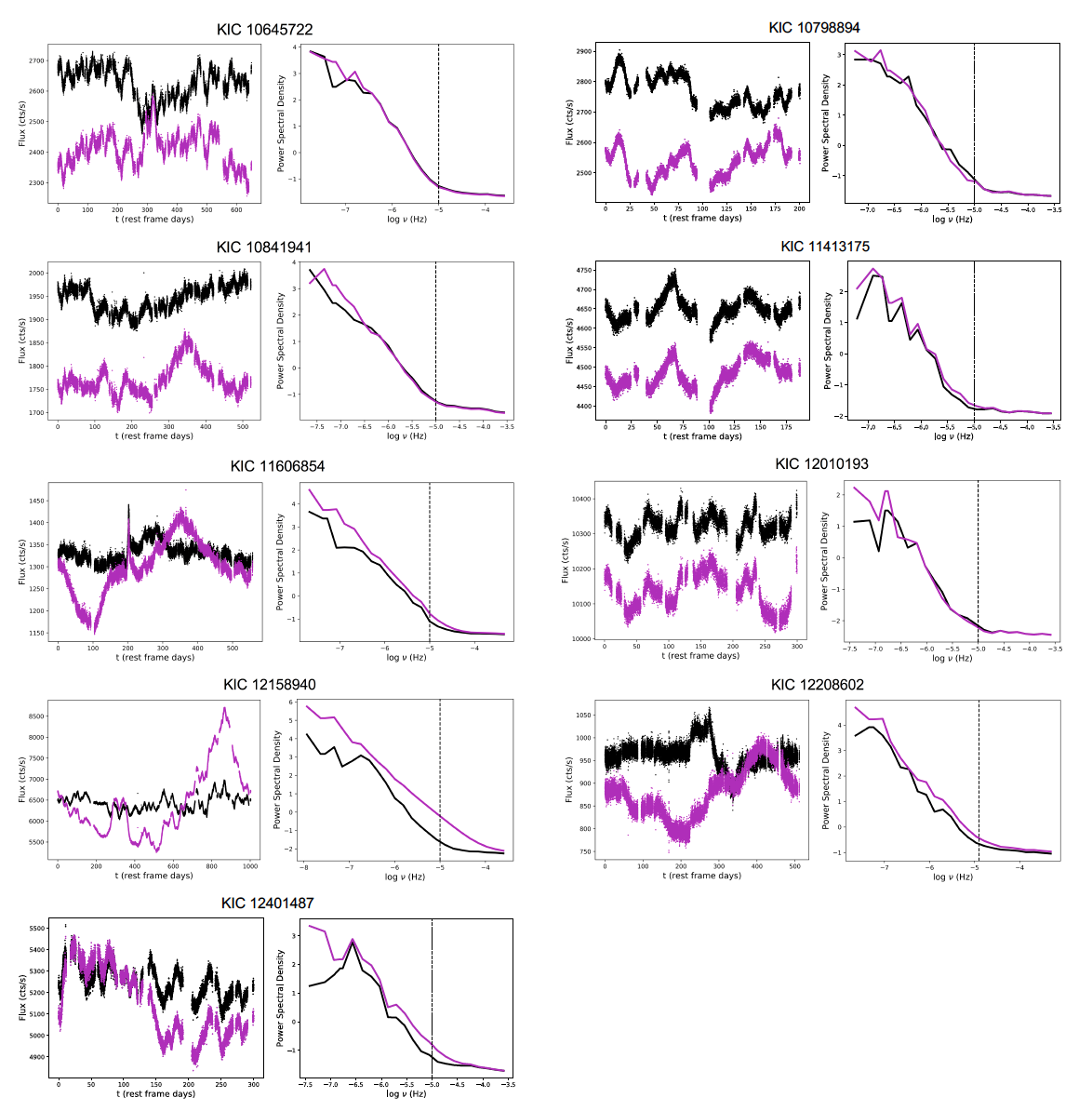}

\caption{Light curves and power spectra before (magenta) and after (black) application of the first two cotrending basis vectors.}
\label{fig:cbvcomp}
\end{figure*}

\begin{figure*}

   \includegraphics[width=\textwidth]{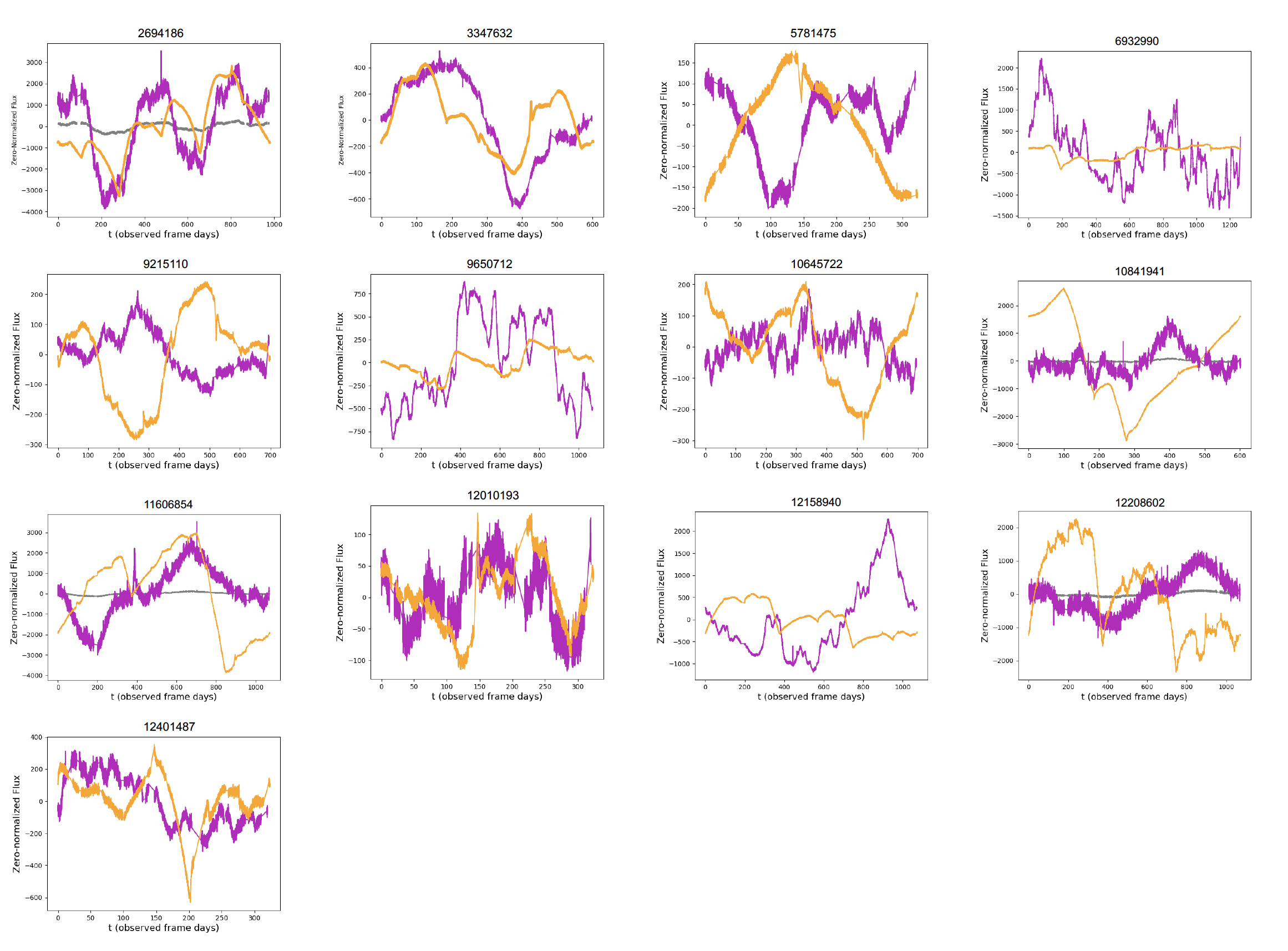}

\caption{Uncorrected light curves of the thirteen objects with light curves and power spectra most affected by the CBV application, as shown in Figure~\ref{fig:cbvcomp} (pink), with the light curve of the nearest star with \emph{Kepler} monitoring (orange). In four cases (KIC~2694186, 10841941, 11606854, and 12208602) the nearest star was much brighter than the object, so we show the object's light curve in grey and then multiply the object's flux by a factor of 10 (then show it in pink) for easier visual comparison. }
\label{fig:starcomp_appendix}
\end{figure*}

There is one further possibility for determining whether or not the behavior removed from the light curve was intrinsic to the source or due to the spacecraft. If there is a non-variable star sufficiently nearby, we may extract its light curve using the exact same method as for the AGN (including thermal corrections, etc. but not creating large custom apertures, as stars are point sources) and compare the long-term behavior. If the same trends are seen in a nearby star, it must be that they are due to spacecraft systematics. However, such systematics can be highly dependent on detector location. If a star is further away than a minute or two of arc, it may suffer from quite different systematic behavior, so this test can only be done if a source is within 1-2~arcmin of a non-variable star with 30-minute cadence monitoring from \emph{Kepler}'s original stellar search target list. For the thirteen sources in Figure~\ref{fig:cbvcomp} with power spectra visibly affected by the CBV application, we have located the nearest star with 30-minute cadence and processed them without CBVs. In some cases, the nearest star was too intrinsically variable to be a reliable calibrator, so we had to go further afield. The comparison of the uncorrected AGN light curves and their nearest monitored stars are shown in Figure~\ref{fig:starcomp_appendix}. We note that in many cases the nearby star's behavior indeed mimics the gross, long-term behavior of the source, indicative of systematics. There are also cases where there do not appear to be strong systematics affecting the star, which correspond to sources like Zw229-015 (KIC~6932990). In two cases (KIC~9215110 and 5781475) the stellar behavior seems to be a reflection of the AGN behavior. This may be due to the systematic moving across the detector, or a coincidence. Occasionally, the nearest star does exhibit strong systematics, but they do not mimic the source behavior at all (KIC~12208602). This is not necessarily surprising, as again, the systematics are highly dependent on source position. The stellar behavior should also not be mistaken for the actual CBV's variation, as the CBVs are constructed using a large ensemble of nearby stars and correct only for their co-varying behavior. However, such comparisons allow us to see for certain that many trends that the CBV correction has removed (like the periodicity in KIC~2694186) are definitively systematic.

The upshot of this analysis is as follows: we see that indeed, the CBVs are both effective in removing otherwise intractable spacecraft systematics and dangerous if the systematic behavior mimics the object's intrinsic variations. If we determine that the two objects for which nearby stellar behavior is opposite of the perceived AGN behavior as an instance in which the CBVs are fitted incorrectly (i.e., removing similar behavior as the ensemble stars but with negative fitting coefficients which ideally would not be permitted by \emph{Kepler}'s CBV tools), then KIC~9215110 and KIC~5781475 are likely to have been wrongly corrected. However, if the reflective stellar behavior is a coincidence, they join many other objects as uncertain. On the basis of the above, we make the following classifications, with KIC~9215110 and KIC~5781475 in parentheses in both categories:\\

\noindent
Not strongly affected: KIC~2837332, 7175757, 7523720, 7610713, 8946433, 9145961, 10798894, 11413175\\
Probably accurately corrected: KIC~2694186, 3347632, 9650712, 11606854, 12010193, 12401487\\
Accuracy of correction undetermined: (KIC~5781475), (9215110),10645722, 10841941, 12158940, 12208602\\
Probably inaccurately corrected: (KIC~5781475), KIC~6932990, (KIC~9215110)\\

\emph{Characteristic timescales:} In cases where significant low-frequency behavior is removed, it is possible that the power spectrum could be artificially flattened and introduce a characteristic timescale. From the analysis in this appendix we can see that 1) this does not always happen, even in sources with rapid high-frequency variability and significant low-frequency modification; 2) even significant flattening does not lead to breaks that pass our statistical test; and 3) because of this, objects with statistically-significant power law breaks are likely to be real, and are simply made easier to detect by this process. A series of examples makes this more clear. In KIC~2694186, there is rapid short-term variability and a steep high-frequency slope. The removal of a large amount of (spurious) power at low frequencies flattens the power spectrum, but the object does not pass our test for the significance of a power spectral break. This is also the case in KIC~3347632, KIC~5781475, KIC~7523720, KIC~10645722, and KIC~10841941. So we can see that removal of long-term trends does not introduce statistically significant breaks as a matter of course, even when significant power is removed. In KIC~6932990 (Zw229-015), the already-known break is detected in our light curve with greater significance than in past studies, which reported only a slightly better fit with a bending power law. In this object, \emph{real} variability was almost certainly removed, but only the existing break was recovered. The uncorrected light curve does not actually pass the \citet{Uttley2002} test, despite being the light curve in which the bending power law was discovered in past papers, so the \citet{Uttley2002} method is stricter. Because of what happened with Zw229-015, there is some basis for accepting breaks that are significant in CBV-treated light curves but insignificant in the original. Of our five other broken power-law objects, both KIC~9215110 and KIC~9650712 pass the \citet{Uttley2002} test either way. The remaining three (KIC~12158940, 12010193 and 12401487) pass the test only after CBV application. In the latter two cases, the behavior removed by the CBV correction is clearly coincident with that of the nearby star, so we can safely assume it is not intrinsic to the object. The major concern is with KIC~12158940. The nearest star is too far away to be reliably used as a calibrator, and in any case does not vary coincidentally with the source, and the light curves are significantly different before and after CBV application both in terms of characteristic timescale and high-frequency slope. It is possible that there is a characteristic timescale in this object's high-frequency variations, and perhaps it would be a coincidence for this to match reasonably well with the mass and the expected orbital timescale of this object (see Figure~\ref{fig:mbhtchar} and Figure~\ref{fig:tcharrkep}), but it is possible that this is not a real break. Accordingly, a hollow symbol is used in plots involving $\tau_\mathrm{char}$~ so that it can be discarded visually if desired. \\

\begin{figure*}
\begin{tabular}{cc}

\includegraphics[width=0.5\textwidth]{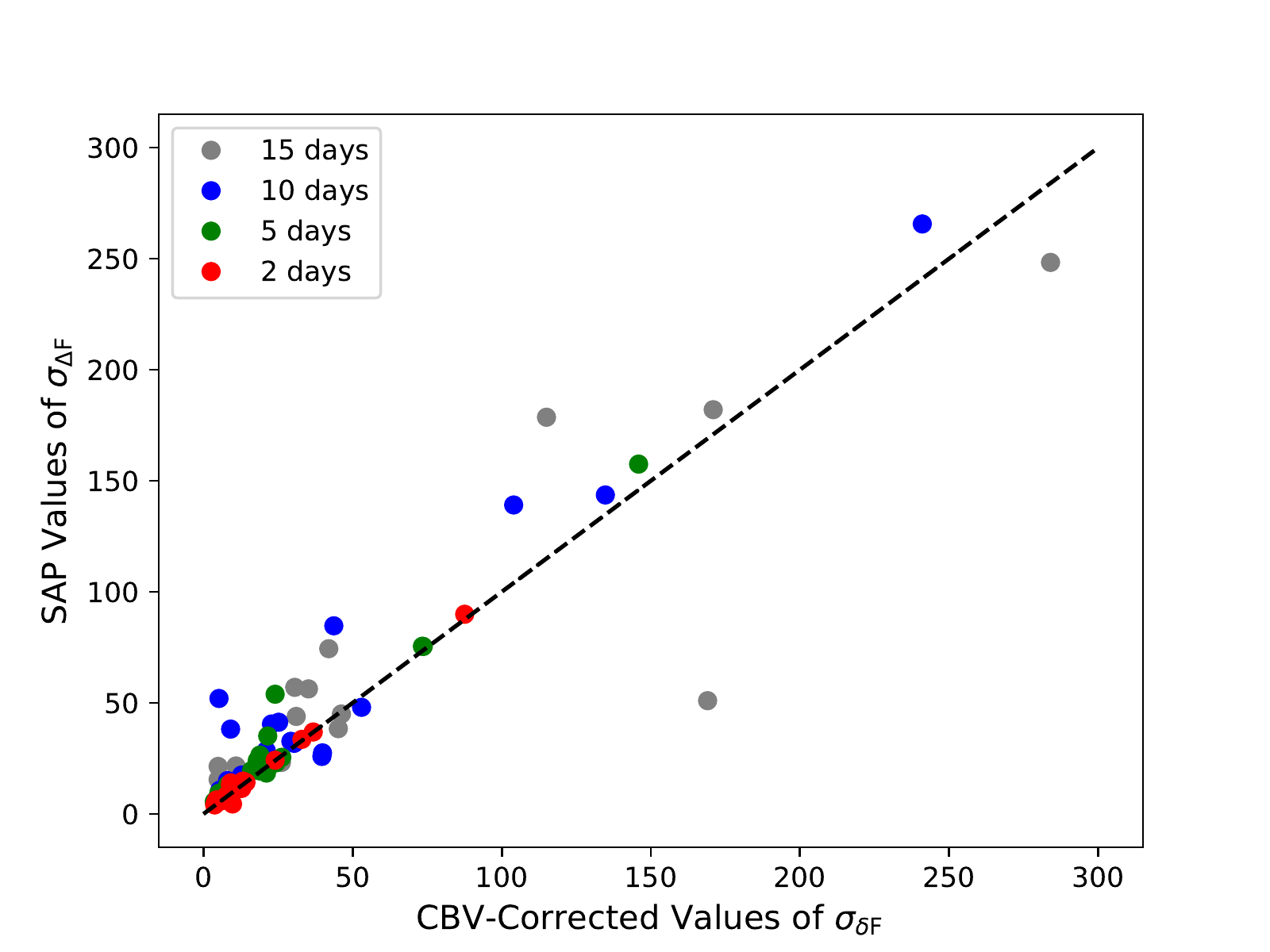} & \includegraphics[width=0.5\textwidth]{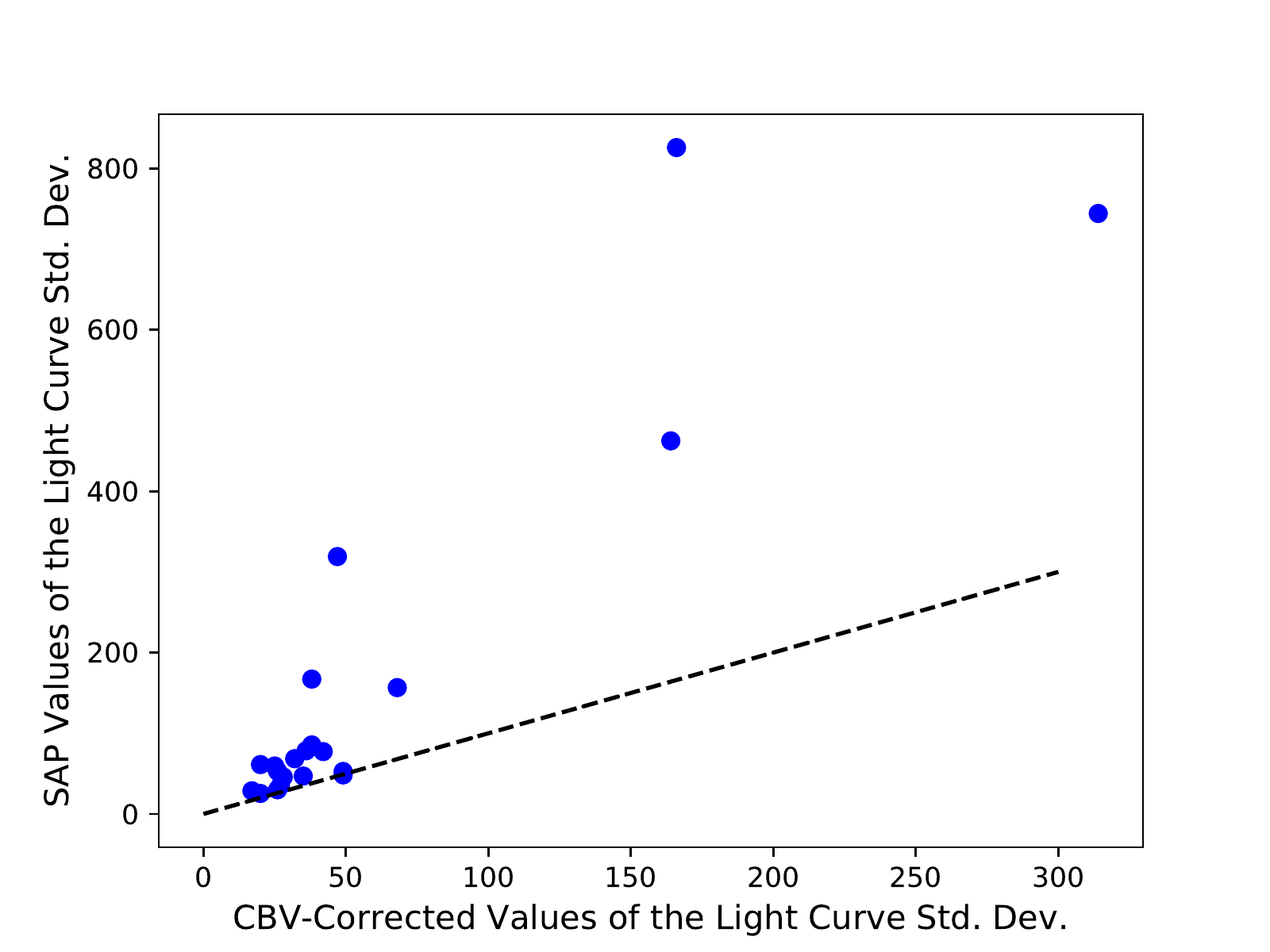} \\

\end{tabular}
\caption{Comparison of the values of variability metrics $\sigma_{\Delta\mathrm{F}}$ (left) and the standard deviation of the 2-day binned light curve (right) between CBV-corrected and uncorrected light curves. Colors in the left plot correspond to the binning of the light curve. }
\label{fig:lcvarcomp}
\end{figure*}

\emph{Variability Metrics:} The metrics we used to measure the variability of the light curve in our search for correlations with physical parameters are described in Section~\ref{mvar}. In short, they are the width of the distribution of the difference between subsequent flux values, $\sigma_{\Delta\mathrm{F}}$, for a light curve binned in 2, 5, 10, and 15 day timescales, and the standard deviation of the flux values in the 2-day binned light curve. Because the trends removed by the CBVs are on the order of hundreds of days, they have little effect on the values of $\sigma_{\Delta\mathrm{F}}$, especially for the 2-day and 5-day binned light curves. Of course, the standard deviation of the entire light curve is significantly higher in the non-CBV corrected light curves. The effects are shown in Figure~\ref{fig:lcvarcomp}. We know that these values are wrong without CBV-correction in many of our light curves (one need only look at instances like KIC~2694186), so it is not prudent to use the uncorrected values of the standard deviations in any parameter searches. Since the 2-day binned $\sigma_{\Delta\mathrm{F}}$ metric is not strongly affected, we leave our findings in Figures~\ref{fig:zstd}, \ref{fig:lbolvar}, and \ref{fig:mbhstdgrid} as-is. The right-hand panel of Figure~\ref{fig:lbolvar} shows the standard deviation relationship with bolometric luminosity, which has the same general sense as the relationship between $\sigma_{\Delta\mathrm{F}}$ and bolometric luminosity. We did not find any other correlations with the light curve standard deviation (as stated in Section~\ref{mvar}), and so none are presented in this work. \\

\emph{Flux Distributions and RMS-Flux Relations}: As has already been demonstrated, the majority of objects appear to have been corrected somewhat accurately or not strongly affected. However, in objects where the correction has either definitely or possibly removed real variability, the flux histograms are perhaps the most clearly affected result, as unlike the high-frequency slopes or short-timescale variability metrics, they take into account large discrepancies in flux over long timescales. Readers should take this into account when evaluating the histograms in Figure~\ref{fig:histo}. Four of the bimodal objects are either not affected or are most likely accurately corrected; the other two are uncertain. The RMS-flux relations remain nonexistent even in the uncorrected light curves (as can generally be seen by eye when inspecting Figure~\ref{fig:cbvcomp}; the sources are not more variable when brighter).

\end{document}